\documentclass[
prd %, twocolumn%
,showpacs, amssymb, superscriptaddress, aps
]{revtex4}
\usepackage{graphicx}
\input{epsf}

\usepackage{amsmath,amssymb}
\usepackage{bm}
\usepackage{times}

\newcommand{\dalm}{\kern1pt\vbox{\hrule height 0.9pt\hbox{\vrule width 0.9pt
\hskip 2.5pt\vbox{\vskip 5.5pt}\hskip 3pt\vrule width 0.3pt}\hrule height 0.3pt}
\kern1pt}

\newcommand{\be}{\begin{eqnarray}}
\newcommand{\ee}{\end{eqnarray}}
\newcommand{\beq}{\begin{eqnarray}}
\newcommand{\eeq}{\end{eqnarray}}

\begin{document}

%\twocolumn[\hsize\textwidth\columnwidth\hsize\csname @twocolumnfals\endcsname

% For two column
%\wideabs{

%\title{Probing geometry of gravitational fields via pulse profile from pulsar}
%\title{Can we distinguish gravitational theories by observing  pulse profile of neutron stars?}
\title{Sensitivity of pulsar light curves to spacetime geometry and efficacy of analytic approximations}

\author{Hajime Sotani}
\email{sotani@yukawa.kyoto-u.ac.jp}
\affiliation{Division of Theoretical Astronomy, National Astronomical Observatory of Japan, 2-21-1 Osawa, Mitaka, Tokyo 181-8588, Japan}

\author{Umpei Miyamoto}
%\email{sotani@yukawa.kyoto-u.ac.jp}
\affiliation{Research and Education Center for Comprehensive Science, Akita Prefectural University, Akita 015-0055, Japan}

\date{\today}

% Abstract
\begin{abstract}
In order to examine the pulse profile from a pulsar, we derive the formula for describing the flux from antipodal hot spots with any static, spherically symmetric spacetime. We find that the pulse profiles are almost independent of the gravitational geometry outside the star when the compactness of neutron stars is low enough, e.g., the stellar mass and radius are $1.4M_\odot$ and 14 km, respectively. On the other hand, the pulse profiles depend strongly on the gravitational geometry when the compactness of neutron stars is so high, e.g., the stellar  mass and radius are $1.8M_\odot$ and 10 km, respectively. Thus, one may probe the spacetime geometry outside the star and even distinguish gravitational theories via the observation of pulse profile with the help of another observations for the stellar compactness, if the compactness of central object is high enough.  We also derive the 1st and 2nd order approximation of the flux with respect to a parameter defined by the radio of the gravitational radius of considered spacetime to the stellar radius. Then, we find that the relative error from full order numerical results in the bending angle becomes $\sim 20-30\%$ with the 1st order and $\sim 5-10\%$ with the 2nd order approximations for a typical neutron star, whose mass and radius are $1.4M_\odot$ and 12 km, respectively. Our results with the 1st order approximation for the Schwarzschild spacetime are different from those obtained in the literature, which suggests that the 1st order approximation has been misunderstood to yield highly accurate prediction.
\end{abstract}

\pacs{04.80.Cc, 95.30.Sf}
%
%%%%%%%%%%%%%%%%%%%%%%%%%%%%%%%%%%%%%%%%%%%%%%%%%
%  04.40.Dg :  Relativistic stars: structure, stability, and oscillations (see also 97.60.-s Late stages of stellar evolution) 
%  04.40.Nr  :  Einstein-Maxwell spacetimes, spacetimes with fluids, radiation or classical fields
%  04.50.Kd : Modified theories of gravity
%  04.70.-s   : Physics of black holes
%  04.80.Cc : Experimental tests of gravitational theories
%  95.30.Sf : Relativity and gravitation (95.30.-k : Fundamental aspects of astrophysics)
%  26.60.Kp : Equations of state of neutron-star matter
%%%%%%%%%%%%%%%%%%%%%%%%%%%%%%%%%%%%%%%%%%%%%%%%%
%]
% For two column
%}

\maketitle

%\baselineskip 24pt
%%%%%%%%%%%%%%%%%%%%%%%%%%%%%%%%%%%%%%%%%%%%%%%%
\section{Introduction}
\label{sec:I}
%%%%%%%%%%%%%%%%%%%%%%%%%%%%%%%%%%%%%%%%%%%%%%%%

Neutron stars, which are formed via core-collapse supernovae, involve various extreme environmental conditions. The density inside neutron stars can significantly exceed the standard nuclear density; the stellar magnetic fields may become stronger than the critical field strength of the quantum electrodynamics; and the gravitational field around/inside the star is quite strong \cite{shapiro-teukolsky}. Thus, the observation of the phenomena associated with neutron stars and/or of the neutron star itself is quite important for understanding the physics under such extreme conditions. Due to the development of observation technology, it is becoming possible to observe such a phenomenon with high precision. In particular, the verification of theories of gravity in a strong gravitational field is one of the most important tasks in modern physics. Up to now, there is no experiments and observations, which indicate a defect of the general relativity, but most of them have been done in a weak gravitational field, such as the solar system. Perchance, the gravitational theory might deviate from the general relativity in the strong field regime. If so, one can probe the theory of gravity via astronomical observations associated with the compact objects (e.g., \cite{Berti2015,DP2003,SK2004,Sotani2014,Sotani2014a}).

The pulse profile from a pulser is one of the useful astronomical information for seeing the gravitational geometry outside the star. The path of the light emitted from the stellar surface can bend due to the general relativistic effect, where the light bending strongly depends on the stellar compactness, i.e., the ratio of the stellar mass to the radius. So, via the observations of pulse profile, one can make a constraint on the mass and radius of the neutron star, which may enable us to determine the equation of state for neutron-star matter \cite{POC2014,Bogdanov2016}. Such an attempt may come true soon by the upcoming Neutron Star Interior Composition Explorer (NICER) X-ray timing mission \cite{NICER}. The observed pulse profile depends on the angle between the magnetic and rotational axes together with the angle between the rotational axis and the direction of observer, assuming the polar cap models. Once these angles are fixed, the pulse profile is determined by the numerical integration \cite{PFC1983}. To easily determine the pulse profile, several approximate relations for the light bending have also been suggested in the Schwarzschild spacetime \cite{LL95,Beloborodov2002,PB06,FFS16}. Furthermore, the rotational effects are also taken into account in the pulse profile \cite{CLM05,PO2014}.

Such analyses have been done in the Schwarzschild spacetime, but the pulse profile may be modified if the gravitational geometry is different from the Schwarzschild spacetime because the light bending should also depend on the gravitational geometry. 
In order to examine how the pulse profile depends on the gravitational geometry, in this paper we calculate the flux radiated from the pulser and compare the profiles of pulse radiated from various geometries. 
In the analysis, we will assume that the way that light couples to the spacetime geometry remains unchanged and light continues to move on null geodesics, even though the theory of gravity is changed. One of the advantages to consider the pulse profile (or the light bending) from a pulser is that such a property can be discussed just via the light path outside the star, assuming the stellar mass and radius. That is, one can avoid the physical uncertainties inside the star. In particular, we consider the neutron star models whose radius and mass are respectively in the range of $10-14$ km and $1.4-1.8M_\odot$, which are typical neutron star models. To see the dependence on the gravitational geometry, we especially adopt the Schwarzschild, the Reissner-Nordstr\"{o}m, and the Garfinkle-Horowitz-Strominger \cite{GHS1991} spacetimes. In this paper, we adopt geometric units, $c=G=1$, where $c$ and $G$ denote the speed of light and the gravitational constant, respectively, and the metric signature is $(-,+,+,+)$.

%%%%%%%%%%%%%%%%%%%%%%%%%%%%%%%%%%%%%%%%%%%%%%%%
\section{Photon trajectory and deflection angle}
\label{sec:II}
%%%%%%%%%%%%%%%%%%%%%%%%%%%%%%%%%%%%%%%%%%%%%%%%

We begin with obtaining the photon trajectory from the stellar surface of a neutron star. The metric describing the static, spherically symmetric spacetime is generally given by 
\begin{equation}
%  ds^2 
	g_{\mu\nu} dx^\mu dx^\nu
	= -A(r)dt^2 + B(r)dr^2 + C(r)\left(d\theta^2 + \sin^2\theta d\psi^2\right).   \label{eq:metric}
\end{equation}
In particular, we focus on the asymptotically flat spacetime, i.e., $A(r)\to 1$, $B(r)\to 1$, and $C(r)\to r^2$ as $r\to \infty$. 
We remark that the most general stationary, spherically symmetric spacetime has only two functional degree of freedom, i.e., one can choose the radial coordinate in such a way that $C(r)=r^2$ \cite{Wald0}. However, we adopt the metric form given by Eq. (\ref{eq:metric}), so that our formalism would be applicable to wider class of stationary, spherically-symmetric spacetimes. In fact, the metric components are difficult to express explicitly with the coordinate system in which $C(r)=r^2$ for a class of static, spherically symmetric spacetimes. In this case, the circumference radius, $r_c$, is given by $r_c^2 :=C(r)$.

On such a spacetime, we consider the photon trajectory radiating from the surface of a compact object with radius $R$ and mass $M$, where $R$ is determined in coordinate $r$. Due to the nature of spherical symmetry, one can choose the coordinate where the photon trajectory is in the plane with $\theta=\pi/2$. The Lagrangian is expressed by
\begin{equation}
   {\cal L} = \frac{1}{2}g_{\mu\nu} u^\mu u^\nu,    \label{eq:LL}
\end{equation}
where $u^\mu := dx^\mu/d\lambda$ is the four-velocity of the photon with an affine parameter $\lambda$. For a massless particle, one can put ${\cal L}=0$. Adopting the metric form given by Eq.~(\ref{eq:metric}) for $\theta=\pi/2$, we obtain that
\begin{equation}
   2{\cal L} = -A\dot{t}^2 + B\dot{r}^2 + C\dot{\psi}^2,  \label{eq:L2}
\end{equation}
where the dot denotes the derivative with respect to the affine parameter. The photon trajectory is determined from the Euler-Lagrange equation, i.e.,
\begin{equation}
   \frac{\partial{\cal L}}{\partial x^\mu} - \frac{d}{d\lambda}\left(\frac{\partial{\cal L}}{\partial \dot{x}^\mu}\right)=0.  \label{eq:EL}
\end{equation}
Since we consider the static spherically symmetric spacetime, $\partial{\cal L}/\partial t = \partial{\cal L}/\partial \psi=0$. Thus, from Eq.~(\ref{eq:EL}), one obtains 
\begin{equation}
  \dot{t} = \frac{e}{A}\ \ {\rm and}\ \ \dot{\psi} = -\frac{\ell}{C},  \label{eq:el}
\end{equation}
where $e$ and $\ell$ are constants corresponding to the energy and angular momentum of photon, respectively. Combining Eqs.~(\ref{eq:L2}) and (\ref{eq:el}) with ${\cal L}=0$, one can get the equation for radial motion of (outward) photon,
\begin{equation}
   \dot{r} = \sqrt{\frac{1}{B}\left(\frac{e^2}{A}-\frac{\ell^2}{C}\right)}. \label{eq:radial}
\end{equation}
The radial dependence of angle is
\begin{equation}
   \frac{d\psi}{dr} = \frac{\dot{\psi}}{\dot{r}} = -\frac{1}{C}\left[\frac{1}{B}\left(\frac{1}{b^2A}-\frac{1}{C}\right)\right]^{-1/2},
\end{equation}
where $b$ is an impact parameter defined by $b :=\ell/e$, and the angle for the observer far from the central object is chosen as $\psi=0$. Thus, the angle at the stellar surface, $r=R$, is given by
\begin{equation}
   \psi(R) = \int_\infty^R \frac{d\psi}{dr} dr = \int_R^\infty \frac{dr}{C}\left[\frac{1}{AB}\left(\frac{1}{b^2}-\frac{A}{C}\right)\right]^{-1/2}.  \label{eq:psi}
\end{equation}

%%%%%%%%%%%%%%%%%%%%%%%%%%%%%%%%%%%
% Figure 1
%%%%%%%%%%%%%%%%%%%%%%%%%%%%%%%%%%%
\begin{figure}
\begin{center}
\includegraphics[scale=0.5]{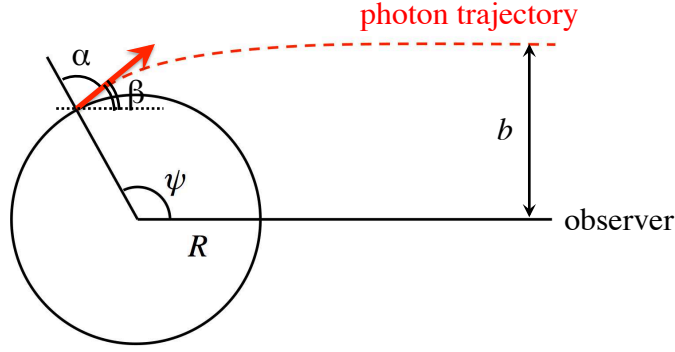} 
\end{center}
\caption{%%
Image of the photon trajectory from the stellar surface with the dashed line, where $R$ and $b$ denote the stellar radius and impact parameter. Photon radiates from the stellar surface with the emission angle $\alpha$ (between the local radial direction and radiating photon direction), where the radiation point on the surface is located with the angle $\psi$ with respect to the direction of observer. In addition, $\beta$ is the bending angle defined as $\psi-\alpha$.
}%%
\label{fig:photon}
\end{figure}
%%%%%%%%%%%%%%%%%%%%%%%%%%%%%%%%%%%

Now, we consider the angle between the local radial direction and the radiating photon direction at the stellar surface, i.e., the emission angle $\alpha$ as shown in Fig.~\ref{fig:photon}. The emission angle is given by $\tan\alpha = [(u^\psi u_\psi)/(u^ru_r)]^{1/2}$ at $r=R$ \cite{Beloborodov2002}, which leads to the relation that
\begin{equation}
   \sin\alpha = b\sqrt{\frac{A(R)}{C(R)}},  \label{eq:alpha}
\end{equation}
where we adopt Eqs. (\ref{eq:el}) and (\ref{eq:radial}) for $u^\psi$ and $u^r$, respectively. 
Using Eqs.~(\ref{eq:psi}) and (\ref{eq:alpha}), one can numerically obtain the relation between $\psi(R)$ and $\alpha$ for given $R$. That is, for given $R$, the value of $b$ is determined from Eq.~(\ref{eq:alpha}) if one chooses a specific value of $\alpha$, while the value of $\psi(R)$ can be determined from Eq.~(\ref{eq:psi}) with such a value of $b$. We remark that $\psi(R)=0$ when $\alpha=0$ by definition. 
On the other hand, one cannot numerically integrate Eq. (\ref{eq:psi}) up to infinity. So, we introduce $R_*$, which is an arbitrary constant being sufficiently large $R_* \gg R$ and we divide the integral interval into two parts, i.e., (1) from $R$ up to $R_*$ and (2) from $R_*$ up to $\infty$. In order to integrate the second part, we assume that $A(r)$, $B(r)$, and $C(r)$ can be expanded in spatial infinity as
\begin{eqnarray}
  A(r) &=& 1+ \frac{a_1}{r} + \frac{a_2}{r^2} + {\cal O}\left(\frac{1}{r^3}\right), \label{eq:ar}  \\
  B(r) &=& 1+ \frac{b_1}{r} + \frac{b_2}{r^2} + {\cal O}\left(\frac{1}{r^3}\right), \label{eq:br}  \\
  C(r) &=& r^2\left[1 + \frac{c_1}{r} + \frac{c_2}{r^2} + {\cal O}\left(\frac{1}{r^3}\right)\right],  \label{eq:cr}
\end{eqnarray}
where $a_1$, $a_2$, $b_1$, $b_2$, $c_1$, and $c_2$ are some constants depending on the considered spacetime. Here, we assume that the metric functions are real and analytic in a neighborhood of $1/r=0$, which excludes the possibility of Yukawa-like terms. Eventually, 
we integrate Eq.~(\ref{eq:psi}) as 
\begin{equation}
  \psi(R) = \int_R^{R_*}\frac{b\sqrt{AB}}{\sqrt{C(C-Ab^2)}}dr + b\left[\frac{1}{R_*} + \frac{a_1 + b_1 + 2c_1}{R_*^2} 
     + {\cal O}\left(\frac{1}{R_*^3}\right)\right].  \label{eq:psi1}
\end{equation}
For considering to the relation between $\psi(R)$ and $\alpha$, the case with $\alpha=\pi/2$ should be calculated separately, because the denominator of integrand in Eq.~(\ref{eq:psi1}) (or Eq.~(\ref{eq:psi})) becomes zero at $r=R$ for $\alpha=\pi/2$ from Eq. (\ref{eq:alpha}), where one cannot integrate numerically. Even so, it is known that this improper integral can be estimated to be finite (see Appendix \ref{sec:a1}). We remark that the stellar radius should be considered as a circumference radius, i.e., $R_c^2 :=C(R)$. Hereafter, we use $R$ as the stellar radius in the coordinate of $r$, and $R_c$ as a corresponding circumference radius.

%%%%%%%%%%%%%%%%%%%%%%%%%%%%%%%%%%%%%%%%%%%%%%%%
\section{Pulse profile}
\label{sec:III}
%%%%%%%%%%%%%%%%%%%%%%%%%%%%%%%%%%%%%%%%%%%%%%%%

We suppose the following situation: the observer is located in the direction of $\bm{D}$ from the central star; the distance between the central star and observer is $D := |{\bm D}|  \gg R$; $\phi$ is an azimuthal angle with respect to the direction of rotation around $\bm{D}$. With this setup, let us consider the photons radiated from the following surface element on the star,
\be
	dS=R^2\sin\psi\ d\psi\ d\phi.
\label{dS}
\ee
Then, these photons are observed in the following solid angle on the sky of observer at impact parameter $b$,
\be
	d\Omega = \frac{ b\ db\ d\phi }{ D^2 }.
\label{dOmega}
\ee
We suppose $0\le\psi\le \pi$ and $0\le\phi\le 2\pi$, even though $\psi$ may be large than $\pi$, depending on the stellar compactness. It should be noticed that the impact parameter $b$ depends on $\psi(R)$ but not on $\phi$. 

The observed radiation flux emitted from the surface element $dS$ is given by 
\be
	dF=Id\Omega,
\label{dF}
\ee
where $I$ is observed radiation intensity~\cite{Rybicki}. The observed frequency $\nu$, red-shifted due to the star's gravity, is related to the frequency at the stellar surface $\nu_0$ as $\nu/\nu_0 = \sqrt{g_{tt}(R)/g_{tt}(\infty)}=\sqrt{A(R)}$ \cite{Wald0}. Assuming the radiation to be thermal (blackbody), $I \propto ({\rm temperature})^4$ and is therefore related to the surface intensity $I_0$ as $I/I_0=\nu^4 / \nu_0^4$~\cite{Misner:1974qy}.
Thus, one obtains 
\be
	I = A(R)^2 I_0.
\label{II0}
\ee
We remark that the quantity of $I_\nu/\nu^3$ should be conserved along null rays, where $I_\nu$ is the specific intensity of radiation at a given frequency $\nu$ \cite{MTW}.
Eliminating $d\Omega$, $d\phi$, and $I$ from Eqs.~\eqref{dS}, \eqref{dOmega}, \eqref{dF} and \eqref{II0}, one obtains 
\begin{equation}
   dF = I_0 \frac{ A(R)C(R) }{ R^2 }  \cos\alpha \frac{d(\cos\alpha)}{d\mu} \frac{dS}{D^2}, 
  \;\;\;
   \mu := \cos\psi. 
 \label{eq:dF0}
\end{equation}

As in Ref. \cite{Beloborodov2002}, we adopt the pointlike spot approximation for simplicity. Namely, the spot area is assumed to be so small that the variables in Eq.~(\ref{eq:dF0}) are constant over the spot. In this case, one can calculate the flux $F=\int dF$ as
\begin{equation}
   F =  I_0 \frac{ s \cos\alpha}{D^2}  \frac{A(R)C(R)}{R^2} \frac{d(\cos\alpha)}{d\mu},
\label{eq:dF1}
\end{equation}
where $s :=\int dS$ is the spot area. Note that in the Newtonian limit, where there is neither bending of light (i.e., $ \alpha = \psi $) nor redshift (i.e., $A(R)=C(R)/R^2 = 1$), the right-hand side of Eq.~\eqref{eq:dF1} simply becomes the surface intensity $I_0$ multiplied by the solid angle of the hot spot viewed from the observer, $s \cos\alpha/D^2$. While $I_0$ can depend on ejection angle $\alpha$,  we hereafter assume the isotropic emission in a local Lorentz frame, i.e., $I_0 = {\rm const.}$ for simplicity, which corresponds to the $\psi$ independence because $\alpha$ and $\psi$ are in one-to-one correspondence. Then, we obtain the following final expression of the observed flux,
\begin{equation}
	  F = F_1 \cos\alpha\frac{d(\cos\alpha)}{d\mu},  
\;\;\;
	F_1 : = I_0 \frac{ s }{D^2}  \frac{A(R)C(R)}{R^2}.
\label{eq:dF2}
\end{equation}

\begin{figure}
\begin{center}
\includegraphics[scale=0.5]{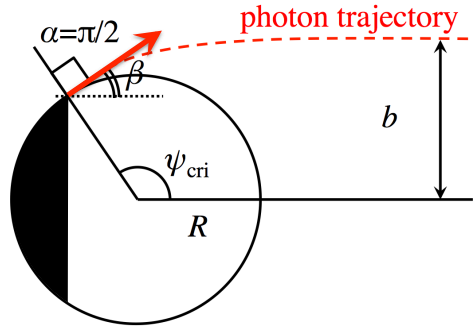} 
\end{center}
\caption{%%
Image for $\psi_{\rm cri}$, where $\alpha=\pi/2$, and invisible zone shown by the shaded region.
}%%
\label{fig:psi_cri}
\end{figure}

As shown in Fig.~\ref{fig:photon}, one might observe the photon radiated from the hot spot even with $\cos\psi<0$ due to the curvature produced by the strong gravity. We remark that the visible surface fraction for the flat spacetime is simply $1/2$, because the visible condition is just $\cos\psi>0$. The critical value of $\psi$, which determines the boundary between the visible and invisible zones, corresponds to the photon orbit with $\alpha=\pi/2$, as shown in Fig.~\ref{fig:psi_cri}. Hereafter, it is referred to as $\psi_{\rm cri}$. Additionally, the visible fraction of stellar surface is given by
\begin{equation}
   \frac{S_{\rm cri}}{4\pi R_c^2} = \frac{1-\cos\psi_{\rm cri}}{2},
\end{equation}
where $S_{\rm cri}$ denotes the area of visible zone. The value of $\psi_{\rm cri}$ increases, as the central object becomes more compact. Eventually, $\psi_{\rm cri}$ becomes $\pi$ with the specific compactness of the central object, where one can observe the hot spot even if it is completely opposite from the observer. Furthermore, $\psi_{\rm cri}$ might become more than $\pi$ if the central object would be so compact. So, one should set to be $S_{\rm cri}=4\pi R_c^2$ for considering the stellar models where $\psi_{\rm cri}$ becomes larger than $\pi$.

%%%%%%%%%%%%%%%%%%%%%%%%%%%%%%%%%%%
% Figure 2
%%%%%%%%%%%%%%%%%%%%%%%%%%%%%%%%%%%
%%%%%%%%%%%%%%%%%%%%%%%%%%%%%%%%%%%

\begin{figure}
\begin{center}
\includegraphics[scale=0.4]{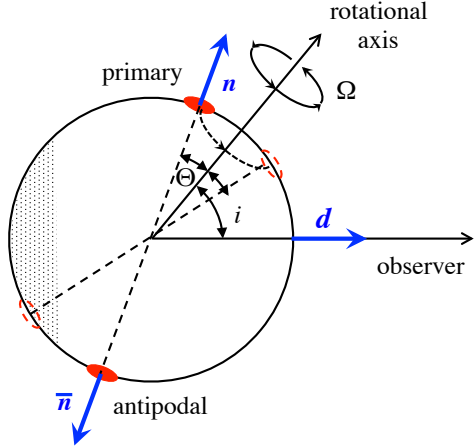} 
\end{center}
\caption{%%
Image of the hot spots on the rotating star with the angular velocity $\omega$. Two hot spots are associated with the magnetic polar caps, where the magnetic axis is inclined to the rotational axis with the angle $\Theta \in [0, \pi/2]$. The unit vector $\bm{d}$ denotes the direction of the observer, while the unit vectors $\bm{n}$ and $\bar{\bm n}$ are the normals on the primary and antipodal spots, respectively. The angle between $\bm{d}$ and the rotational axis is $i \in [0, \pi/2] $. 
}%%
\label{fig:pulsar}
\end{figure}

Now, as shown in Fig.~\ref{fig:pulsar}, we consider a neutron star with two antipodal hot spots, which may be associated with the polar caps of the stellar magnetic field. We basically adopt the same notations for the various angles as in Ref. \cite{Beloborodov2002}. The angle between the direction of the observer from the center of star and the rotational axis is $ i \in [0, \pi/2]$. The angle between the rotational and magnetic axes is $ \Theta \in [0, \pi/2]$, where we identify that the hot spot closer to the observer is ``primary" and the other spot is ``antipodal". With the unit vector pointing toward the observer $\bm{d}:=\bm{D}/D$ and the normal vector at the primary spot $\bm{n}$, the value of $\mu=\cos\psi$ is determined by $\mu = \bm{n} \cdot \bm{d}$. With the angular velocity of the pulsar $\omega$, one gets
\begin{equation}
   \mu(t)=\sin i\sin\Theta\cos(\omega t) + \cos i\cos\Theta,  \label{eq:mut_p}
\end{equation}
where we particularly choose $t=0$ when $\mu$ becomes maximum, i.e., the primary hot spot comes closest to the observer. 
To derive Eq. (\ref{eq:mut_p}), by putting the $z$ axis being the same as the rotational axis and by choosing a Cartesian coordinate system in such a way that $\bm{d}$ should be on $z$-$x$ plane, we used that $\bm{n}$ and $\bm{d}$ can be expressed as
\begin{gather}
  \bm{d} = (\sin i, 0, \cos i), \\
  \bm{n} = (\sin\Theta\cos(\omega t), \sin \Theta\sin(\omega t), \cos\Theta).
\end{gather}
From Fig.~\ref{fig:pulsar}, one can see that $\psi$ is in the range of 
\be
	\psi_{\rm min}\le\psi\le\psi_{\rm max},
\;\;\;
	\psi_{\rm min} := |i-\Theta|,
\;\;\;
	\psi_{\rm max} := i+\Theta.
\ee
That is, $\mu$ is in the range of
\be
	\mu_{\rm min}\le \mu\le \mu_{\rm max},
\;\;\;
	\mu_{\rm min} := \cos\psi_{\rm max},
%= \cos(i+\Theta),
\;\;\;
	\mu_{\rm max} := \cos\psi_{\rm min}.
%=\cos(|i-\Theta|).
\ee
In the same way, the value of $\bar{\mu}$, which is for the antipodal spot, is given by $\bar\mu=\bm{\bar{n}} \cdot \bm{d}$, where $\bm{\bar{n}}$ is the normal vector at the antipodal spot. Since $\bm{\bar{n}}=-\bm{n}$, one obtains $\bar\mu(t)=-\mu(t)$ and  $\bar{\mu}(t) \in [ -\mu_{\rm max}, -\mu_{\rm min} ]$.

%%%%%%%%%%%%%%%%%%%%%%%%%%%%%%%%%%%
% Figure 3
%%%%%%%%%%%%%%%%%%%%%%%%%%%%%%%%%%%
%%%%%%%%%%%%%%%%%%%%%%%%%%%%%%%%%%%

\begin{figure}
\begin{center}
\includegraphics[scale=0.5]{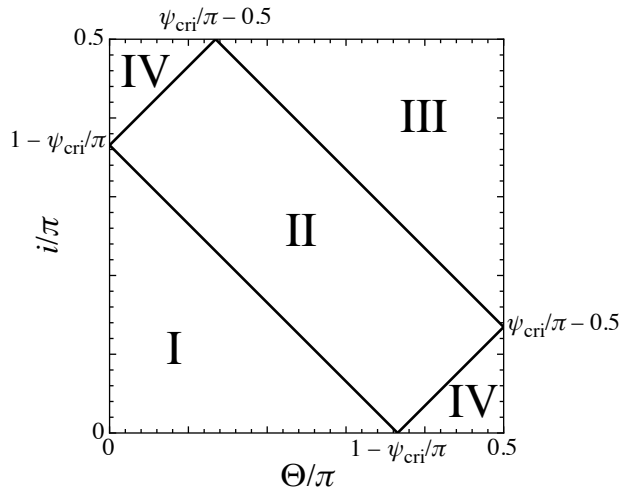} 
\end{center}
\caption{%%
The classification whether the hot spots are observed or not depending on angles $\Theta$ and $i$ for the neutron star model with $R_c=14$ km and $M=1.4M_\odot$ in the Schwarzschild spacetime, where $\psi_{\rm cri}=0.635\pi$. The regions denoted by I, II, III, and IV, correspond to the situations of I, II, III, and IV explained in the text.
}%%
\label{fig:itheta0}
\end{figure}

In the flat spacetime, only one of two hot spots can be observed at any instant (if the extension of spot area is neglected). In the curved spacetime, one may simultaneously observe the both hot spots due to the light bending. In fact, depending on the combination of $i$ and $\Theta$, one can consider following four situations depending on the visibility of two hot spots \cite{Beloborodov2002}. 
\begin{itemize}
\item[{(I)}]
$\mu_{\rm min}> -\cos\psi_{\rm cri}$: only the primary hot spot is observed at any time.
\item[{(II)}]
$\cos\psi_{\rm cri} < \mu_{\rm min}<- \cos\psi_{\rm cri}$: the primary hot spot is observed at any time, while the antipodal hot spot is also observed sometime.
\item[{(III)}]
$\mu_{\rm min}< \cos\psi_{\rm cri}$, $\mu_{\rm max}>-\cos\psi_{\rm cri}$: the primary hot spot is not observed sometime.
\item[{(IV)}]
$\cos\psi_{\rm cri}<\mu_{\rm min}$, $\mu_{\rm max}<-\cos\psi_{\rm cri}$: the both hot spots are observed at any time. 
\end{itemize}
Such a classification is visualized in Fig.~\ref{fig:itheta0}, which is a result for the neutron star model with $R_c=14$ km and $M=1.4M_\odot$ in the Schwarzschild spacetime, where $\psi_{\rm cri}=0.635\pi$.

%%%%%%%%%%%%%%%%%%%%%%%%%%%%%%%%%%%
% Figure 4
%%%%%%%%%%%%%%%%%%%%%%%%%%%%%%%%%%%
%%%%%%%%%%%%%%%%%%%%%%%%%%%%%%%%%%%

\begin{figure*}
\begin{center}
\begin{tabular}{cc}
\includegraphics[scale=0.5]{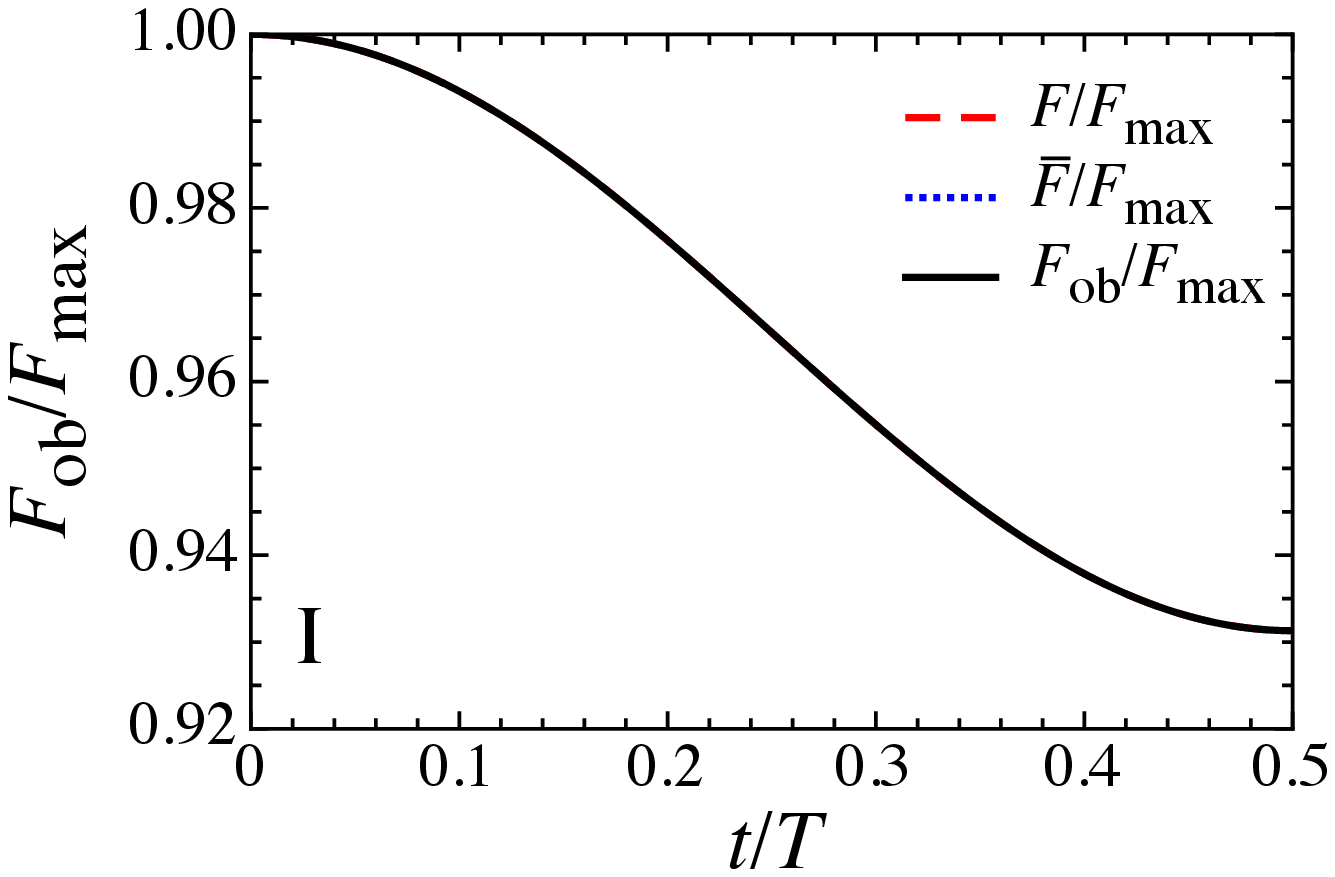} &
\includegraphics[scale=0.5]{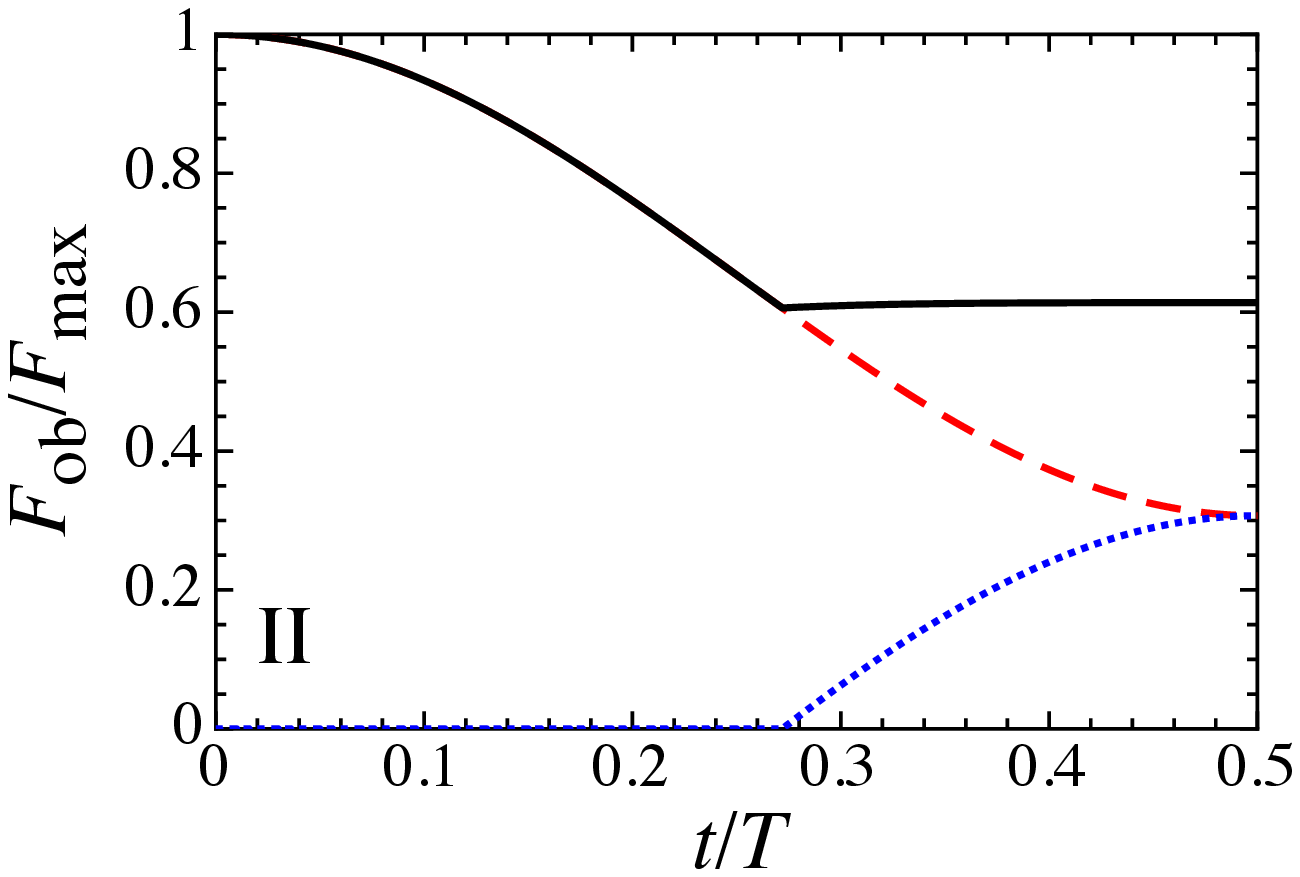} \\ 
\includegraphics[scale=0.5]{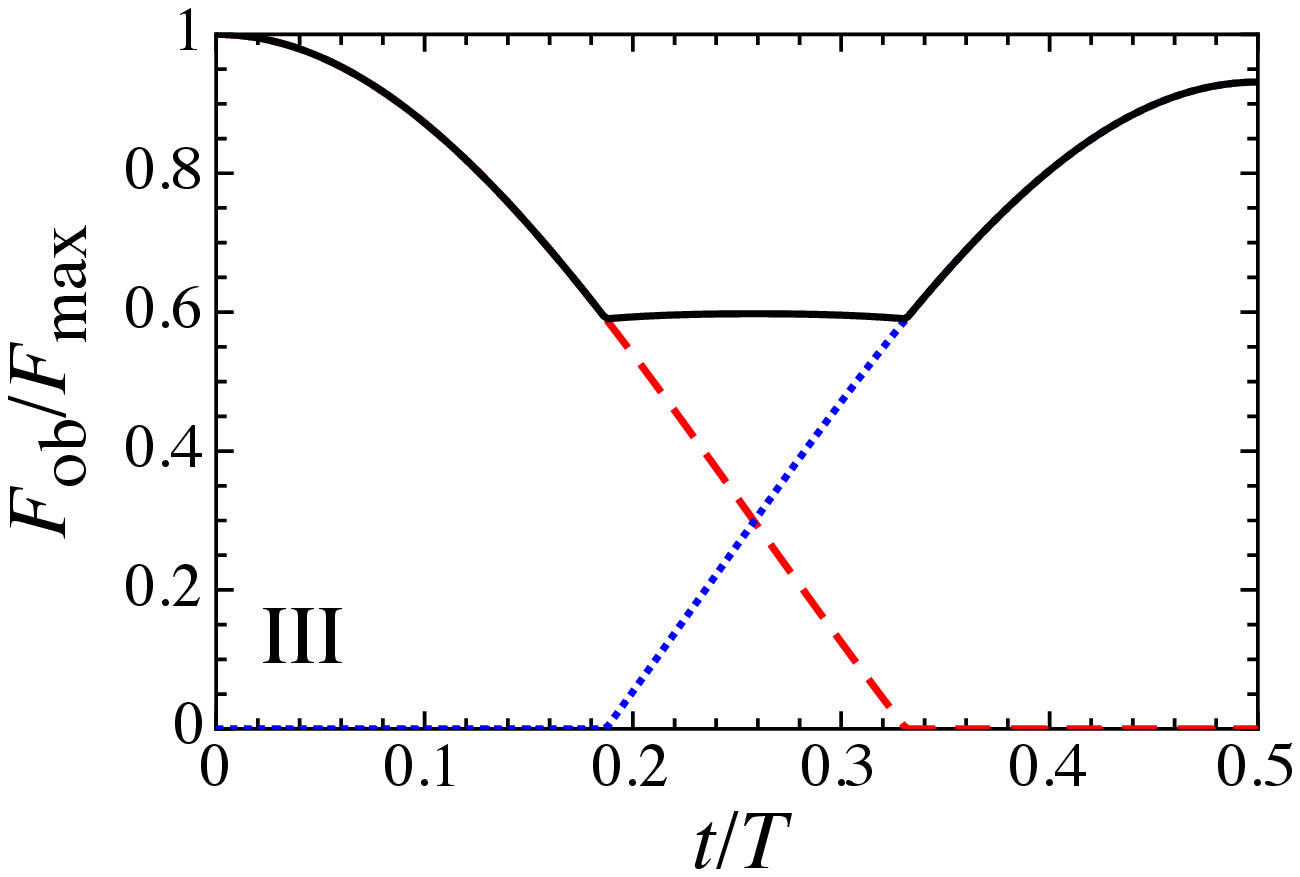} &
\includegraphics[scale=0.5]{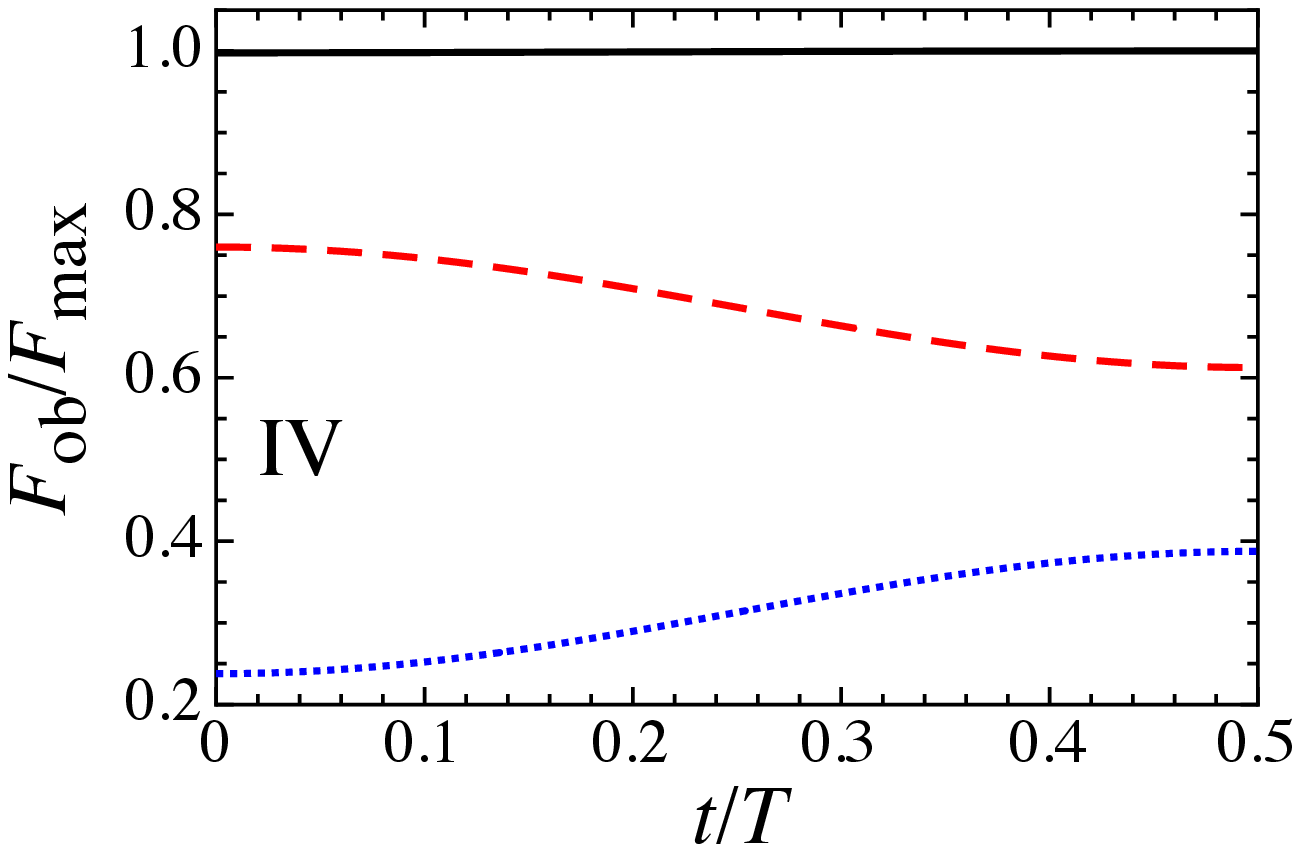} 
\end{tabular}
\end{center}
\caption{%%
Pulse profile from the neutron star with $R_c=14$ km and $M=1.4M_\odot$ in the Schwarzschild spacetime, where I, II, III, and IV correspond to $(\Theta/\pi,i/\pi)=(0.1,0.05)$, $(0.3,0.2)$, $(0.45,0.4)$, and $(0.45,0.02)$, respectively. The dashed, dotted, and solid lines denote the flux from the primary hot spot $F$, the flux from the antipodal hot spot $\bar{F}$, and the observed flux $F_{\rm ob} := F + \bar{F}$, which are normalized by the observed maximum flux $F_{\rm max}$. 
}%%
\label{fig:pulse}
\end{figure*}

Since $\psi$ would vary with time as in Eq.~(\ref{eq:mut_p}), one observes a pulse profile from the neutron star, depending on the angles $\Theta$ and $i$. It should be noticed that, due to the symmetry between the angles of $\Theta$ and $i$ in Eq.~(\ref{eq:mut_p}), the pulse shape with $(\Theta,i)=(\theta_1,\theta_2)$ is the same as that with $(\Theta,i)=(\theta_2,\theta_1)$ for $0\le\theta_1\le\pi/2$ and $0\le\theta_2\le\pi/2$. In addition, from the symmetry of system, one can expect that the pulse shape is periodic in $0\le t/T\le 1$ with the rotational period $T=2\pi/\omega$ and that the amplitude of the shape at $t/T$ for $0.5\le t/T\le 1$ is the same as that at $1-t/T$. Thus, hereafter we focus on the pulse shape for $0\le t/T\le 0.5$. As an example, we show the pulse profile for the neutron star with $R_c=14$ km and $M=1.4M_\odot$ in the Schwarzschild spacetime in Fig.~\ref{fig:pulse}, where the cases of I, II, III, and IV correspond to the results with $(\Theta/\pi,i/\pi)=(0.1,0.05)$, $(0.3,0.2)$, $(0.45,0.4)$, and $(0.45,0.02)$, respectively. We remark that the flux from the primary hot spot (the dashed line) completely agrees with the observed flux (the solid line) for the case of I because the flux from the antipodal hot spot cannot be observed in any time for this case.  
We argue that pulse profile observed in a specific range of wavelength, e.g., in X-ray observation, would be the same with that obtained in this paper, provided the photons observed come from the hot spots. The reason is twofold: since the pulse profile obtained in this paper is that of the flux integrated over the frequency, the pulse contains any wavelength of photons; the photon trajectory is independent of wavelength within the validity of geometric-optics approximation, which we used.

%%%%%%%%%%%%%%%%%%%%%%%%%%%%%%%%%%%
% Figure 5
%%%%%%%%%%%%%%%%%%%%%%%%%%%%%%%%%%%
%%%%%%%%%%%%%%%%%%%%%%%%%%%%%%%%%%%

%%%%%%%%%%%%%%%%%%%%%%%%%%%%%%%%%%%%%%%%%%%%%%%%
%\section{Comparison with the different spacetime}
\section{Comparison among various spacetime models}
\label{sec:V}
%%%%%%%%%%%%%%%%%%%%%%%%%%%%%%%%%%%%%%%%%%%%%%%%

With the some astronomical observations, the stellar radius and mass might be fixed. In such a situation, one could test the gravitational geometry outside the star via the observation of the shape of the pulse profile, if it depends on the geometry. In this section, we consider how the pulse profiles depend on the gravitational geometry outside the star, varying the angles $\Theta$ and $i$ for the specific stellar models. 
For this purpose, in particular, we consider three cases of spacetime outside the star, i.e., the Schwarzschil spacetime, Reissner-Nordstr\"{o}m spacetime, and the Garfinkle-Horowitz-Strominger spacetime \cite{GHS1991}. These spacetimes are static, spherically symmetric, and asymptotically flat. The coefficients in the asymptotically behavior given in Eqs.\ (\ref{eq:ar}) -- (\ref{eq:cr}) are shown in Table \ref{tab:coefficients}. As a neutron star model, we consider the objects with $R_c=10-14$ km and $M=1.4-1.8M_\odot$. For considering the light bending, the stellar compactness is more important than the stellar mass and radius themselves. So, we particularly focus on three stellar models with $(M,R_c)=(1.4M_\odot, 14 {\rm km})$, $(1.6M_\odot, 12 {\rm km})$, and $(1.8M_\odot, 10 {\rm km})$ as the representatives of neutron star with low, middle, and high compactness, where the corresponding compactness is $M/R_c=0.148$, $0.197$, and $0.266$, respectively.

%%%%%%%%%%%%%%%%%%%%%%%%%%%%%%%%%%%
% Table 1
%%%%%%%%%%%%%%%%%%%%%%%%%%%%%%%%%%%
\begin{table}
\centering
\caption{
The expansion coefficients in Eqs.\ (\ref{eq:ar}) -- (\ref{eq:cr}) for Schwarzschild (S), Reissner-Nordstr\"{o}m (RN), and Garfinkle-Horowitz-Strominger (GHS) solutions.
}
\begin{tabular}{c|cccccc}
\hline\hline
  spacetime & $a_1$ & $a_2$ & $b_1$ & $b_2$ & $c_1$ & $c_2$ \\
\hline
  S       & $-2M$ & 0         & $2M$ & $4M^2$         & 0                                        & 0       \\
  RN    & $-2M$ & $Q^2$ & $2M$ & $4M^2-Q^2$ & 0                                        & 0       \\  
  GHS & $-2M$ & 0         & $2M$ & $4M^2$         & $-Q^2 e^{-2\varphi_0}/M$ & 0       \\ 
\hline\hline
\end{tabular}
\label{tab:coefficients}
\end{table}
%%%%%%%%%%%%%%%%%%%%%%%%%%%%%%%%%%%

The metric functions of the Schwarzschild spacetime are 
\begin{equation}
  A(r) = 1- \frac{2M}{r},\  B(r) = \frac{1}{A(r)},\   C(r) = r^2.
\end{equation}
In this case, the coordinate $r$ corresponds to the circumference radius, i.e., $R_c=R$.
The metric form of the Reissner-Nordstr\"{o}m spacetime is
\begin{equation}
  A(r) = 1- \frac{2M}{r} + \frac{Q^2}{r^2},\  B(r) = \frac{1}{A(r)},\   C(r) = r^2.
\end{equation}
Here, $Q$ denotes the electric charge of the central object in the range of $0\le Q/M\le 1$. As in the Schwarzschild spacetime, the radial coordinate $r$ agrees with the circumference radius, i.e., $R_c=R$.
As another example, 
we consider the Garfinkle-Horowitz-Strominger spacetime \cite{GHS1991}. This is a solution for static charged black holes in string theory. The line element for this spacetime is given by
\begin{equation}
  A(r) = 1 - \frac{2M}{r},\  B(r) = \frac{1}{A(r)},\  C(r) = r\left(r-\frac{Q^2 e^{-2\varphi_0}}{M}\right),
\end{equation}
where $Q$ and $\varphi_0$ denote the magnetic charge and the asymptotic value of the dilaton field \cite{GHS1991}, respectively. In this spacetime, the dilaton field $\varphi(r)$ is given by 
\begin{equation}
   e^{-2\varphi} = e^{-2\varphi_0}\left[1-\frac{Q^2 e^{-2\varphi_0}}{Mr}\right],
\end{equation}   
together with a purely magnetic Maxwell field such as $F=Q\sin\theta d\theta\wedge d\psi$. We remark that, unlike the case of the Reissner-Nordstr\"{o}m spacetime, $Q/M$ in the Garfinkle-Horowitz-Strominger spacetime can be in the range of $0\le Q/M\le \sqrt{2}e^{-\varphi_0}$ \cite{GHS1991}. In this paper, we simply adopt $\varphi_0=0$.
Since the radial coordinate $r$ is associated with the circumference radius $r_c$ via $C(r)=r_c^2$, the stellar radius $R$ in coordinate $r$ is expressed by the corresponding circumference radius $R_c$ as
\begin{equation}
  R=\sqrt{R_c^2+\frac{Q^4}{4M^2}}+\frac{Q^2}{2M},
\end{equation}
which leads to the relation of
\begin{equation}
  u \equiv \frac{r_g}{R} = 2 \left[\sqrt{\frac{R_c^2}{M^2}+\frac{Q^4}{4M^4}}+\frac{Q^2}{2M^2}\right]^{-1}.
\end{equation}

\begin{figure*}
\begin{center}
\begin{tabular}{c}
\includegraphics[scale=0.5]{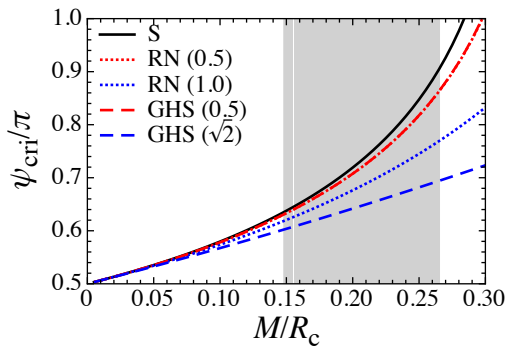}
\end{tabular}
\end{center}
\caption{%%
$\psi_{\rm cri}$ as a function of the stellar compactness $M/R_{\rm c}$ for different spacetime, i.e., the Schwarzschild spacetime (S) (solid line), the Reissner-Nordstr\"{o}m spacetime (RN) with $Q/M=0.5$ and $1.0$ (dotted lines), and the Garfinkle-Horowitz-Strominger spacetime (GHS) with $Q/M=0.5$ and $\sqrt{2}$ (dashed lines). We remark that the results for the Reissner-Nordstr\"{o}m spacetime with $Q/M=0.5$ are almost the same as the results for the Garfinkle-Horowitz-Strominger spacetime with $Q/M=0.5$. The shaded region denotes the possible value of $M/R_{\rm c}$ for the stellar models with $R_{\rm c}=10-14$ km and $M=1.4-1.8M_\odot$. 
}%%
\label{fig:psic-MR}
\end{figure*}

In Fig.~\ref{fig:psic-MR} the critical value of $\psi$, which is an important property for dividing into the classes of the observation of the hot spots, is shown as a function of the stellar compactness with different geometries. 
The shaded region denotes the allowed compactness by the stellar model with the mass and radius in the range of  $M=1.4-1.8M_\odot$ and $R_c=10-14$ km, i.e., $0.148\le M/R_c\le 0.266$. We remark that the left and right boundaries of the shaded region correspond to the stellar models with $(M/R_c)=(1.4M_\odot,14\ {\rm km})$ and $(1.8M_\odot, 10\ {\rm km})$. From this figure, one can observe that deviation from the Schwarzschild spacetime increases with the stellar compactness and with the value of $Q/M$. In practice, compared with the Schwarzschild spacetime, the value of $\psi_{\rm cri}$ becomes $2.61\%$ and $15.14\%$ smaller for the stellar models with $M/R_c=0.148$ and $0.266$ in the Reissner-Nordstr\"{o}m spacetime with $Q/M=1.0$, while $5.14\%$ and $23.46\%$ smaller in the Garfinkle-Horowitz-Strominger spacetime with $Q/M=\sqrt{2}$.

%%%%%%%%%%%%%%%%%%%%%%%%%%%%%%%%%%%
% Figure 6
%%%%%%%%%%%%%%%%%%%%%%%%%%%%%%%%%%%
%%%%%%%%%%%%%%%%%%%%%%%%%%%%%%%%%%%

\begin{figure*}
\begin{center}
\begin{tabular}{cc}
\includegraphics[scale=0.5]{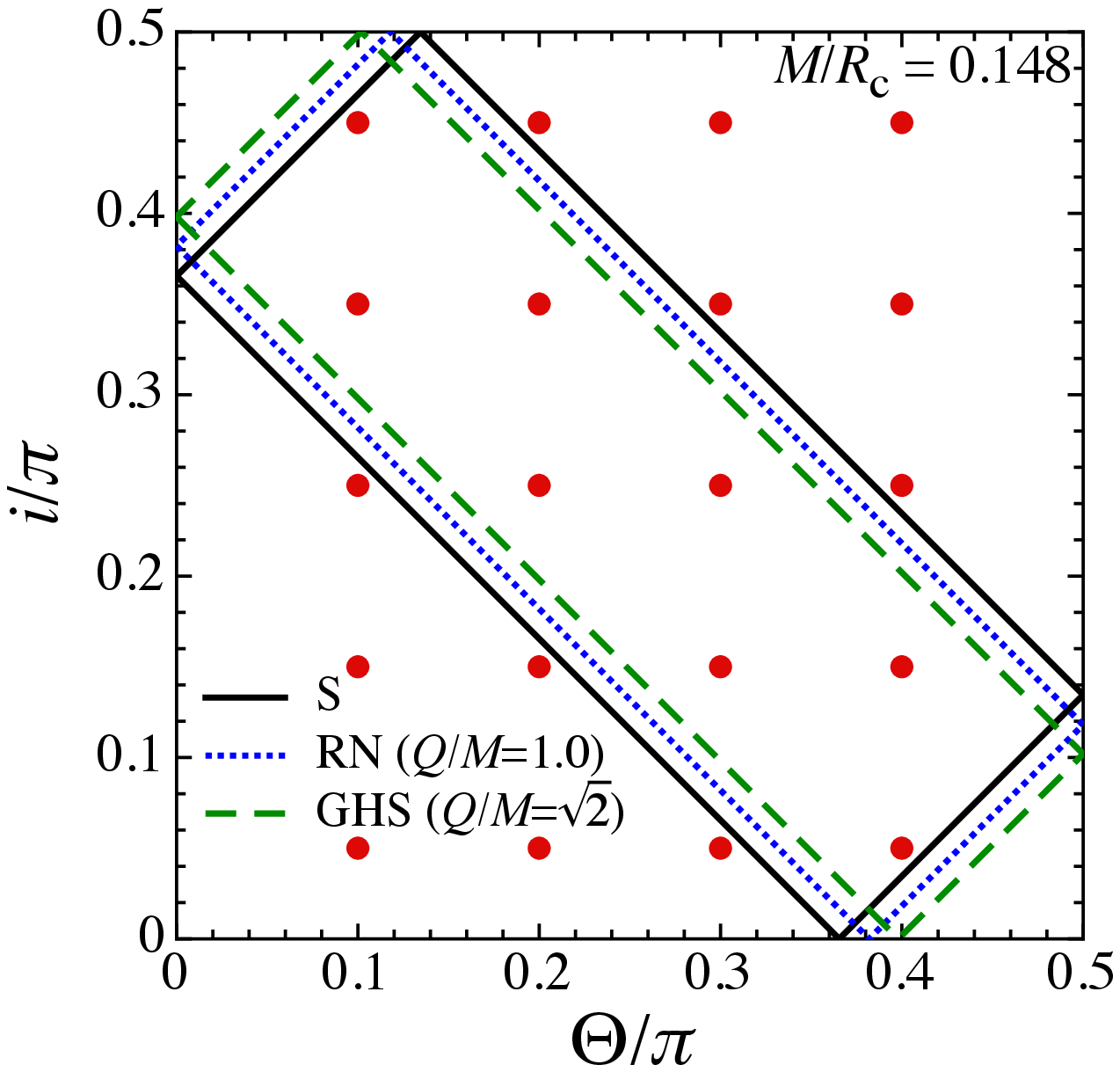} &
\includegraphics[scale=0.5]{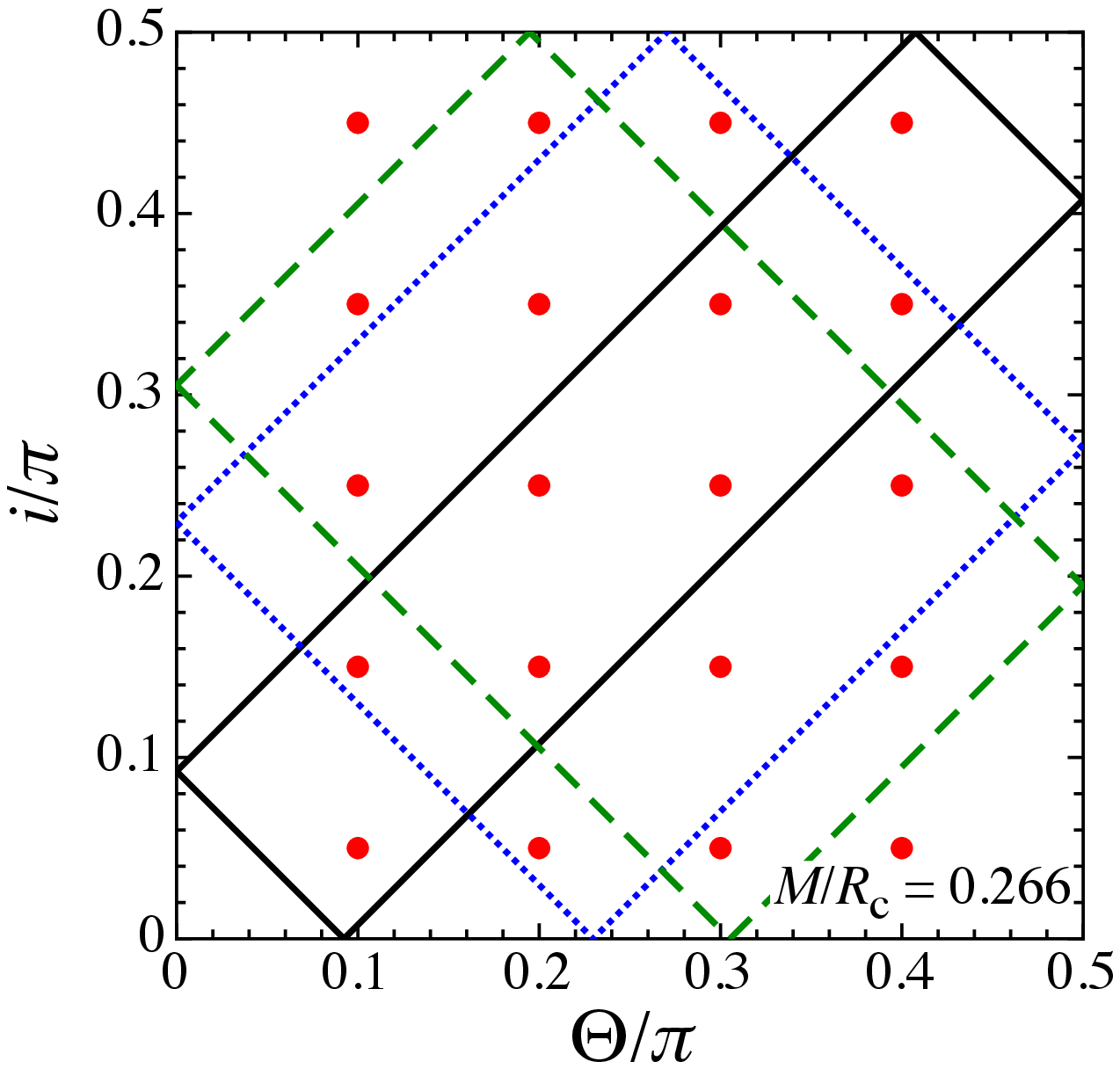}
\end{tabular}
\end{center}
\caption{%%
Classification for the stellar models with $M/R_c=0.148$ and $0.266$ for various spacetimes, where the solid, dotted, and dashed lines denote the boundary of the classification with the Schwarzschild spacetime (S), the Reissner-Nordstr\"{o}m spacetime (RN) with $Q/M=1.0$, and the Garfinkle-Horowitz-Strominger spacetime (GHS) with $Q/M=\sqrt{2}$. The dots in the figure denote the specific angles of $\Theta$ and $i$, with which the pulse profiles are shown in Figs. \ref{fig:pulse-comM14} and \ref{fig:pulse-comM18}.
}%%
\label{fig:class-e}
\end{figure*}

With the value of $\psi_{\rm cri}$ depending on the gravitational geometry, the classification whether the two hot spots are visible or not is shown in Fig.~\ref{fig:class-e} for the given stellar mass and radius, where the solid, dotted, and dashed lines denote the boundary of the classification with the Schwarzschild spacetime, the Reissner-Nordstr\"{o}m spacetime with $Q/M=1.0$, and the Garfinkle-Horowitz-Strominger spacetime with $Q/M=\sqrt{2}$. The deviation in $\psi_{\rm cri}$ shown in Fig.~\ref{fig:psic-MR} is also significantly visible in this figure especially for the stellar model with $M/R_c=0.266$.

%%%%%%%%%%%%%%%%%%%%%%%%%%%%%%%%%%%
% Figure 7
%%%%%%%%%%%%%%%%%%%%%%%%%%%%%%%%%%%
%%%%%%%%%%%%%%%%%%%%%%%%%%%%%%%%%%%

\begin{figure*}
\begin{center}
\begin{tabular}{c}
\includegraphics[scale=0.4]{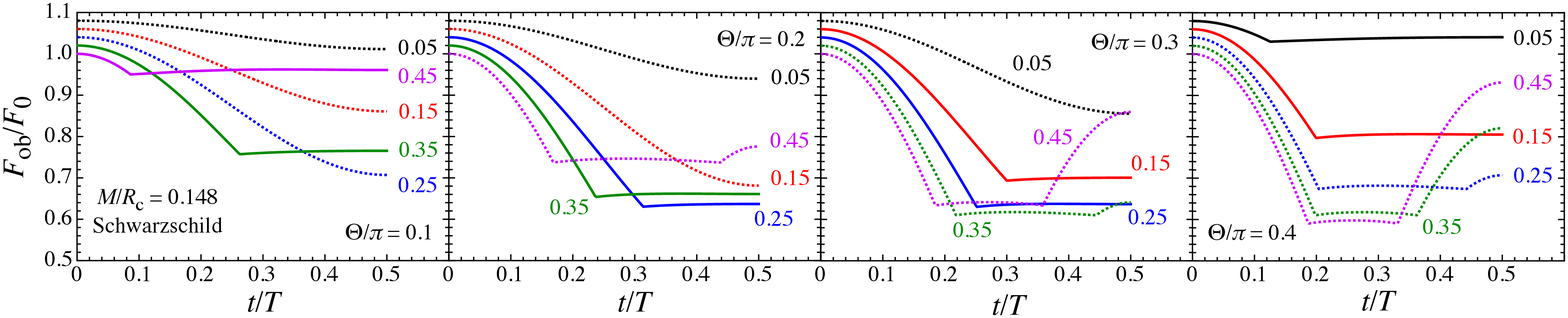} \\
\includegraphics[scale=0.4]{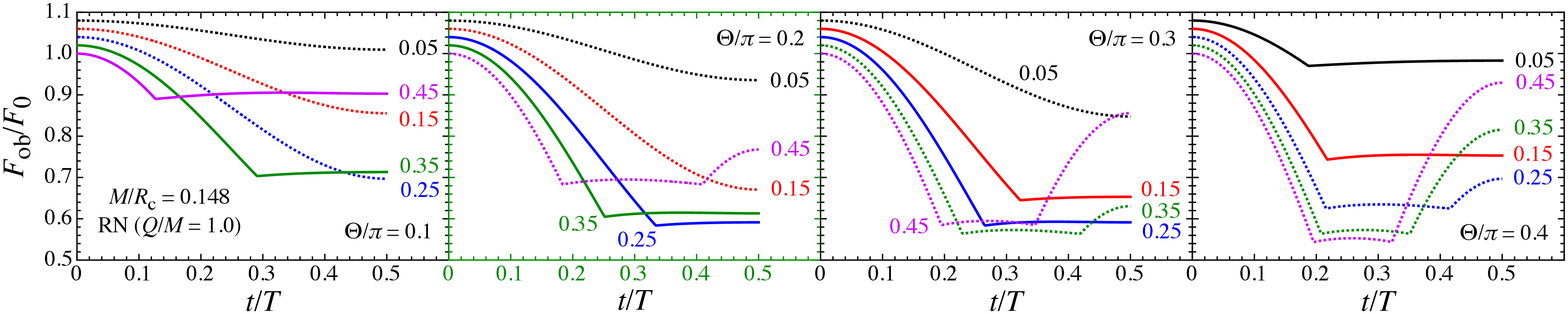} \\
\includegraphics[scale=0.4]{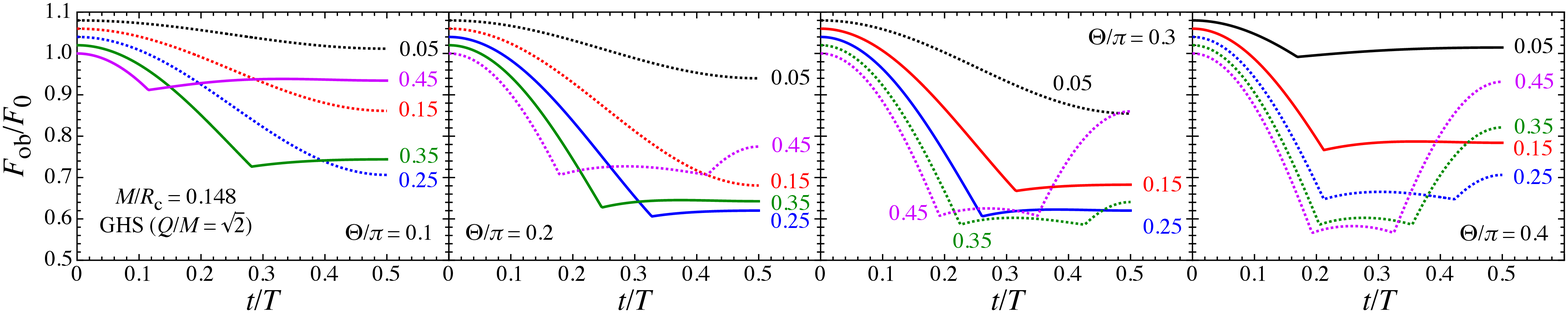} 
\end{tabular}
\end{center}
\caption{%%
Pulse profiles for the stellar models with $M/R_c=0.148$ are shown as a function of $t/T$ for various angles $\Theta$ and $i$ with different spacetimes. The upper, middle, and lower panels respectively correspond to the results for the Schwarzschild spacetime, the Reissner-Nordstr\"{o}m spacetime (RN) with $Q/M=1.0$, and the Garfinkle-Horowitz-Strominger spacetime (GHS) with $Q/M=\sqrt{2}$. For each spacetime, the panels from left to right are results with $\Theta/\pi=0.1$, 0.2, 0.3, and 0.4. In each panel, the different lines denote the results for $i/\pi=0.05$, 015, 0.25, 0.35, and 0.45. In addition, the different type of lines corresponds to the different class whether the two hot spots are observed or not as shown in Fig.~\ref{fig:itheta0}, where the solid and dashed lines correspond to the class II and IV, respectively, while the dotted lines correspond to the class I or III. The amplitude of pulse profiles are normalized by the amplitude at $t/T=0$, denoted by $F_0$, and shifted a little in order to easily distinguish the different lines.
}%%
\label{fig:pulse-comM14}
\end{figure*}

\begin{figure*}
\begin{center}
\begin{tabular}{c}
\includegraphics[scale=0.4]{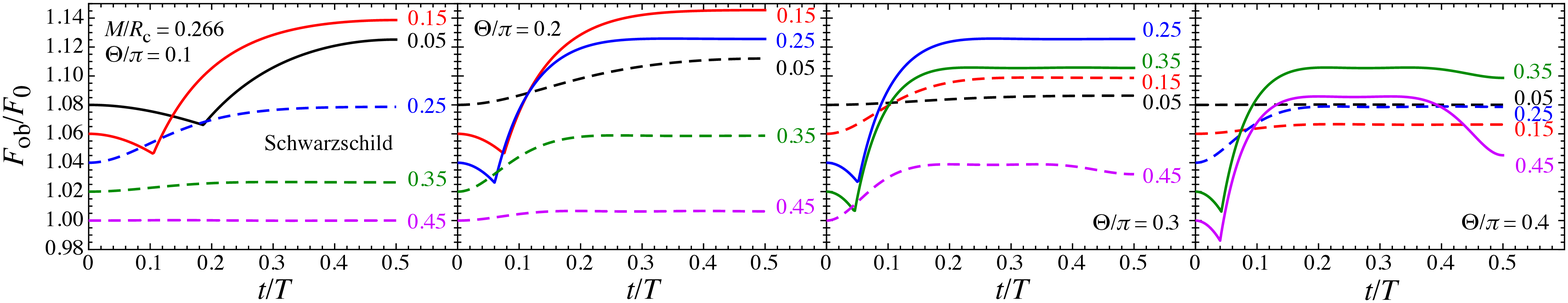} \\
\includegraphics[scale=0.4]{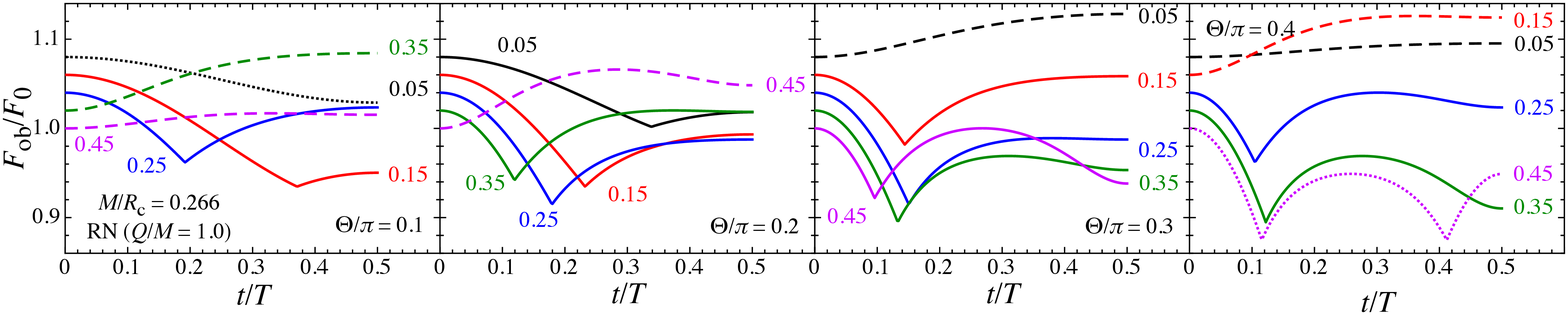} \\
\includegraphics[scale=0.4]{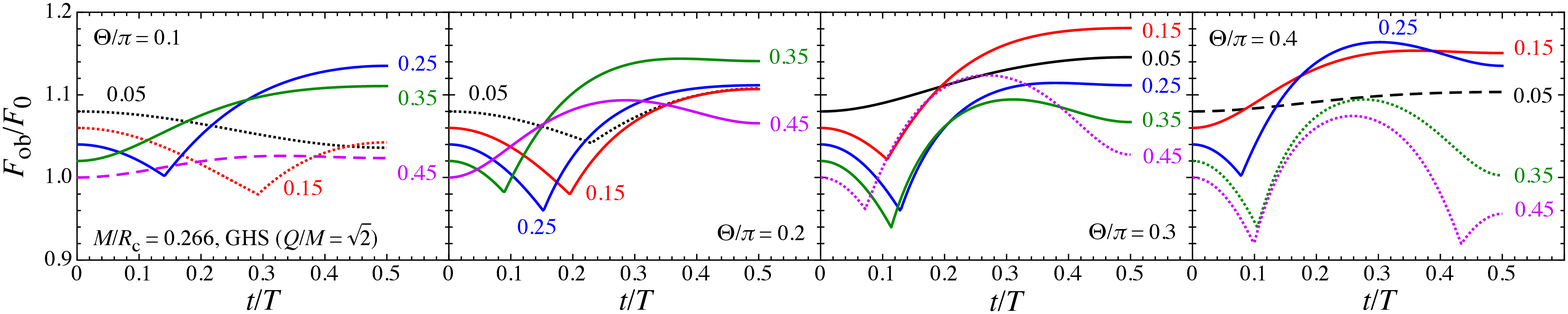} 
\end{tabular}
\end{center}
\caption{%%
Same as Fig.~\ref{fig:pulse-comM14}, but for the stellar model with $M/R_c=0.266$.
}%%
\label{fig:pulse-comM18}
\end{figure*}

Finally, adopting various combinations of angles $\Theta$ and $i$ denoted in Fig.~\ref{fig:class-e} with the dots, we show the pulse profiles in Fig.~\ref{fig:pulse-comM14} for the stellar model with $M/R_c=0.148$ and in Fig.~\ref{fig:pulse-comM18} for that with $M/R_c=0.266$. The amplitude for each model is normalized by that at $t/T=0$ and shifted a little in order to easily distinguish the different lines. In each figure, the upper, middle, and lower panels correspond to the results with the Schwarzschild spacetime, the Reissner-Nordstr\"{o}m spacetime with $Q/M=1.0$, and the Garfinkle-Horowitz-Strominger spacetime with $Q/M=\sqrt{2}$. In the both figures, the solid and dashed lines denote the pulse profiles in the class II and IV, respectively, while the dotted lines denote those in the class I or III. From Fig.~\ref{fig:pulse-comM14}, one can see that the pulse profiles with any angles are almost independent of the gravitational geometry for the stellar model with $M/R_c=0.148$, where the profile expected for the Schwarzschild spacetime is almost the same as those for the Reissner-Nordstr\"{o}m spacetime and for the Garfinkle-Horowitz-Strominger spacetime even in the extreme cases. However, for the stellar model with $M/R_c=0.266$ as in Fig.~\ref{fig:pulse-comM18}, the shapes of the pulse profiles completely depend on the gravitational geometry. For example, the cases with $(\Theta/\pi,i/\pi)=(0.3,0.25)$ and $(0.3,0.35)$ correspond to the class II independently of the geometry as shown in Fig.~\ref{fig:class-e}, but the shapes for the Reissner-Nordstr\"{o}m spacetime and with the Garfinkle-Horowitz-Strominger spacetime are significantly different from that for the Schwarzschild spacetime. That is, at least, one may distinguish whether the gravitational geometry outside the star is the Schwarzschild spacetime or the extreme Reissner-Nordstr\"{o}m/Garfinkle-Horowitz-Strominger spacetimes via the observation of pulse profiles from the pulsar, when the stellar compactness is known to be large enough with the help of another observations of mass and radius.

%%%%%%%%%%%%%%%%%%%%%%%%%%%%%%%%%%%
% Figure 8
%%%%%%%%%%%%%%%%%%%%%%%%%%%%%%%%%%%
%%%%%%%%%%%%%%%%%%%%%%%%%%%%%%%%%%%

%%%%%%%%%%%%%%%%%%%%%%%%%%%%%%%%%%%
% Figure 9
%%%%%%%%%%%%%%%%%%%%%%%%%%%%%%%%%%%
%%%%%%%%%%%%%%%%%%%%%%%%%%%%%%%%%%%

%%%%%%%%%%%%%%%%%%%%%%%%%%%%%%%%%%%%%%%%%%%%%%%%
\section{Conclusion}
\label{sec:VI}
%%%%%%%%%%%%%%%%%%%%%%%%%%%%%%%%%%%%%%%%%%%%%%%%

Since the profiles of pulse radiated from the neutron star depend on the gravitational geometry outside the star, one may probe the gravitational geometry via the observation of pulse profiles. In this paper, we consider a neutron star model with two antipodal hot spots, which may be associated with the polar caps of the stellar magnetic field. We derive the formula for describing the pulse profiles with any metric for static, spherically symmetric spacetime, and also derive the approximate formula in the linear and 2nd order of parameter $u$ (Appendix \ref{sec:IV}), which is defined by the ratio of the gravitational radius of considered spacetime to the stellar radius. 
The pulse profiles can be obtained by numerical integration. In order to examine the dependence of the pulse profiles on the gravitational geometry, we particularly adopt three spacetimes, i.e., the Schwarzschild, the Reissner-Nordstr\"{o}m, and the Garfinkle-Horowitz-Strominger spacetimes. Then, by systematically varying the stellar mass and radius (which lead to various values of stellar compactness or $u$), the angle between the rotational and magnetic axes, and the angle between the direction to the observer and the rotational axis, we examine the pulse profiles from the neuron stars.

In particular, we examine the pulse profiles with various angles for the stellar models with $M/R_c=0.148$ and $0.266$. For the stellar model with $M/R_c=0.148$, the pulse profiles with the Schwarzschild spacetime are completely similar to those with the Reissner-Nordstr\"{o}m  and the Garfinkle-Horowitz-Strominger spacetimes even for the extreme cases. On the other hand, for the stellar model with $M/R_c=0.266$, the pulse profiles with the Schwarzschild spacetime are significantly different from those with the extreme Reissner-Nordstr\"{o}m and the extreme Garfinkle-Horowitz-Strominger spacetimes. That is, if the stellar compactness is high enough, the pulse profiles recognizably depend on the gravitational geometry outside the star, which would enable us to probe the geometry and/or gravitational theory assumed by observing pulse profile with the help of another observations determining the stellar radius and mass.

Additionally, to estimate the validity of the approximate relations for given gravitational geometry,  as shown in Appendix \ref{sec:V}, we check the relative error in the bending angle estimated with the approximate relations and that with the full order numerical integration, and we find that it becomes $\sim 20-30\%$ for the 1st order and $\sim 5-10\%$ for the 2nd order approximations, adopting the typical neutron star model with $M=1.4M_\odot$ and $R_c=12$ km, where $R_c$ is the circumference stellar radius. We notice that our results with the 1st order approximation for the Schwarzschild spacetime seem to be different from those obtained by the previous well-known approximation \cite{Beloborodov2002}, which predicts unnaturally accurate results even thought it is the 1st order approximation. This suggests that the previous approximation might be wrong. We also find the existence of the jump in the pulse profiles estimated with the 1st order approximation for any geometry we adopted at the moment when the antipodal spot comes into the visible zone. The pulse profiles estimated with the 2nd order approximation seem to be qualitatively better to express the full order ones.

Here, let us mention the validity of setup and assumptions that we have supposed in this paper. Firstly, we do not argue that the black-hole solutions we adopted as the spacetime metrics outside the star, the Reissner-Nordstr\"{o}m and Garfinkle-Horowitz-Strominger spacetimes, are realistic from astrophysical viewpoint. The reason we adopted them are that they are simple analytic solutions to serve as the rigorous first steps to further general analysis. Since there are many modified theories of gravity, the application of the analysis in this paper to them would be important to verify the validity of modify theories of gravity from astrophysical observations. Secondly, we have neglected the effects caused by the rotation of star. As the spin increases, the effects of rotation gradually begin to affect the radiation around the spin frequency of a few hundred Hz. Such effects are the Doppler shifts and aberration, frame dragging, quadrupole moment, the oblateness of surface, and so on (see, e.g., \cite{PO2014}). Thus, for the pulsars with a rapid rotation above the a few hundred Hz frequency, the rotational effects could be comparable with the effects stemming from the difference of gravitational theories, %and black-hole solutions, 
which force us to numerically calculate the emission and propagation of radiation in the rotating backgrounds.

%\newpage
%%%%%%%%%%%%%%%%%%%%%%%%%%%%%%%%%%%%%%%%%%%%%%%%
\acknowledgments
%%%%%%%%%%%%%%%%%%%%%%%%%%%%%%%%%%%%%%%%%%%%%%%%
HS is grateful to T. Kawashima for giving valuable comments. This work was supported in part by Grant-in-Aid for Scientific Research (C) through Grant No. 17K05458 (HS) and No. 15K05086 (UM) provided by JSPS.

\appendix
%%%%%%%%%%%%%%%%%%%%%%%%%%%%%%%%%%%%%%%%%%%%%%%%
\section{Integration of Eq.~(\ref{eq:psi1}) with $\alpha=\pi/2$}
\label{sec:a1}
%%%%%%%%%%%%%%%%%%%%%%%%%%%%%%%%%%%%%%%%%%%%%%%%

In this appendix, we show how to calculate $\psi(R)$ given by Eq.~(\ref{eq:psi1}) with $\alpha=\pi/2$, where $b^2=C(R)/A(R)$ from Eq.~(\ref{eq:alpha}). Introducing a new variable $z$ defined by $z=1-R/r$, i.e., $r(z)=R/(1-z)$, Eq.~(\ref{eq:psi1}) can be transformed as
\begin{align}
  \psi(R) &= \psi_*(R)
                 + b\left[\frac{1}{R_*} + \frac{a_1 + b_1 + 2c_1}{R_*^2} + {\cal O}\left(\frac{1}{R_*^3}\right)\right], \label{eq:psi2}\\
  \psi_*(R) &\equiv \int_0^{z_*}\frac{b\sqrt{AB}}{\sqrt{C(C-Ab^2)}}\frac{R}{(1-z)^2}dz, \nonumber \\
              &= \int_0^{z_*}{\cal F}(z,R) \sqrt{AB}dz,  \label{eq:psi_A1}
\end{align}
where $z_*$ is the constant defined by $z_*=1-R/R_*$ with $R_*$ in Eq.~(\ref{eq:psi1}), while ${\cal F}(z,R)$ is given by 
\begin{equation}
  {\cal F}(z,R) \equiv \frac{br^2}{R\sqrt{C(C-Ab^2)}}.
\end{equation}
As mentioned in text, the function of ${\cal F}(z,R)$ diverges at $z=0$, i.e., $r=R$, while $\sqrt{AB}$ is regular for any values of $z$. In the vicinity of $z=0$, the function of ${\cal F}(z,R)$ can be expressed as
\begin{equation}
  {\cal F}(z,R) \simeq \frac{1}{\sqrt{h_1z + h_2z^2}}\equiv {\cal F}_0(z,R).  \label{eq:F0}
\end{equation}
Here, $h_1$ and $h_2$ are appropriate functions of $R$ such as
\begin{align}
  h_1 &= \frac{1}{R}\left(A_0C_0' - A_0'C_0\right), \label{eq:h1} \\
  h_2 &= -\frac{3}{R}\left(A_0C_0' - A_0'C_0\right) + \frac{A_0C_0'^2}{C_0} + \frac{A_0C_0''}{2}
               -\frac{A_0''C_0}{2} - A_0'C_0',  \label{eq:h2}
\end{align}
where the variables with the subscript $0$ denote the corresponding values at $r=R$ or $z=0$, and the prime denotes the derivative with respect to $r$. Thus, $\psi_*(R)$ is a finite value if $h_1\neq 0$.

Now, we consider the radial motion of photon, which is subject to Eq.~(\ref{eq:radial}), i.e., 
\begin{equation}
   AB\dot{r}^2 + V(r) =1,
\end{equation}
where $V(r)$ is an effective potential given by $V(r)=Ab^2/C$. The radius of the photosphere, $R_{\rm ph}$, is determined by solving the equation of $dV/dr=0$ for $d^2V/dr^2<0$, from which one can get the relation that $A'C-AC' = 0$ at $r=R_{\rm ph}$. That is, $h_1$ in Eq.~(\ref{eq:F0}) becomes zero only if $R=R_{\rm ph}$. Since the radius of compact object should be larger than $R_{\rm ph}$, it can be considered that $\psi_*(R)$ is a finite value for the radiation photon from the surface of compact objects.

Therefore, the value of $\psi_*(R)$ can be calculated as
\begin{equation}
   \psi_*(R) = \psi_D(R) + \psi_R(R),
\end{equation}
where
\begin{align}
  \psi_D(R) &= \sqrt{A_0B_0}\int_0^{z_*} {\cal F}_0(z,R)dz, \\
  \psi_R(R) &= \int_0^{z_*} {\cal G}(z,R)dz, \\
  {\cal G}(z,R) &\equiv  {\cal F}(z,R)\sqrt{AB} -  {\cal F}_0(z,R)\sqrt{A_0B_0}.
\end{align}
As in Ref. \cite{Bozza2002,SM2015}, $\psi_D$(R) can be analytically integrated as
\begin{align}
  \psi_D(R) &= 2\sqrt{\frac{A_0B_0}{h_2}}\log\frac{\sqrt{h_2z_*}+\sqrt{h_1+h_2z_*}}{\sqrt{h_1}}  \ \ \ {\rm for}\ \ \  h_2 > 0, \\
  \psi_D(R) &= 2\sqrt{\frac{A_0B_0z_*}{h_1}}  \ \ \ {\rm for}\ \ \  h_2 = 0, \\
  \psi_D(R) &= -\sqrt{\frac{A_0B_0}{|h_2|}}\left[\arcsin\left(\frac{2h_2z_*+h_1}{h_1}\right) - \frac{\pi}{2} \right] \ \ \ {\rm for}\ \ \ h_2 < 0,
  \label{eq:psiD}
\end{align}
On the other hand, in the vicinity of $z=0$, ${\cal G}(z,R)$ can be expanded as 
\begin{align}
  {\cal G}{(z,R)} &\simeq \left[\sqrt{AB}-\sqrt{A_0B_0}\right]{\cal F}_0(z,R) \nonumber \\
       &=  \frac{d\sqrt{AB}}{dz}\bigg|_{z=0}\sqrt{\frac{z}{h_1}} + {\cal O}(z).
\end{align}
Thus, ${\cal G}{(0,R)}=0$, i.e.,
\begin{equation}
   \psi_R(R) = \int_{\epsilon_1}^{z_*}{\cal G}(z,R)dz,
\end{equation}
where $\epsilon_1$ is an appropreate constant such as $\epsilon_1\ll 1$. 
At last, $\psi(R)$ for $\alpha=\pi/2$ is calculated via
\begin{equation}
  \psi(R) = \psi_D(R) + b\left[\frac{1}{R_*} + \frac{a_1 + b_1 + 2c_1}{R_*^2}\right]
       + \int_{\epsilon_1}^{z_*}{\cal G}(z,R)dz.
\end{equation}

%%%%%%%%%%%%%%%%%%%%%%%%%%%%%%%%%%%
% Table 1
%%%%%%%%%%%%%%%%%%%%%%%%%%%%%%%%%%%
\begin{table}
\centering
\caption{
The expansion coefficients $h_1$ and $h_2$ given by Eqs.\ (\ref{eq:h1}) and (\ref{eq:h2}) for Schwarzschild (S), Reissner-Nordstr\"{o}m (RN), and Garfinkle-Horowitz-Strominger (GHS) solutions, where $\eta$ denotes $Q^2e^{-2\varphi_0}/M$.
}
\begin{tabular}{c|ccc}
\hline\hline
  spacetime & $h_1$ & & $h_2$  \\
\hline
  S       & $2-6M/R$                   & & $-1+6M/R$                       \\
  RN    & $2-6M/R+4Q^2/R^2$ & & $-1+6M/R-6Q^2/R^2$       \\  
  GHS & $2-6M/R-\eta(1-4M/R)/R$ & & $-5+14M/R+3\eta(1-4M/R)/R+(1-2M/R)(2R-\eta)^2/R/(R-\eta)$       \\ 
\hline\hline
\end{tabular}
\label{tab:h1h2}
\end{table}
%%%%%%%%%%%%%%%%%%%%%%%%%%%%%%%%%%%

%%%%%%%%%%%%%%%%%%%%%%%%%%%%%%%%%%%%%%%%%%%%%%%%
\section{Approximate relations}
\label{sec:IV}
%%%%%%%%%%%%%%%%%%%%%%%%%%%%%%%%%%%%%%%%%%%%%%%%

In this appendix, we derive the approximate relation between $\psi(R)$ and $\alpha$ by expanding Eqs.~(\ref{eq:psi}) and (\ref{eq:alpha}) with a small parameter $u := r_g/R$ up to the second order of $u$, where $r_g$ denotes the gravitational radius of considered spacetime. The approximate relation in the Schwarzschild spacetime up to the linear order of $u$ has been derived by Beloborodov \cite{Beloborodov2002}. Since the metric functions are expressed as $A(R)=1+(a_1/r_g)u+(a_2/r_g^2)u^2+{\cal O}(u^3)$ and $C(R)=R^2[1+(c_1/r_g)u+(c_2/r_g^2)u^2+{\cal O}(u^3)]$, from Eq.~(\ref{eq:alpha}) one can derive that
\begin{equation}
   \sin\alpha = \frac{b}{R}\left[1+\frac{a_1-c_1}{2r_g}u + \left(-a_1^2+4a_2+3c_1^2-4c_2-2a_1c_1\right)\frac{u^2}{8r_g^2} + {\cal O}(u^3)\right].
\end{equation}
In the similar way, considering the expansion of $\psi(R)$ up to the second order of $u$, one can get the relation 
\begin{equation}
   \psi(R) = \alpha + \psi_1 u + \psi_2 u^2 + {\cal O}(u^3), \label{eq:psi_ex}
\end{equation}
where
\begin{align}
 \psi_1 =& -\frac{a_1-b_1}{2r_g}\tan{\left(\frac{\alpha}{2}\right)}, \\
 \psi_2 =& -\frac{1}{16r_g^2\sin\alpha}\bigg[4 (a_1 - b_1) (a_1 - c_1)
      - \left(-4a_1^2 + 8a_2 + b_1^2 - 4b_2 + 2b_1 c_1 + c_1^2 - 4c_2\right)\cos\alpha  \nonumber \\
     & + \left\{-8a_1^2 + 8a_2 + b_1^2 - 4b_2 - 2 b_1 c_1 + c_1^2 + 4 a_1 (b_1 + c_1) - 4 c_2\right\} \frac{\alpha}{\sin\alpha}
       \bigg]. 
\end{align}
We remark that, since $\psi_2$ can be expanded with small $\alpha$ as
\begin{align}
  \psi_2 =&  \frac{1}{24r_g^2}\left(5a_1^2 - 8a_2 - a_1 b_1 - b_1^2 + 4b_2 - a_1 c_1 - b_1 c_1 - c_1^2 + 4c_2\right)\alpha \nonumber \\
               & + \frac{1}{1440r_g^2}\left(49a_1^2 - 64a_2 - 17a_1 b_1 - 8b_1^2 + 32b_2 - 17a_1 c_1 + b_1 c_1 - 8c_1^2 + 32c_2\right)\alpha^3 
               + {\cal O}(\alpha^4),
\end{align}
the approximate relation of $\psi(R)$ expressed by Eq.~(\ref{eq:psi_ex}) still gives us zero for $\alpha=0$. Then, from Eq.~(\ref{eq:dF2}) together with Eq.~(\ref{eq:psi_ex}), one can calculate the flux radiating from the primary hot spot, $F$, as 
\begin{equation}
   F = F_1 \frac{\sin\alpha\cos\alpha}{\sin\psi}\left(\frac{d\psi}{d\alpha}\right)^{-1}.
\end{equation} 

In particular, only taking into account the linear order of $u$, one can get the following relation from Eq.~(\ref{eq:psi_ex}),
\begin{equation}
  \frac{1-\cos\alpha}{1-\cos\psi(R)} = 1 + \frac{a_1-b_1}{2r_g} u + {\cal O}(u^2).  \label{eq:Belo}
\end{equation}
This is equivalent to Eq.~(1) in Ref. \cite{Beloborodov2002}, if one considers the Schwarzschild spacetime. With Eqs.~(\ref{eq:dF2}) and (\ref{eq:Belo}), one can get 
\begin{equation}
   F = F_1 \left(1+\frac{a_1-b_1}{2R} \right) \left[\left(1+\frac{a_1-b_1}{2R}\right)\cos\psi - \frac{a_1-b_1}{2R}\right].
\end{equation} 
The flux from the antipodal hot spot $\bar{F}$ is calculated by replacing $\psi$ by $\psi+\pi$, i.e.,
\begin{equation}
   \bar{F} = F_1 \left(1+\frac{a_1-b_1}{2R} \right) \left[-\left(1+\frac{a_1-b_1}{2R}\right)\cos\psi - \frac{a_1-b_1}{2R}\right].
\end{equation} 
Whenever the both hot spots are observed simultaneously, the observed flux is given by
\begin{equation}
   F_{\rm ob} := F + \bar{F} = F_1 \left(1+\frac{a_1-b_1}{2R} \right) \frac{b_1-a_1}{R}.   \label{eq:Fob_1}
\end{equation} 
That is, the observed flux obtained from the 1st order approximation of $u$ has no dependence on $\psi$ to be constant in time.

%%%%%%%%%%%%%%%%%%%%%%%%%%%%%%%%%%%%%%%%%%%%%%%%
%\section{Applications}
\section{Applications of approximate relations to various spacetime models}
\label{sec:V}
%%%%%%%%%%%%%%%%%%%%%%%%%%%%%%%%%%%%%%%%%%%%%%%%

Now, we apply the formulas derived in the previous sections to specific examples of spacetime. In particular, we consider three cases as a spacetime outside the star, i.e., the Schwarzschil spacetime, Reissner-Nordstr\"{o}m spacetime, and the Garfinkle-Horowitz-Strominger spacetime \cite{GHS1991}. As a neutron star model, again we consider the objects with $R_c=10-14$ km and $M=1.4-1.8M_\odot$.

%%%%%%%%%%%%%%%%%%%%%%%%%%%%%%%%%%%%%%%%%%%%%%%%
\subsection{Schwarzschild spacetime}
\label{sec:V-a}
%%%%%%%%%%%%%%%%%%%%%%%%%%%%%%%%%%%%%%%%%%%%%%%%

\begin{figure*}
\begin{center}
\begin{tabular}{cc}
\includegraphics[scale=0.5]{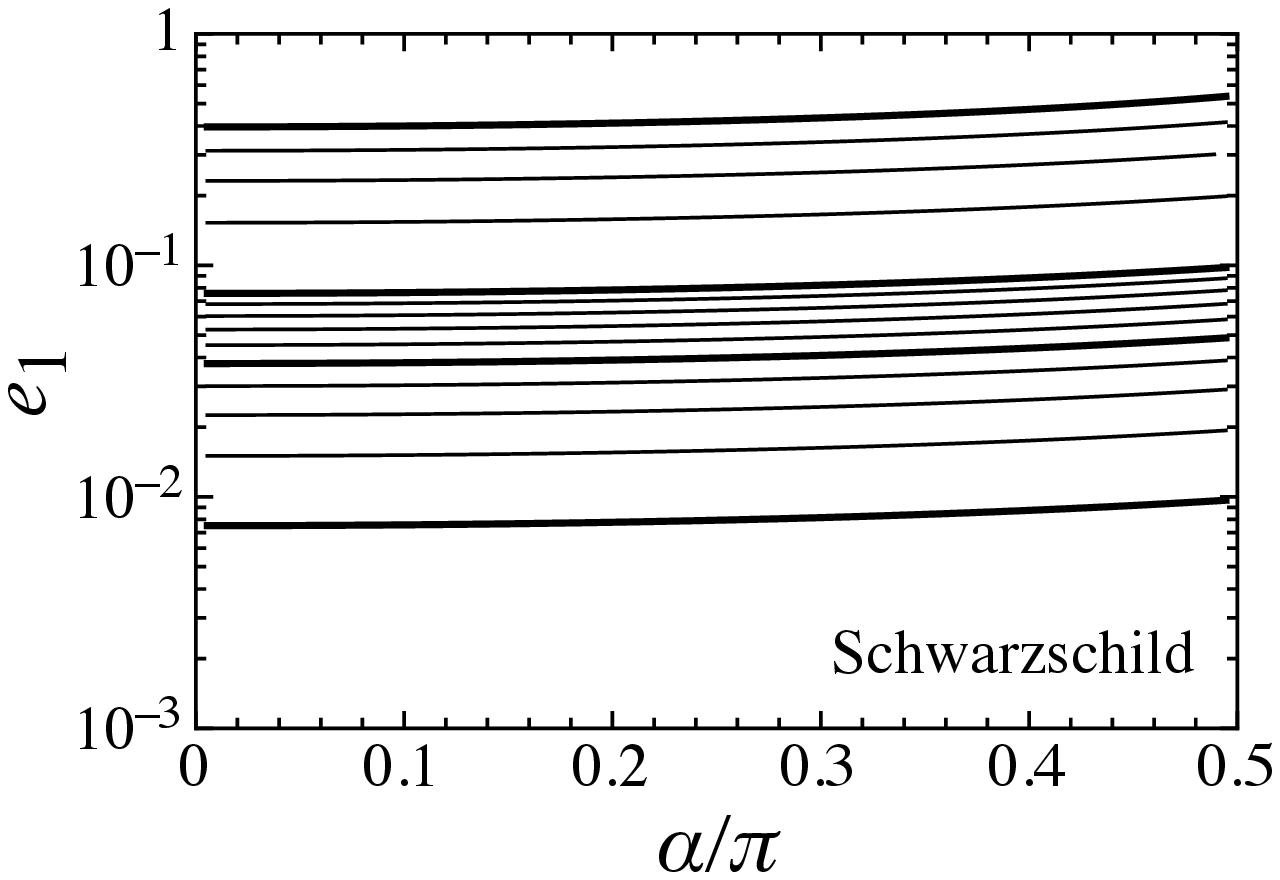} &
\includegraphics[scale=0.5]{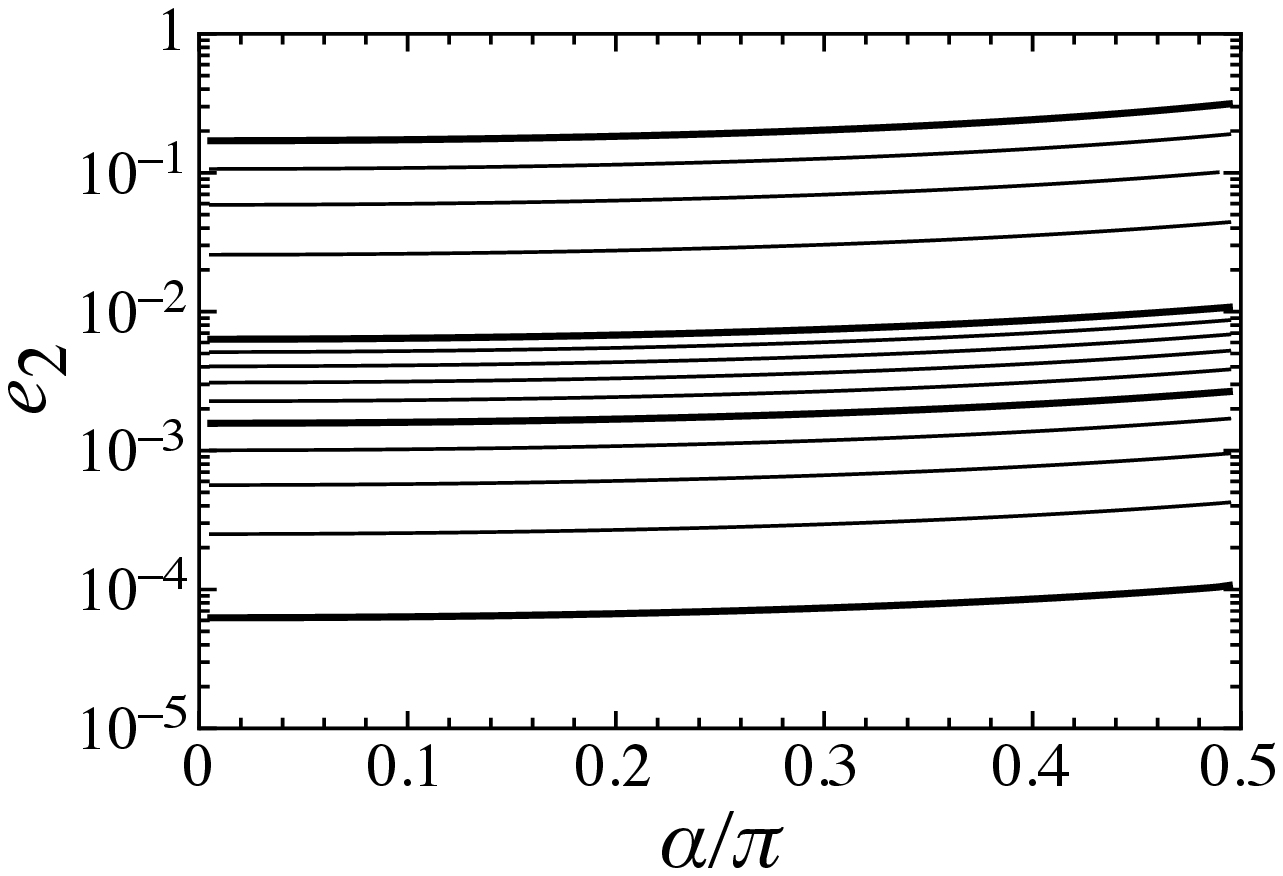} 
\end{tabular}
\end{center}
\caption{%%
Relative error in the bending angle $\beta=\psi-\alpha$ with various sets of $(\alpha,u)$ for the Schwarzschild spacetime. The left and right panels respectively correspond to $e_1$ and $e_2$ defined as $e_1=(\beta_f-\beta_1)/\beta_f$ and  $e_2=(\beta_f-\beta_2)/\beta_f$, where $\beta_f$ is the bending angle calculated with Eq.~(\ref{eq:psi}), while $\beta_1$ and $\beta_2$ are calculated with Eq.~(\ref{eq:psi_ex}) up to the linear order of $u$ and Eq.~(\ref{eq:psi_ex}) up to the second order of $u$, respectively. In the figure, the lines denote the values of $e_1$ and $e_2$ with the fixed value of $u$, i.e., $u=0.01$, 0.02, 0.03, 0.04, 0.05, 0.06, 0.07, 0.08, 0.09, 0.1, 0.2, 0.3, 0.4, and 0.5 in order from the bottom.
}%%
\label{fig:aeS}
\end{figure*}

The gravitational radius is given by $r_g=2M$. We examine the accuracy of the approximate relation given by Eq.~(\ref{eq:psi_ex}). For this purpose, we calculate the bending angle $\beta := \psi-\alpha$ with various sets of $(\alpha,u)$. In Fig.~\ref{fig:aeS}, we show the relative error of $\beta$ with fixed value of $u$ as a function of $\alpha$. The left and right panels correspond to the relative error $e_1$ and $e_2$ defined by 
\be
	e_1:= \frac{ \beta_f-\beta_1}{ \beta_f },
	\;\;\;
	e_2 := \frac{ \beta_f-\beta_2 }{ \beta_f },
\ee where $\beta_f$ is the bending angle calculated with Eq.~(\ref{eq:psi}), while $\beta_1$ and $\beta_2$ are calculated with Eq.~(\ref{eq:psi_ex}) up to the linear order of $u$ and Eq.~(\ref{eq:psi_ex}) up to the second order of $u$, respectively. Since typical mass and radius of a neutron star are $M\simeq 1.4M_\odot$ and $R_c\simeq 12$ km, which leads to $u=2M/R_c\simeq 0.345$, the accuracy of the approximate relation [Eq.~(\ref{eq:psi_ex})] in the bending angle is only $\sim 32\%$ if one takes into account only linear order of $u$ and $\sim 11\%$ even if one takes into account up to the second order of $u$. We notice that we cannot reproduce the result obtained by Beloborodov, i.e., Fig.~2 in \cite{Beloborodov2002}, where he concluded that the relative error is at most $3\%$ with $u=1/3$ even though he took into account only linear order of $u$. Here, two authors in the present paper independently performed numerical integrations to obtain the data in Fig.~\ref{fig:aeS} with completely different scheme and obtained the same results. So, while we could not identify the reason why our results are different from those obtained by Beloborodov \cite{Beloborodov2002}, we believe that our results are correct.

%%%%%%%%%%%%%%%%%%%%%%%%%%%%%%%%%%%
% Figure 10
%%%%%%%%%%%%%%%%%%%%%%%%%%%%%%%%%%%
%%%%%%%%%%%%%%%%%%%%%%%%%%%%%%%%%%%

\begin{figure*}
\begin{center}
\begin{tabular}{cc}
\includegraphics[scale=0.5]{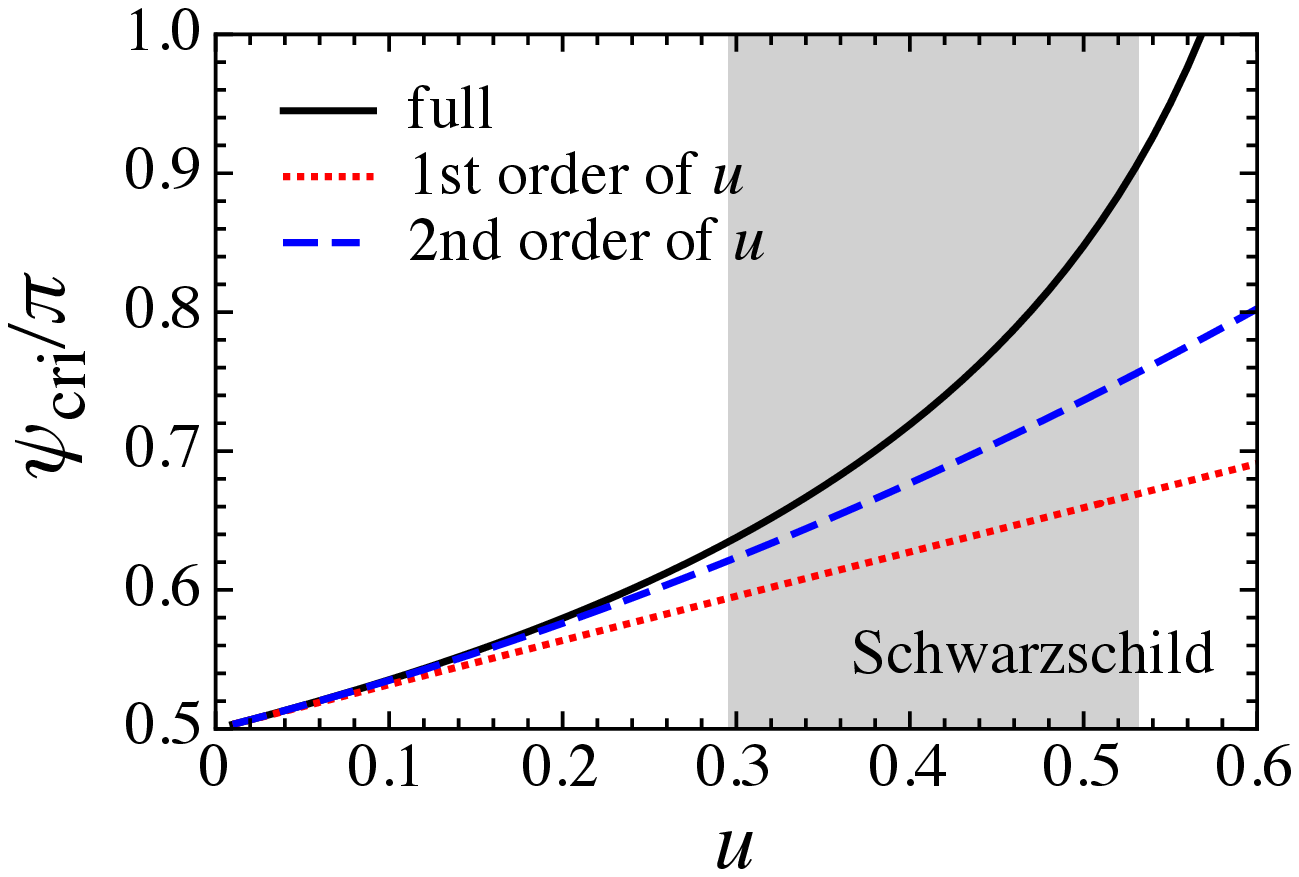} &
\includegraphics[scale=0.5]{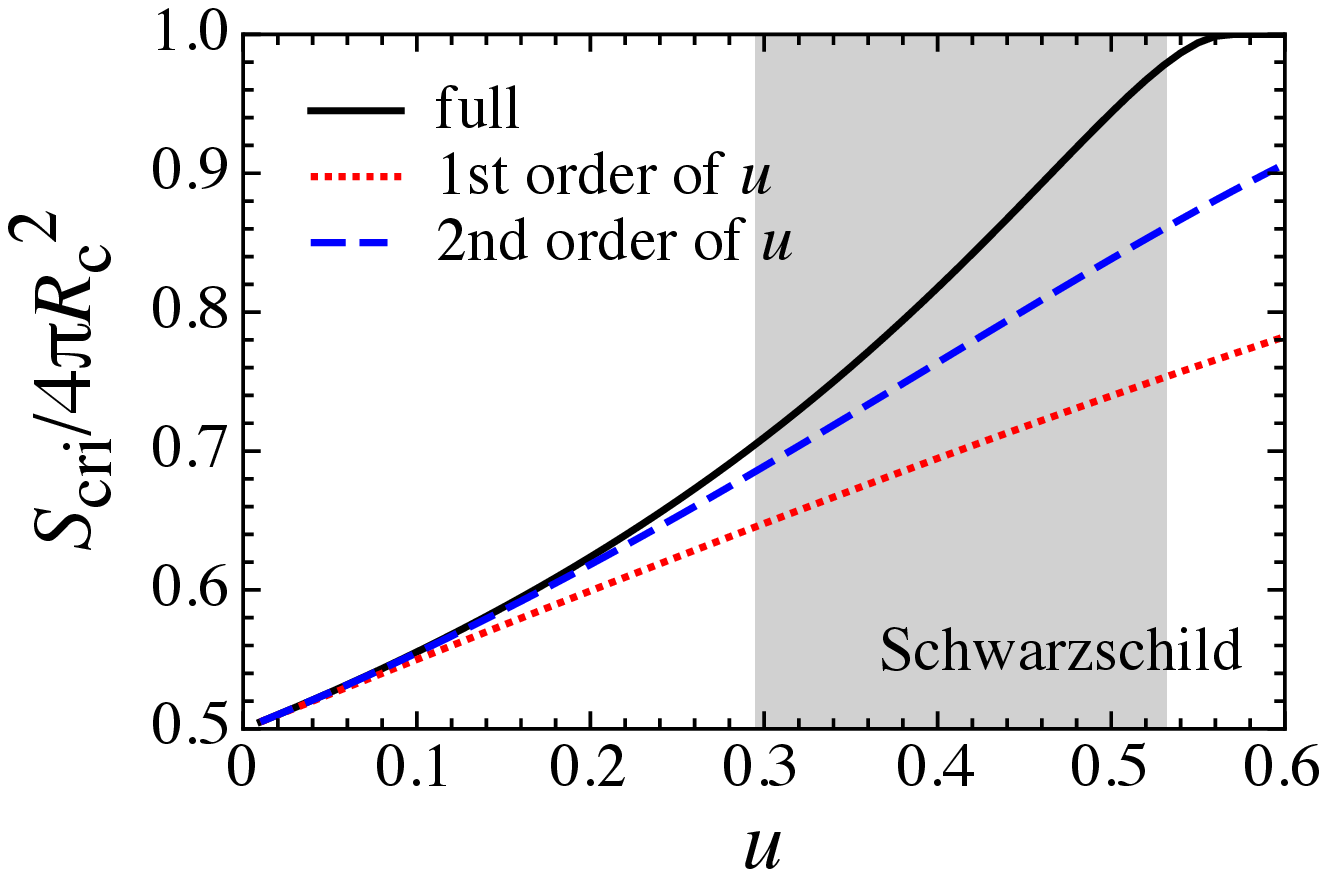} 
\end{tabular}
\end{center}
\caption{%%
The left and right panels are the value of $\psi_{\rm cri}$ and the visible fraction of the stellar surface $S_{\rm cri}/4\pi R_c^2$, respectively, as a function of $u$. In the both panels, the solid, dotted, and dashed lines respectively correspond to the full order numerical values, approximate values obtained from Eq.~(\ref{eq:psi_ex}) up to the 1st order of $u$, and those from Eq.~(\ref{eq:psi_ex}) up to the 2nd order of $u$. The shaded region corresponds to the possible value of $u$ for the stellar models with $R_c=10-14$ km and $M=1.4-1.8M_\odot$.
}%%
\label{fig:critical_S}
\end{figure*}

In Fig.~\ref{fig:critical_S}, we show the value of $\psi_{\rm cri}$ as a function of $u$ in the left panel, while the visible fraction of the stellar surface is in the right panel. In the both panels, we show the results obtained by the full order numerical integration (solid line), by the 1st order approximation of $u$ (dotted line), and by the 2nd order approximation of $u$ (dashed line). In addition, the shaded region denotes that of $u$ for the neutron star models with $R_c=10-14$ km and $M=1.4-1.8M_\odot$, which leads to that $u$ becomes in the range of $0.295\le u\le0.532$. From this figure, we find that $\psi_{\rm cri}$ obtained from the full order numerical integration can be $\pi$ for the stellar model with larger compactness, such as $u\ge 0.568$. On the other hand, the value of $\psi_{\rm cri}$ obtained with the approximate relation up to the 1st and 2nd order of $u$ cannot reach $\pi$. As expected, such a deviation from the full order value becomes large with stellar compactness defined by $M/R_c = u/2$.

%%%%%%%%%%%%%%%%%%%%%%%%%%%%%%%%%%%
% Figure 11
%%%%%%%%%%%%%%%%%%%%%%%%%%%%%%%%%%%
%%%%%%%%%%%%%%%%%%%%%%%%%%%%%%%%%%%

\begin{figure*}
\begin{center}
\begin{tabular}{ccc}
\includegraphics[scale=0.4]{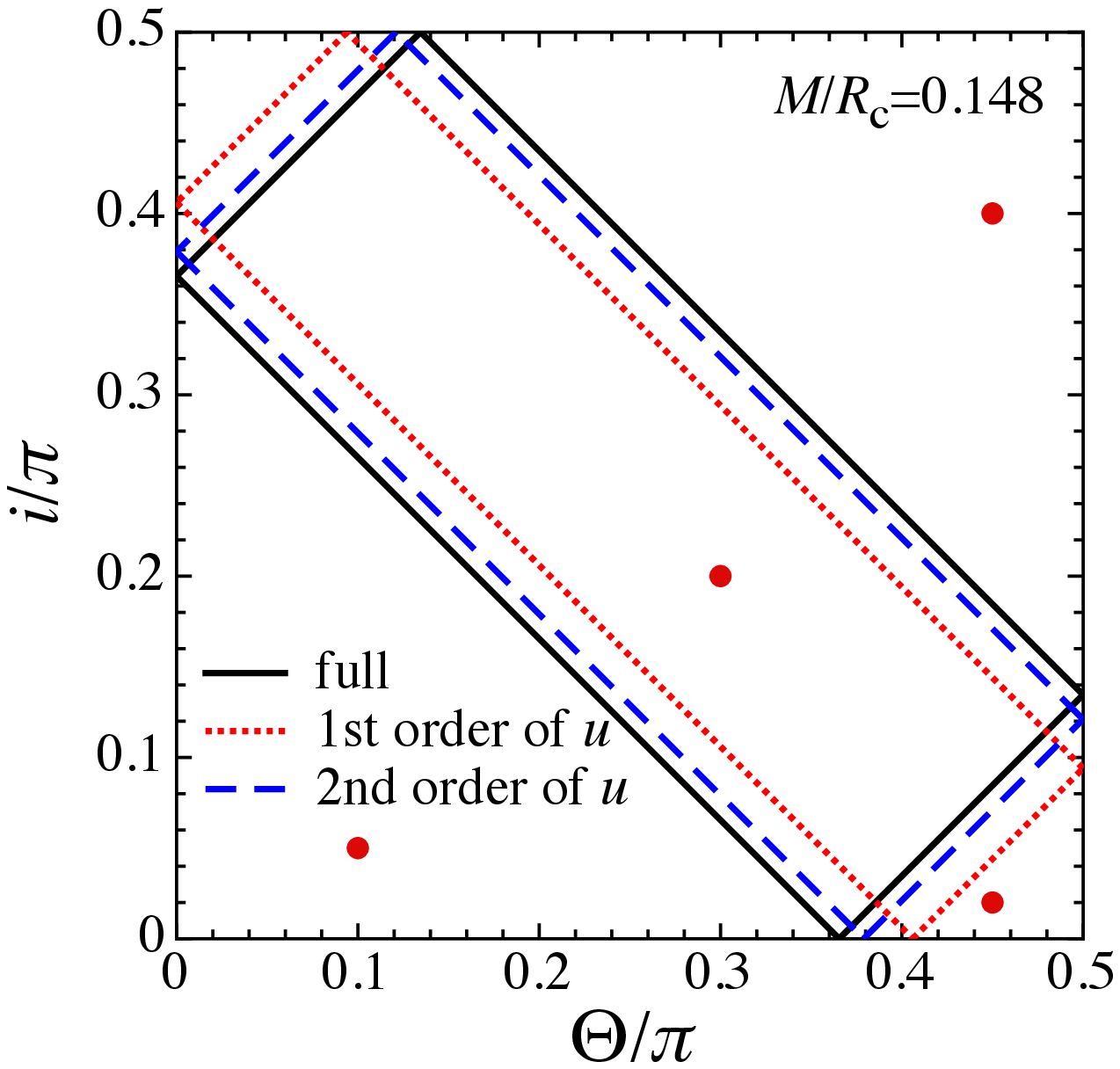} &
\includegraphics[scale=0.4]{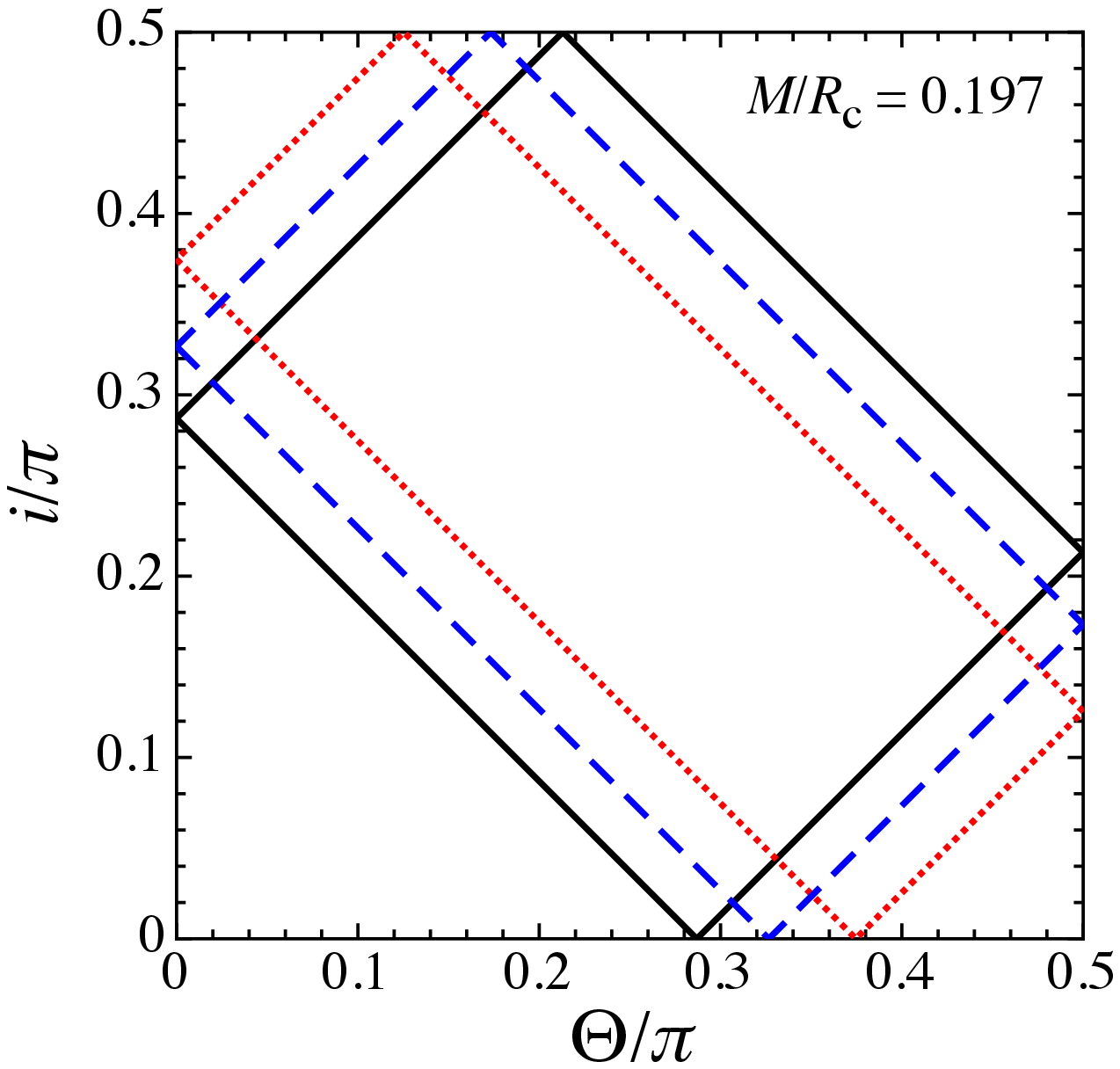} &
\includegraphics[scale=0.4]{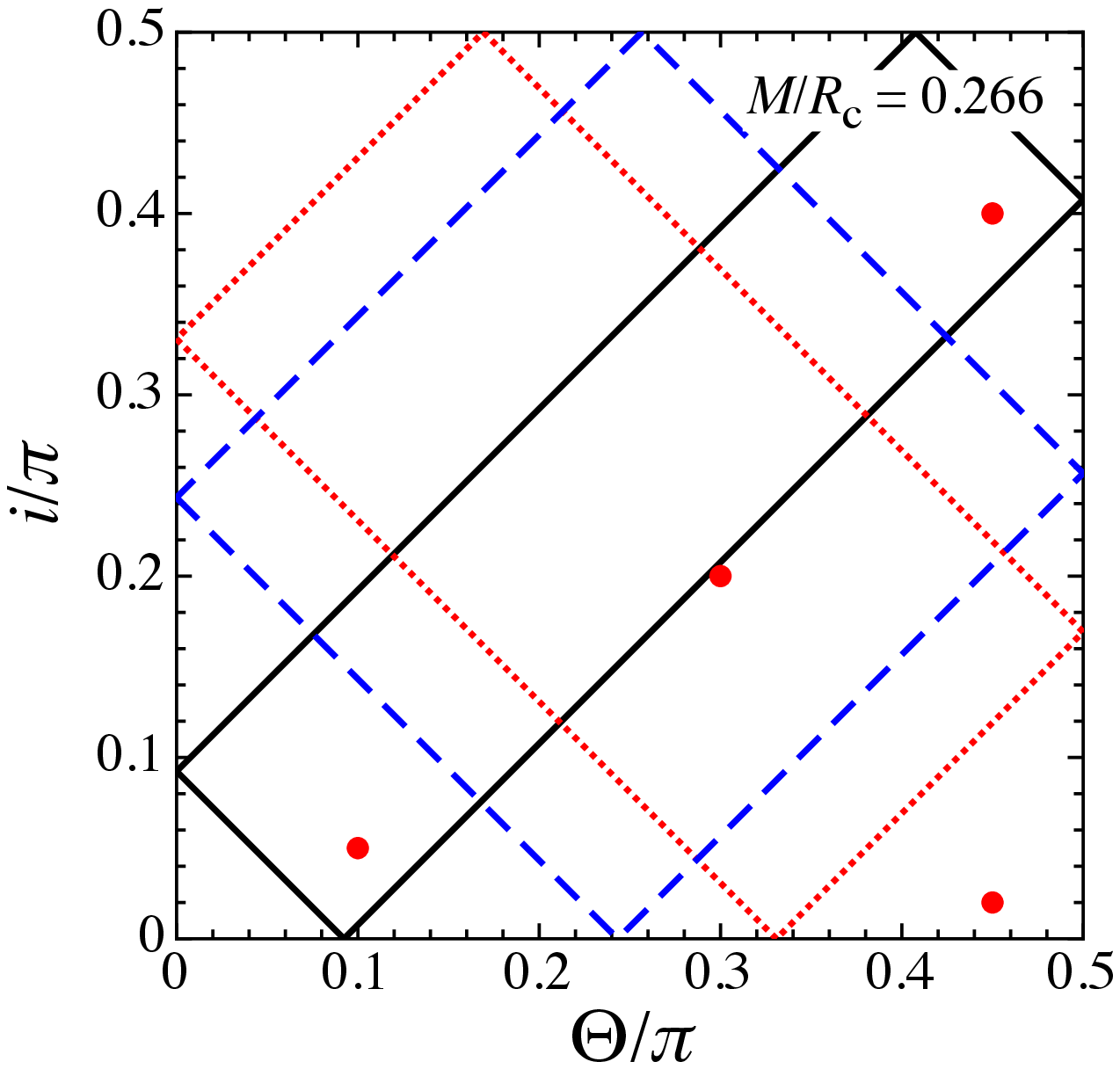} 
\end{tabular}
\end{center}
\caption{%%
The classification whether the two hot spots are observed or not is shown as a dependence on the angle $\Theta$ and $i$ for the Schwarzschild spacetime, as in Fig.~\ref{fig:itheta0}. The left, middle, and right panels correspond to the stellar models with $M/R_c=0.148$, $0.197$, $0.266$, respectively. In the figure, the solid, dotted, and dashed lines denote the boundaries of classification with $\psi_{\rm cri}$ obtained by the full order numerical integration, the 1st order approximation of $u$, and the 2nd order approximation of $u$. The dots in the left and right panels denote the stellar models with which the pulse profiles are shown in Fig.~\ref{fig:pulse-S}.
}%%
\label{fig:ithetaS}
\end{figure*}

Using the values of $\psi_{\rm cri}$ obtained by the full order numerical integration and by the 1st order and 2nd order approximations, in Fig.~\ref{fig:ithetaS} we show divide the region of $(\Theta, i)$ into I, II, III, and IV regions for the stellar models with $M/R_c=0.148$ in the left panel, $0.197$ in the middle panel, and $0.266$ in the right panel, respectively, where the solid, dotted, and dashed lines denote the results obtained from the full order integration, the 1st order approximation, and the 2nd order approximation, respectively. From this figure, one can observe that the results with the approximate relations are not so bad for the stellar model with lower compactness (the left panel in Fig.~\ref{fig:ithetaS}), while those are not acceptable for the stellar model with higher compactness (the right panel in Fig.~\ref{fig:ithetaS}).

%%%%%%%%%%%%%%%%%%%%%%%%%%%%%%%%%%%
% Figure 12
%%%%%%%%%%%%%%%%%%%%%%%%%%%%%%%%%%%
%%%%%%%%%%%%%%%%%%%%%%%%%%%%%%%%%%%

\begin{figure*}
\begin{center}
\begin{tabular}{ccc}
\includegraphics[scale=0.5]{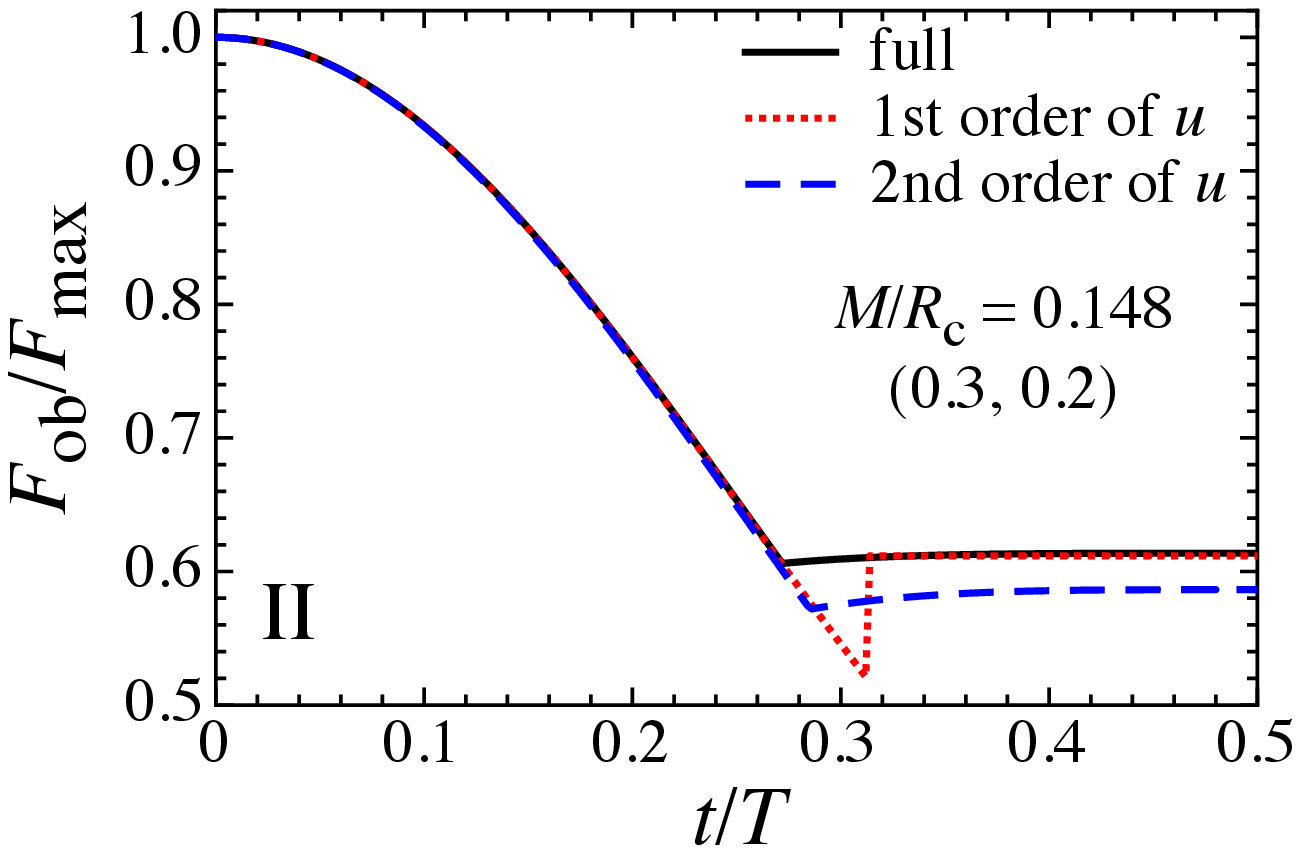} &
\includegraphics[scale=0.5]{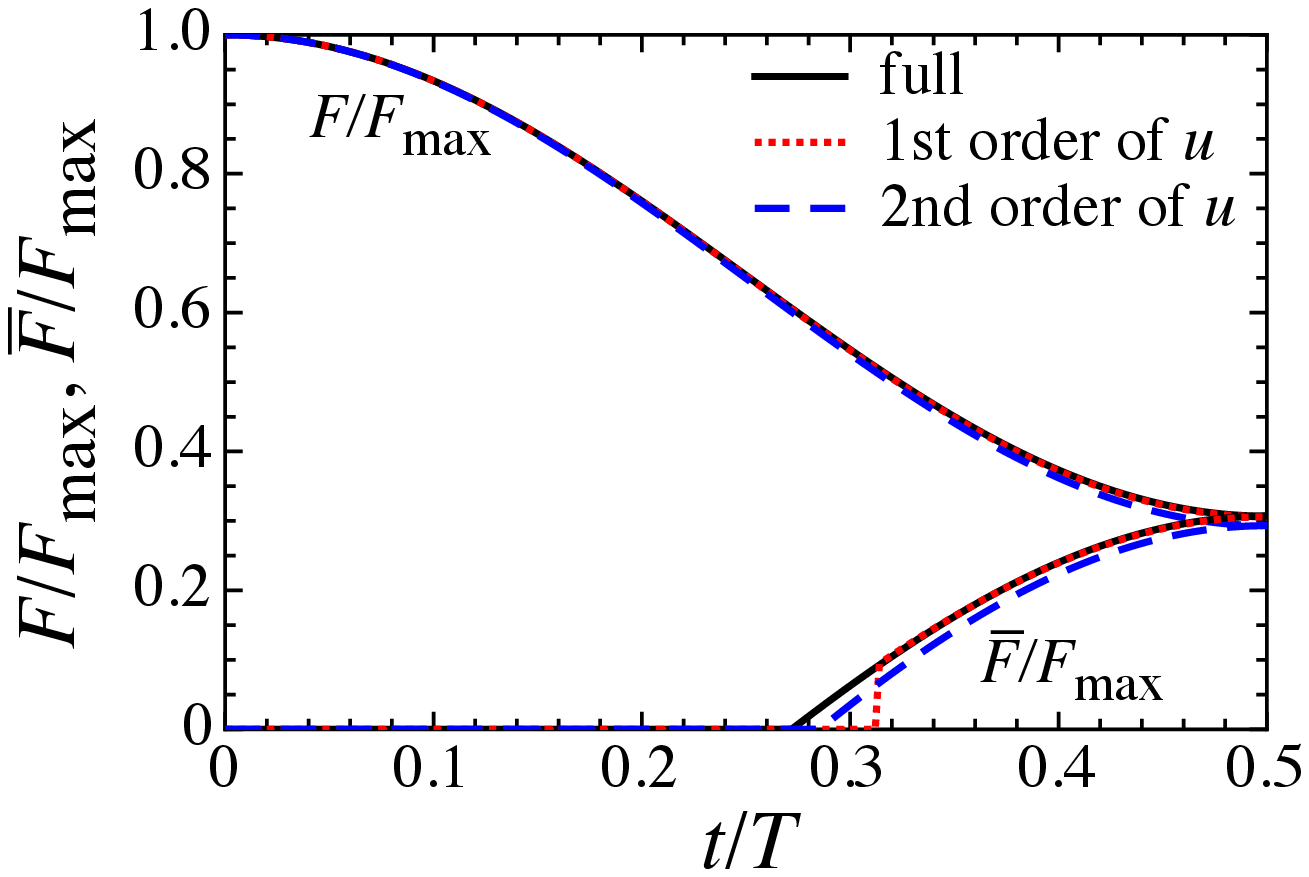} 
\end{tabular}
\end{center}
\caption{%%
The observed flux $F_{\rm ob}$ normalized by the maximum flux $F_{\rm max}$ is shown in the left panel, while the flux from the primary $F$ and the antipodal hot spots $\bar{F}$ normalized by $F_{\rm max}$ are shown in the right panel. The results are for the stellar model with $M/R_c=0.148$ and for $(\Theta/\pi,i/\pi)=(0.3,0.2)$. The solid, dotted, and dashed lines denote the results obtained from the full order numerical integration, the 1st order approximation of $u$, and the 2nd order of approximation of $u$.
}%%
\label{fig:pulse-M14SII}
\end{figure*}

In Fig.~\ref{fig:pulse-M14SII}, we consider the pulse profiles from the stellar models denoted in Fig.~\ref{fig:ithetaS} with the dots. In particular, we show the results for the stellar model with $M/R_c=0.148$, adopting $(\Theta/\pi,i/\pi)=(0.3,0.2)$, where the left panel shows the observational flax $F_{\rm ob}$ normalized by the maximum flax $F_{\rm max}$ and the right panel shows the flux radiated from the primary spot $F$ and from the antipodal spot $\bar{F}$ normalized by $F_{\rm max}$. We remark that this model correspond to the class II in Fig.~\ref{fig:itheta0}, as shown in the left panel of Fig.~\ref{fig:ithetaS}. In the both panels, the solid, dotted, and dashed lines denote the results obtained by the full order numerical integration, the 1st order approximation, and the 2nd order approximation. As shown in Fig.~\ref{fig:critical_S}, the values of $\psi_{\rm cri}$ with the approximate relations of $u$ are estimated lower than that with the full order numerical integration, which leads to that the flux radiated from the antipodal spot is more difficult to observe. From the right panel of Fig.~\ref{fig:pulse-M14SII}, one can see this point, i.e., $\bar{F}$ obtained with the approximate relation of $u$ appears later. In the case with the 1st order approximation of $u$, since the observed flux becomes constant given by Eq.~(\ref{eq:Fob_1}) once the antipodal spot becomes visible, $F_{\rm ob}$ exhibits a jump at the time when $\bar{\mu}=\cos\psi_{\rm cri}^{(1)}$ where $\psi_{\rm cri}^{(1)}$ denotes the value of $\psi_{\rm cir}$ obtained with the 1st order approximation of $u$. On the other hand, the shape of the pulse profiles with the 2nd order approximation might be better than that with the 1st order approximation in the sense of the pulse profile without jump.

%%%%%%%%%%%%%%%%%%%%%%%%%%%%%%%%%%%
% Figure 13
%%%%%%%%%%%%%%%%%%%%%%%%%%%%%%%%%%%
%%%%%%%%%%%%%%%%%%%%%%%%%%%%%%%%%%%

\begin{figure*}
\begin{center}
\begin{tabular}{c}
\includegraphics[scale=0.4]{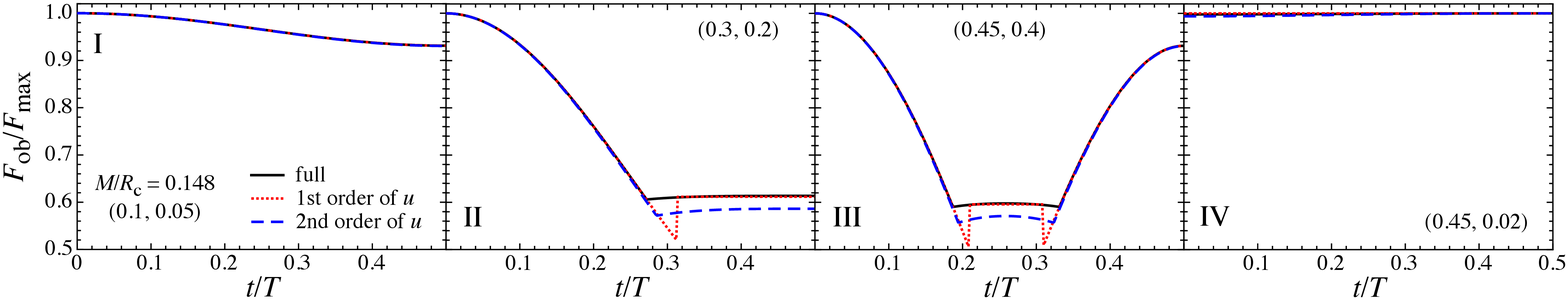} \\
\includegraphics[scale=0.4]{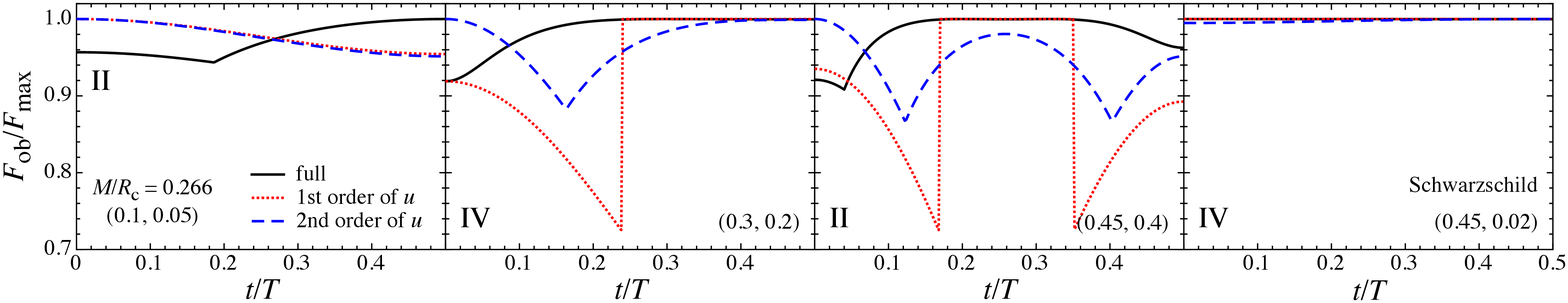}
\end{tabular}
\end{center}
\caption{%%
The observed flux normalized by the maximum flux with various angles of $\Theta/\pi$ and $i/\pi$ for two stellar models. The upper and lower panels correspond to the pulse profiles from the stellar models with $M/R_c=0.148$ and $0.266$. The panels from left to right correspond to the profiles with $(\Theta/\pi,i/\pi)=(0.1,0.05)$, $(0.3,0.2)$, $(0.45,0.4)$, and $(0.45,0.02)$, which denote in Fig.~\ref{fig:ithetaS} with the dots. The solid, dotted, and dashed lines denote the results obtained by the full order numerical integration, the 1st order approximation of $u$, and the 2nd order approximation of $u$. In the panels, I, II, III, and IV denote the classification as shown in Fig.~\ref{fig:itheta0} with the results obtained by the full order numerical integration. 
}%%
\label{fig:pulse-S}
\end{figure*}

In Fig.~\ref{fig:pulse-S}, we show the observed flux $F_{\rm ob}$ normalized by $F_{\rm max}$ for the stellar model with $M/R_c=0.148$ in the upper panel and with $M/R_c=0.266$ in the lower panel. The panels from left to right for each stellar model respectively correspond to the results for $(\Theta/\pi,i/\pi)=(0.1,0.05)$, $(0.3,0.2)$, $(0.45,0.4)$, and $(0.45,0.02)$. We remark that Fig.~\ref{fig:pulse-M14SII} corresponds to the second panel from left in the upper row. From this figure, one can see that the pulse profiles for the stellar model with higher compactness are quite difficult to reproduce with the approximation of $u$. In fact, as shown in Fig.~\ref{fig:ithetaS}, the classification itself whether the two hot spots are observed or not with the approximation of $u$ becomes different from that with the full order numerical integration, e.g., the results for the stellar model with $M/R_c=0.266$ except for the rightmost panel.

%%%%%%%%%%%%%%%%%%%%%%%%%%%%%%%%%%%
% Figure 14
%%%%%%%%%%%%%%%%%%%%%%%%%%%%%%%%%%%
%%%%%%%%%%%%%%%%%%%%%%%%%%%%%%%%%%%

%%%%%%%%%%%%%%%%%%%%%%%%%%%%%%%%%%%%%%%%%%%%%%%%
\subsection{Reissner-Nordstr\"{o}m spacetime}
\label{sec:V-b}
%%%%%%%%%%%%%%%%%%%%%%%%%%%%%%%%%%%%%%%%%%%%%%%%

\begin{figure*}
\begin{center}
\begin{tabular}{cc}
\includegraphics[scale=0.5]{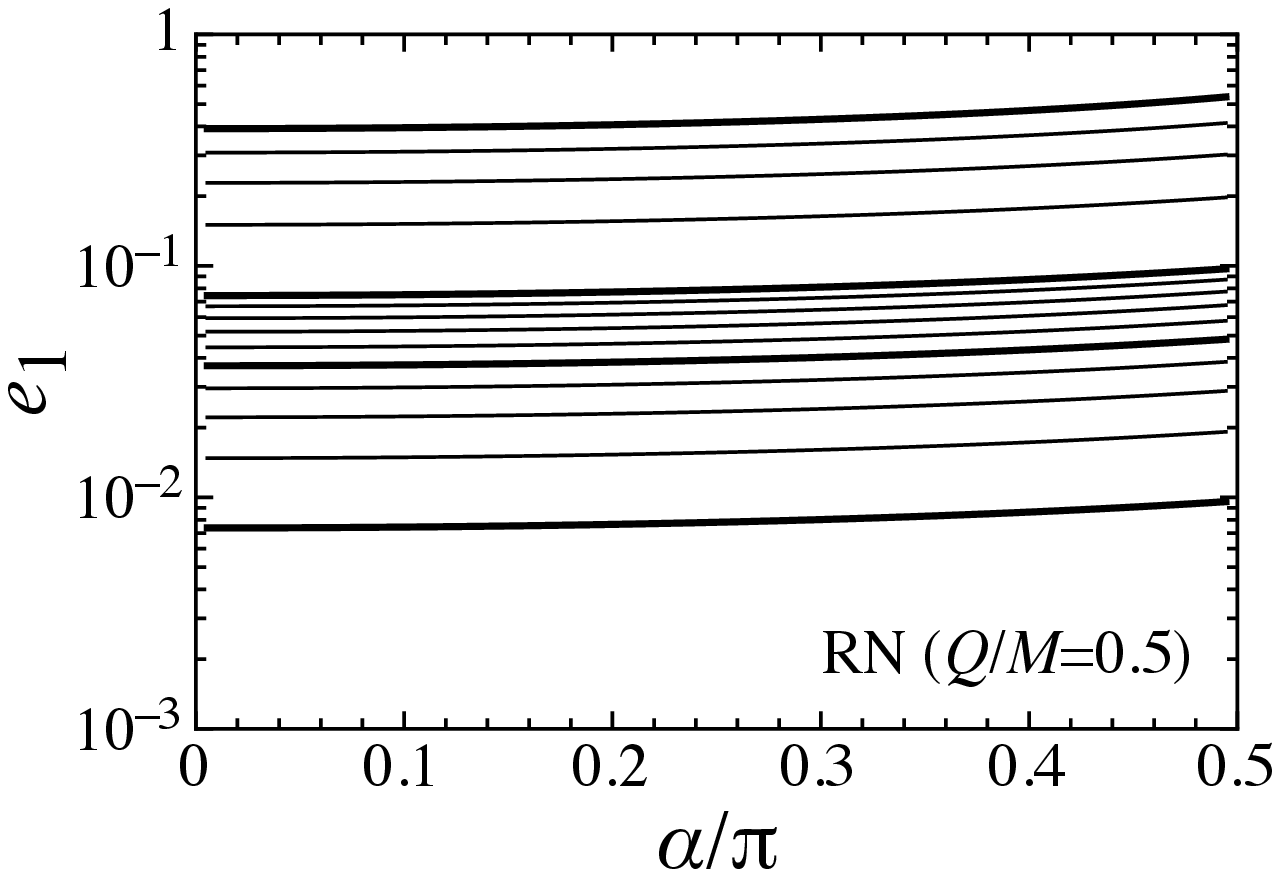} &
\includegraphics[scale=0.5]{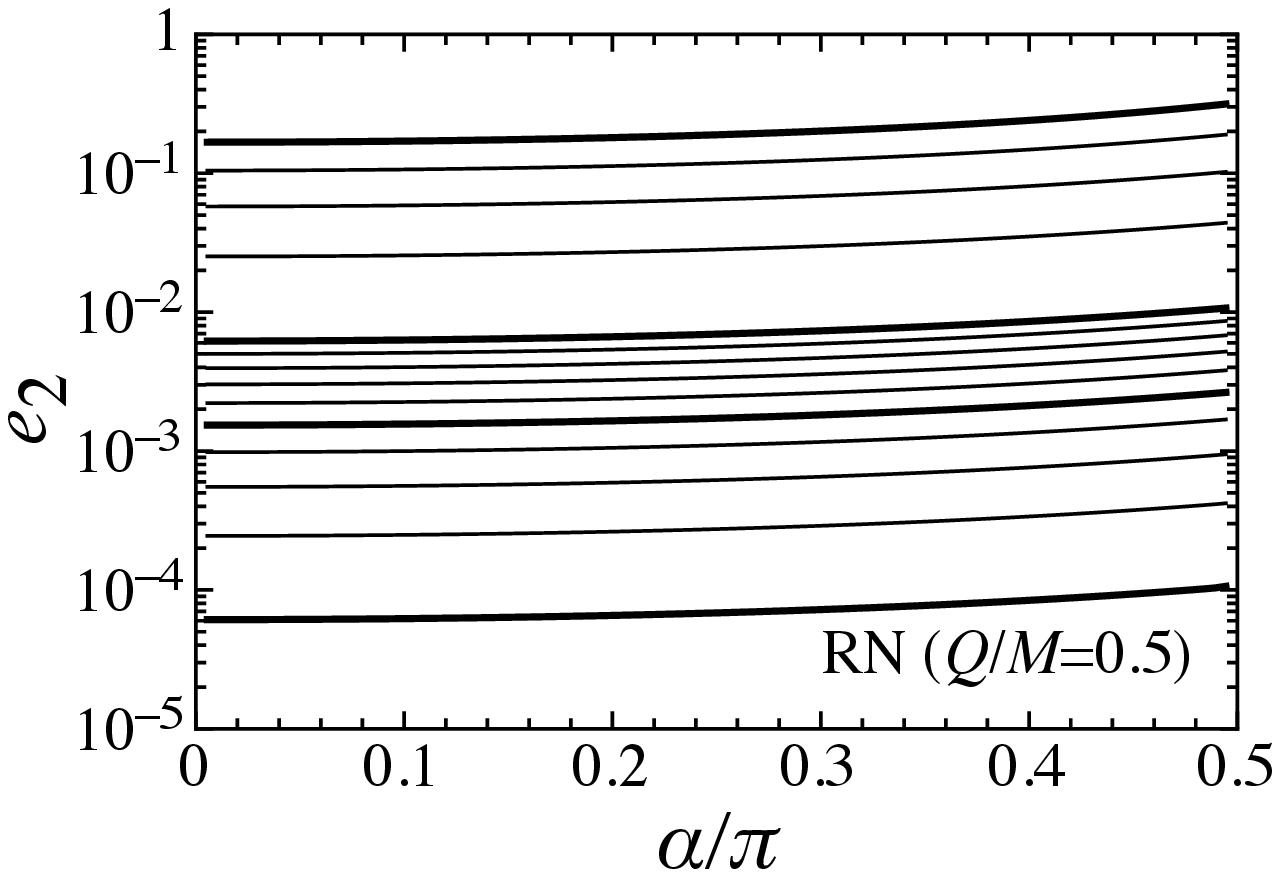} \\
\includegraphics[scale=0.5]{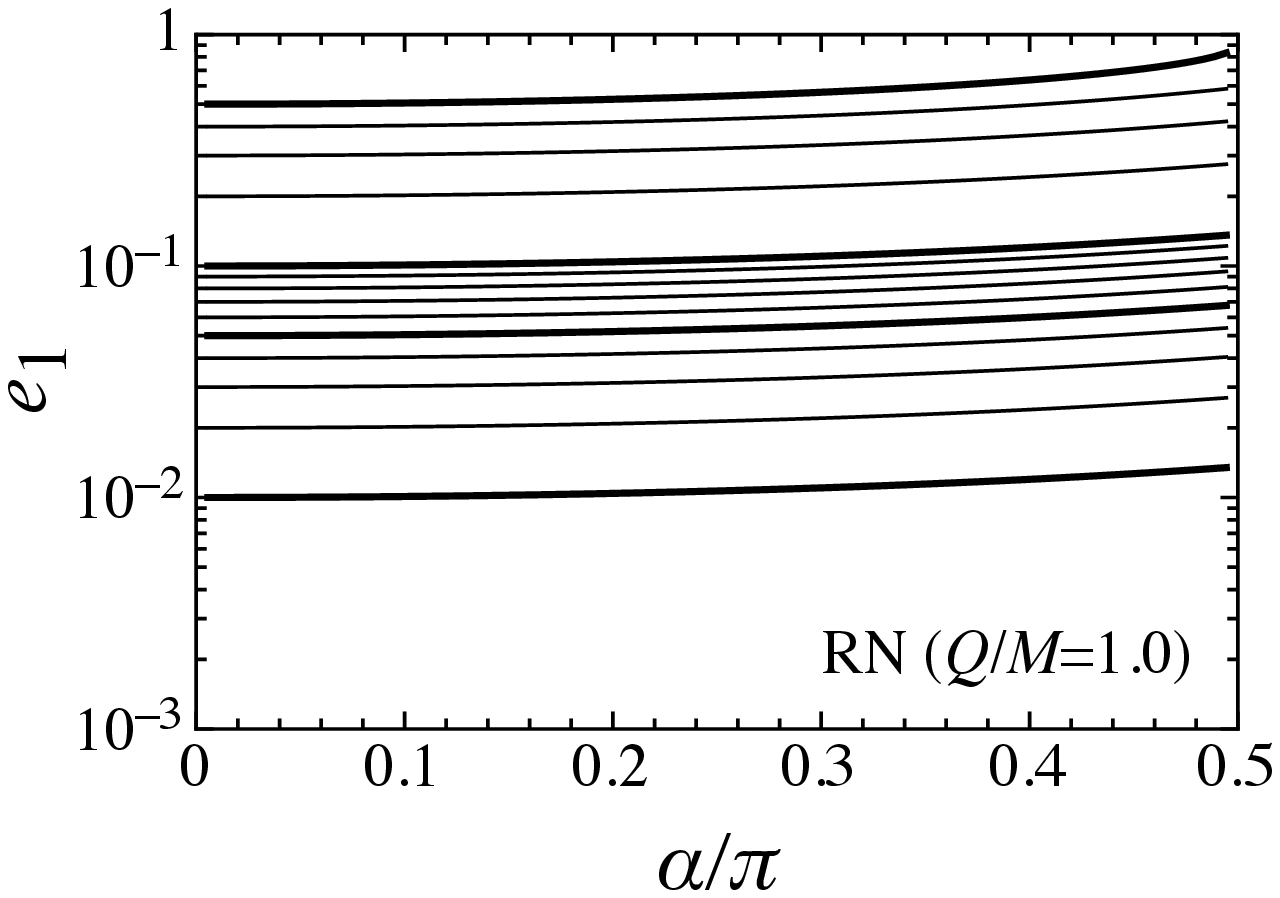} &
\includegraphics[scale=0.5]{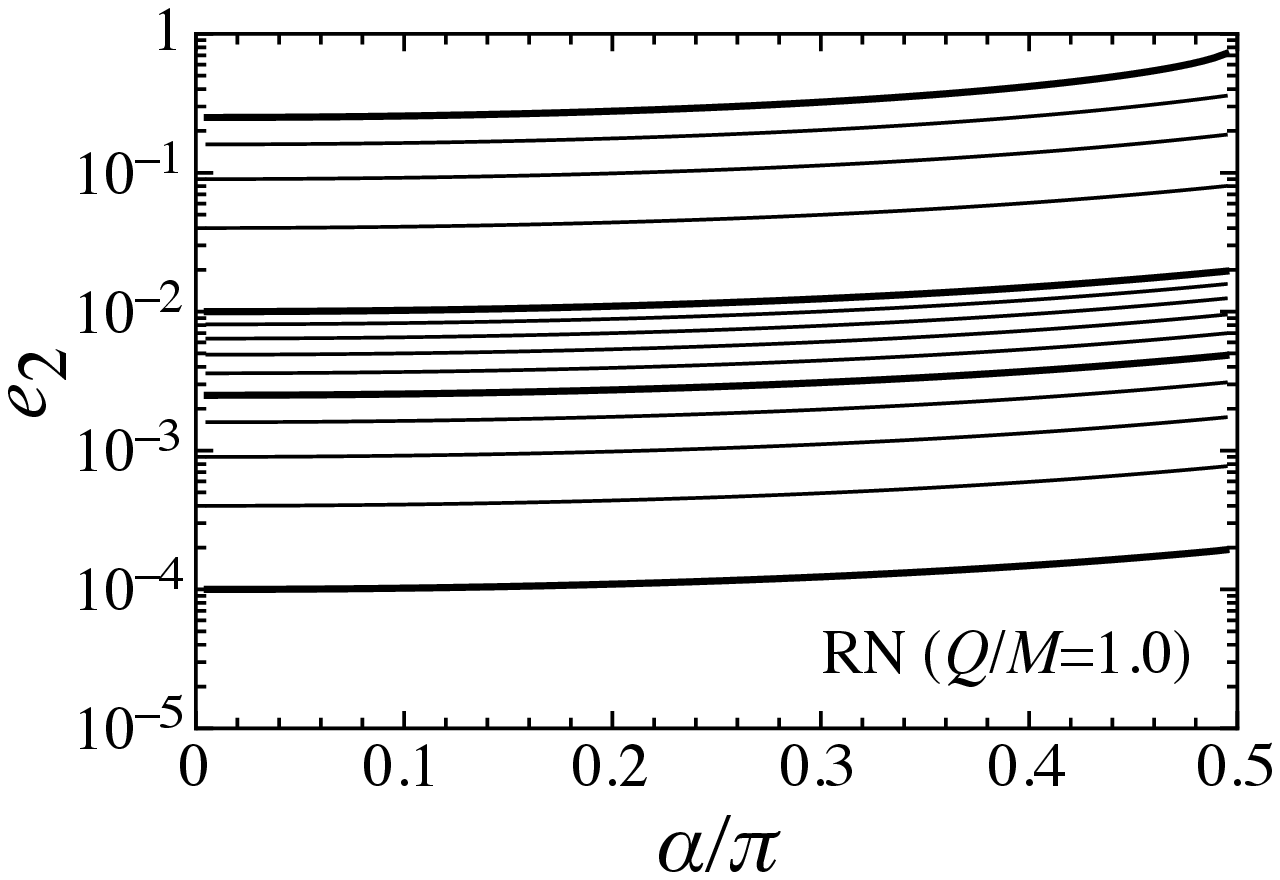} \\
\end{tabular}
\end{center}
\caption{%%
Same as Fig.~\ref{fig:aeS}, but for the Reissner-Nordstr\"{o}m spacetime. The upper and lower panels correspond to the cases for $Q/M=0.5$ and 1.0, respectively.
}%%
\label{fig:aeRN}
\end{figure*}

The gravitational radius is given by $r_g=M+\sqrt{M^2-Q^2}$. In Fig.~\ref{fig:aeRN}, we show the relative error in $\beta = \psi-\alpha$ with the fixed value of $u$ as a function of $\alpha$, which are obtained with the 1st order approximation ($e_1$) and with the 2nd order approximation ($e_2$), comparing with the result of the full order integration. The upper and lower panels correspond to the results for the case with $Q/M=0.5$ and 1.0, respectively. As in the Schwarzschild spacetime, the relative error, $e_1$ and $e_2$, have weak dependence on the angle $\alpha$ once the value of $u$ is fixed. For a typical neutron star model with $M=1.4M_\odot$ and $R_c=12$ km, which corresponds to $u=0.321$ for $Q/M=0.5$ and $u=0.172$ for $Q/M=1.0$, $e_1\sim 28\%$ and $e_2\sim 9\%$ for $Q/M=0.5$, while $e_1\sim 20\%$ and $e_2\sim 4\%$ for $Q/M=1.0$. Thus, for a given stellar model in the Reissner-Nordstr\"{o}m spacetime, the relative error becomes smaller with the charge $Q/M$.

%%%%%%%%%%%%%%%%%%%%%%%%%%%%%%%%%%%
% Figure 15
%%%%%%%%%%%%%%%%%%%%%%%%%%%%%%%%%%%
%%%%%%%%%%%%%%%%%%%%%%%%%%%%%%%%%%%

\begin{figure*}
\begin{center}
\begin{tabular}{cc}
\includegraphics[scale=0.5]{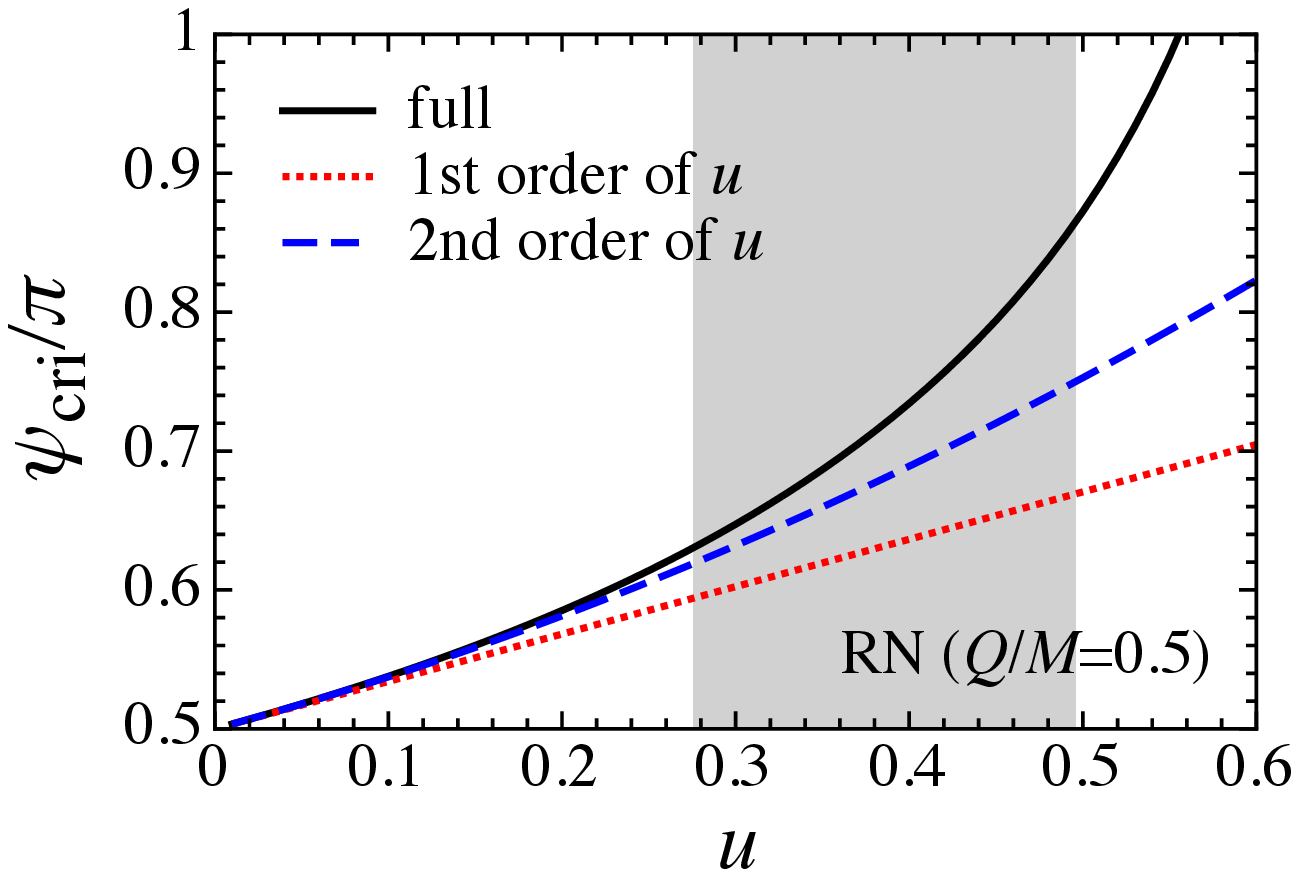} &
\includegraphics[scale=0.5]{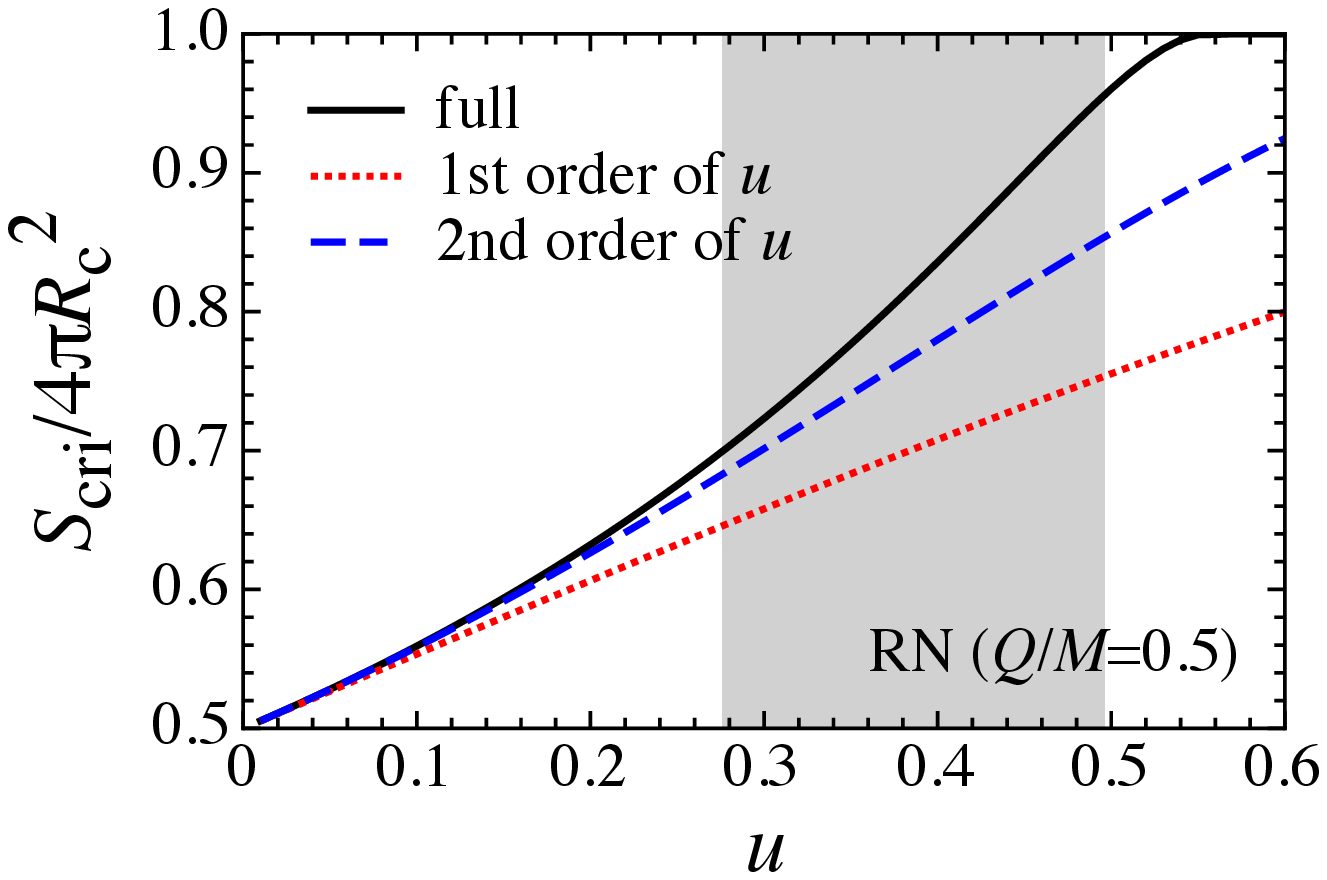} \\ 
\includegraphics[scale=0.5]{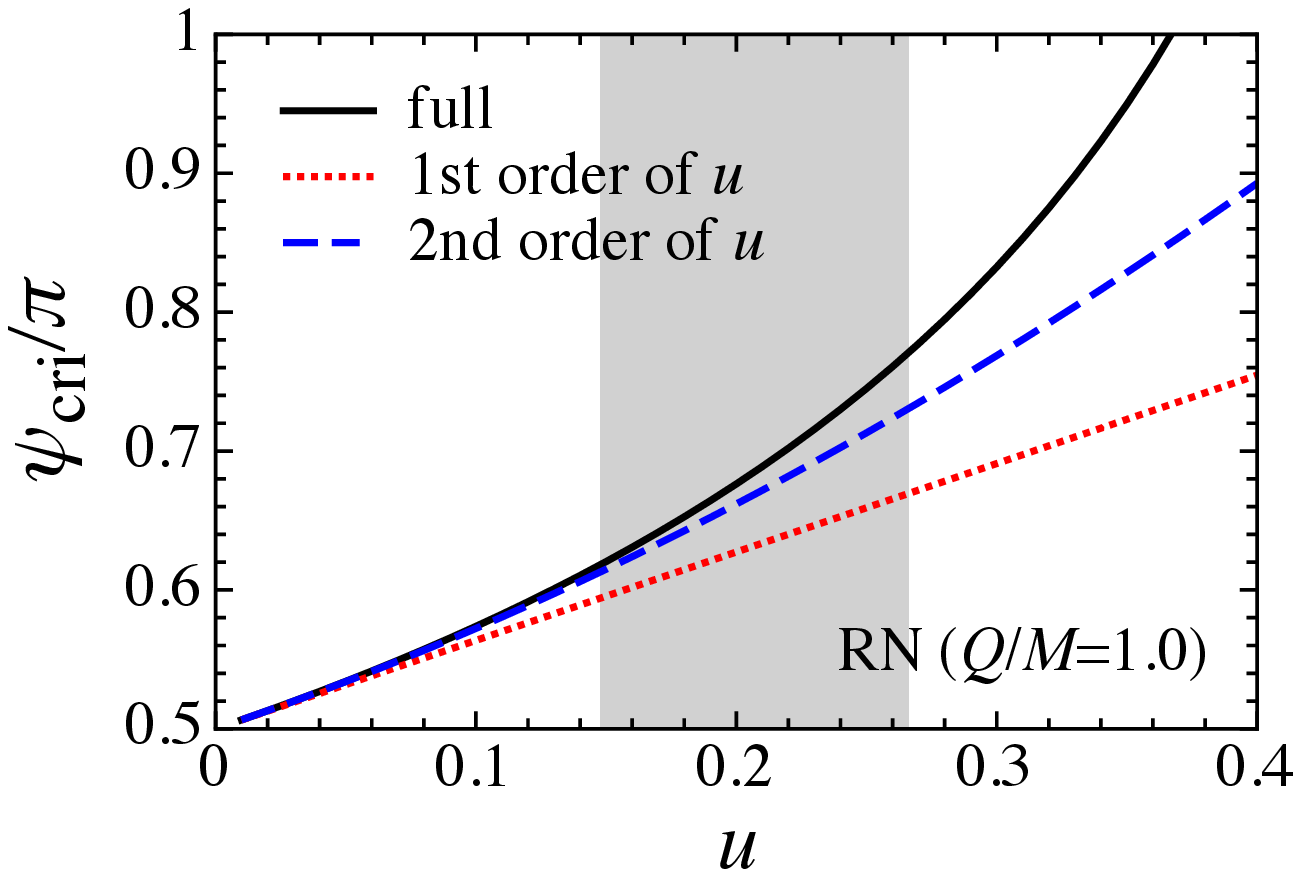} &
\includegraphics[scale=0.5]{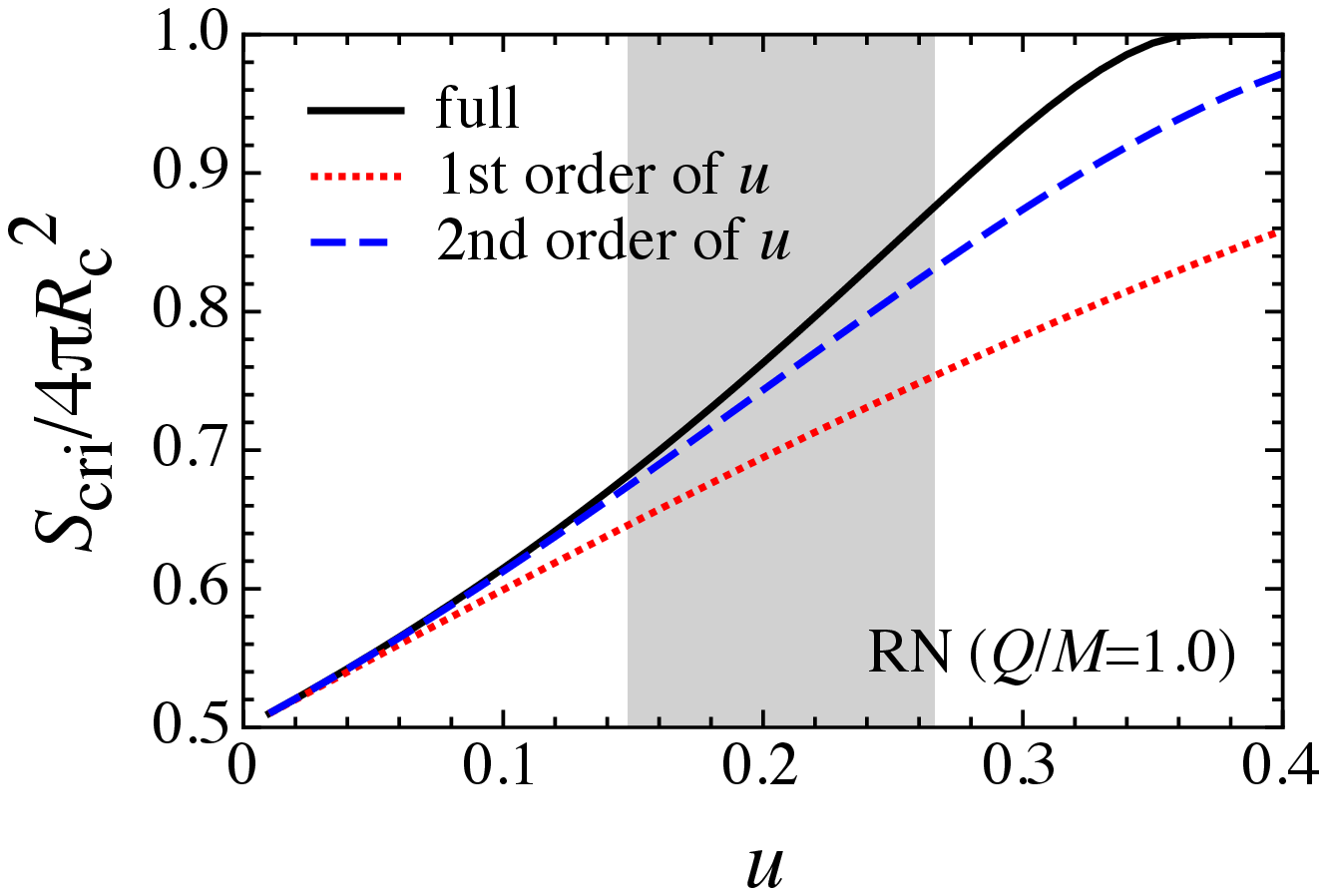} 
\end{tabular}
\end{center}
\caption{%%
Same as Fig.~\ref{fig:critical_S}, but for the Reissner-Nordstr\"{o}m spacetime. The upper and lower panels correspond to the cases for $Q/M=0.5$ and 1.0, respectively.
}%%
\label{fig:critical_RN}
\end{figure*}

In the case with $Q/M=0.5$ and $1.0$, the critical value of $\psi$ when the $\alpha$ is $\pi/2$ and the visible fraction of the stellar surface are shown as a function of $u$ in Fig.~\ref{fig:critical_RN}. In this figure, the solid, dotted, and dashed lines correspond to the results with the full order numerical integration, the 1st order approximation, and the 2nd order approximation, while the shaded region denotes the that of $u$ for the neutron star models with $R_c=10-14$ km and $M=1.4-1.8M_\odot$, i.e., $u$ becomes in the range of $0.276\le u\le0.496$ for $Q/M=0.5$, where $r_g=(2+\sqrt{3})M/2$, and $0.148\le u\le0.266$ for $Q/M=1.0$, where $r_g=M$. We remark that the critical value of $u$, where $\psi_{\rm cri}$ is $\pi$, is 0.555 and 0.367 for $Q/M=0.5$ and $1.0$, respectively. As in the Schwarzschild spacetime, $\psi_{\rm cri}$ obtained with the 1st and 2nd order approximations deviates more from $\psi_{\rm cri}$ obtained with the full order integration, as $u$ (or the stellar compactness) increases.

%%%%%%%%%%%%%%%%%%%%%%%%%%%%%%%%%%%
% Figure 16
%%%%%%%%%%%%%%%%%%%%%%%%%%%%%%%%%%%
%%%%%%%%%%%%%%%%%%%%%%%%%%%%%%%%%%%

\begin{figure*}
\begin{center}
\begin{tabular}{ccc}
\includegraphics[scale=0.4]{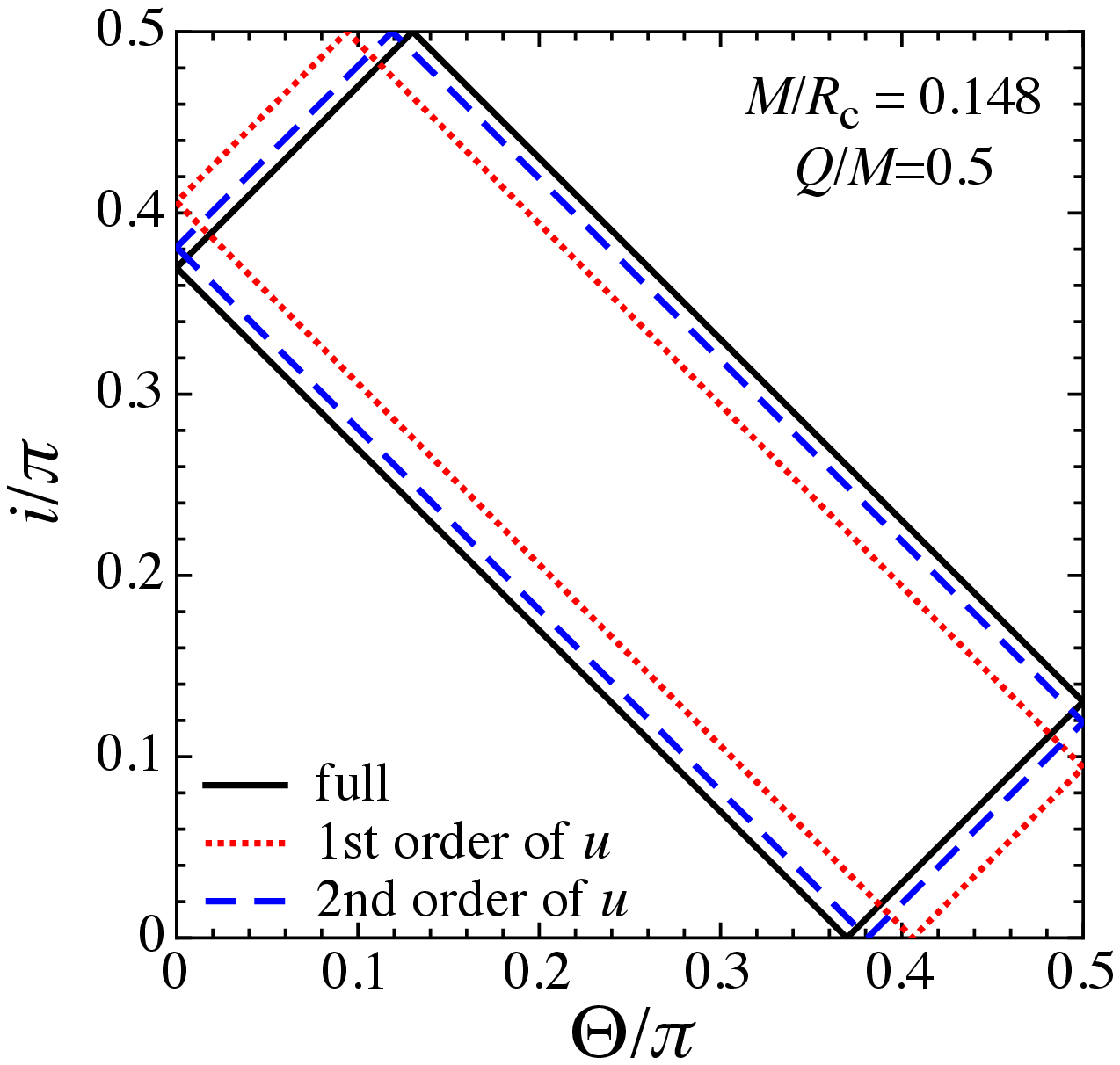} &
\includegraphics[scale=0.4]{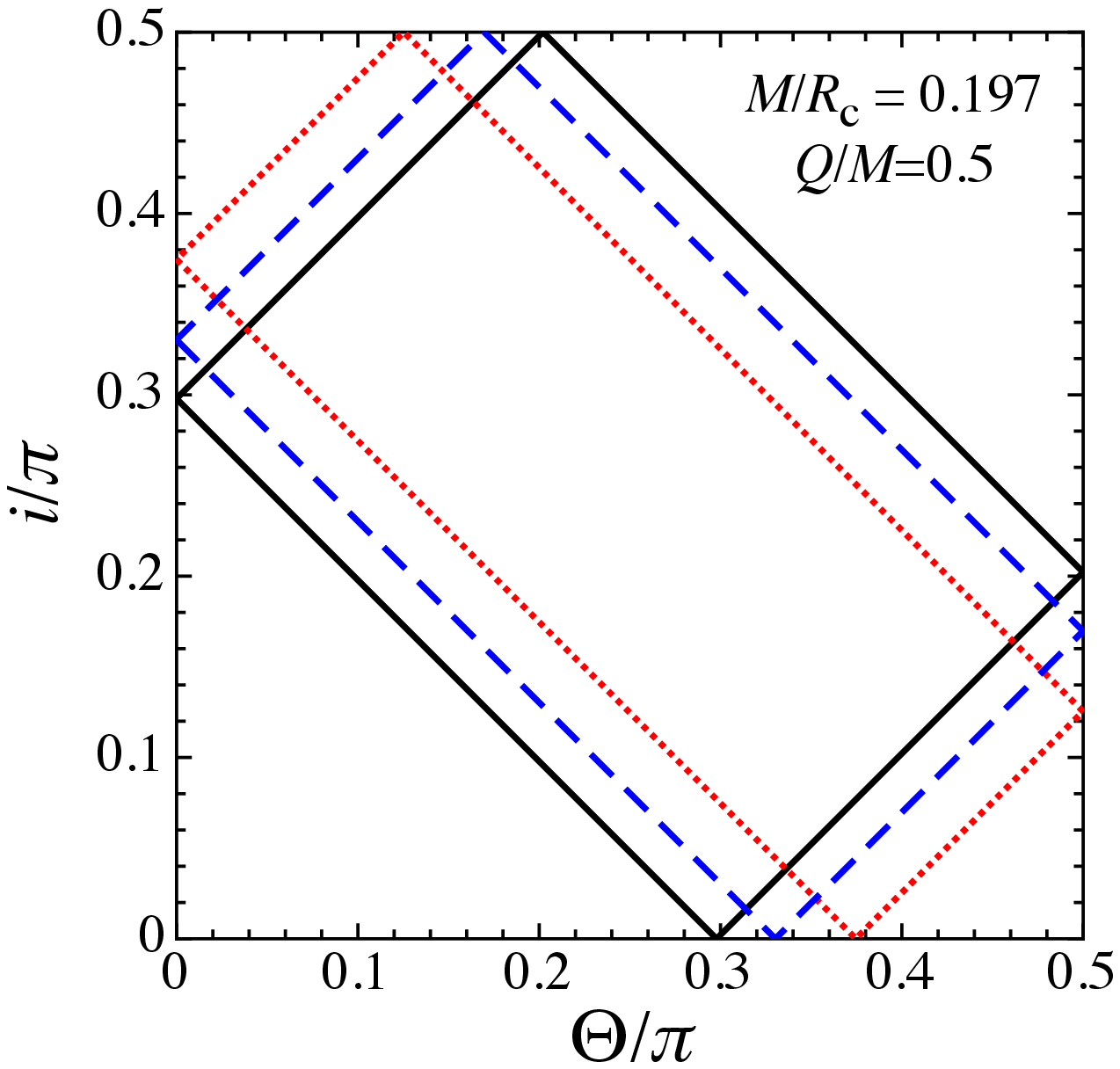} &
\includegraphics[scale=0.4]{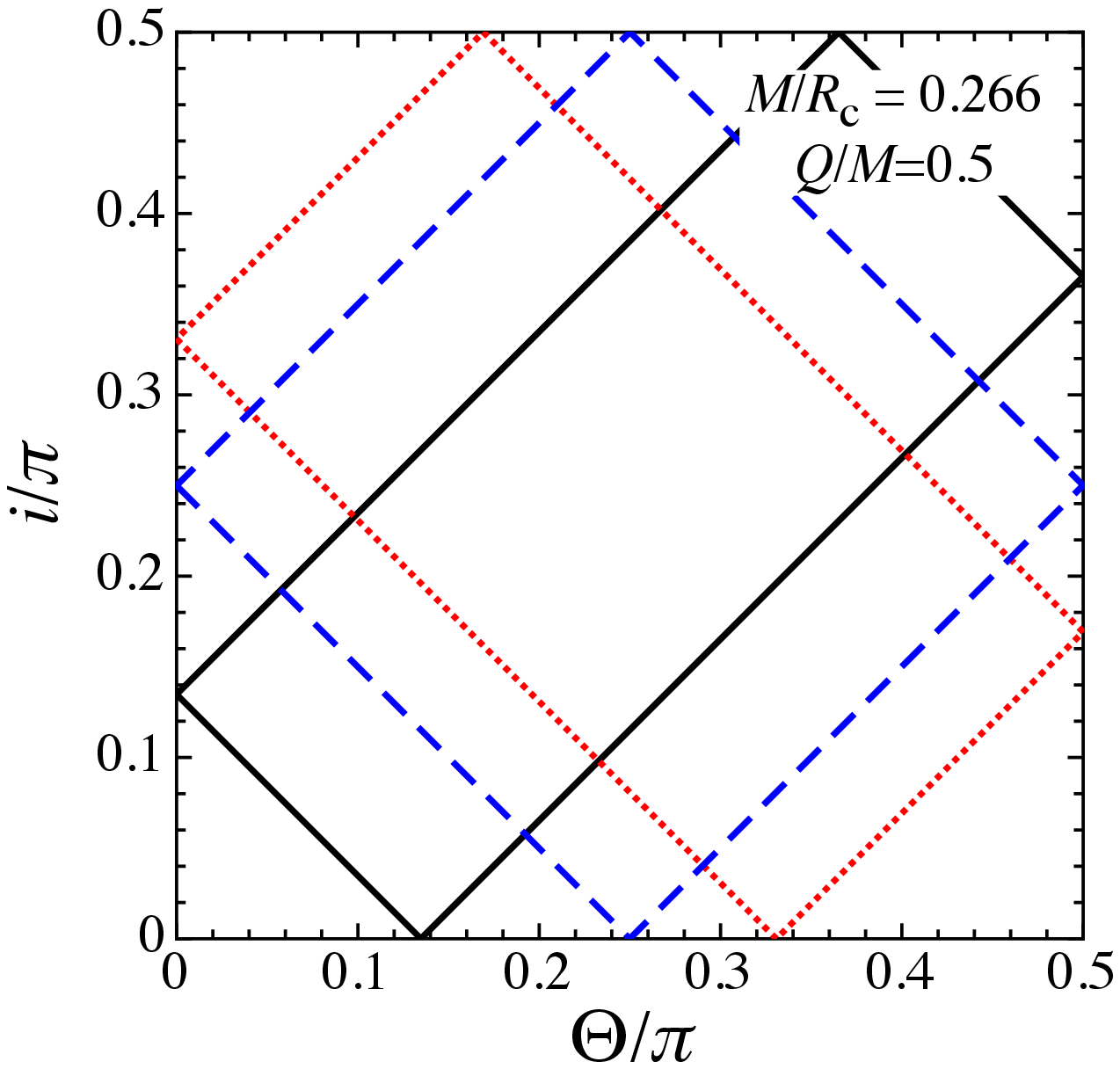} \\
\includegraphics[scale=0.4]{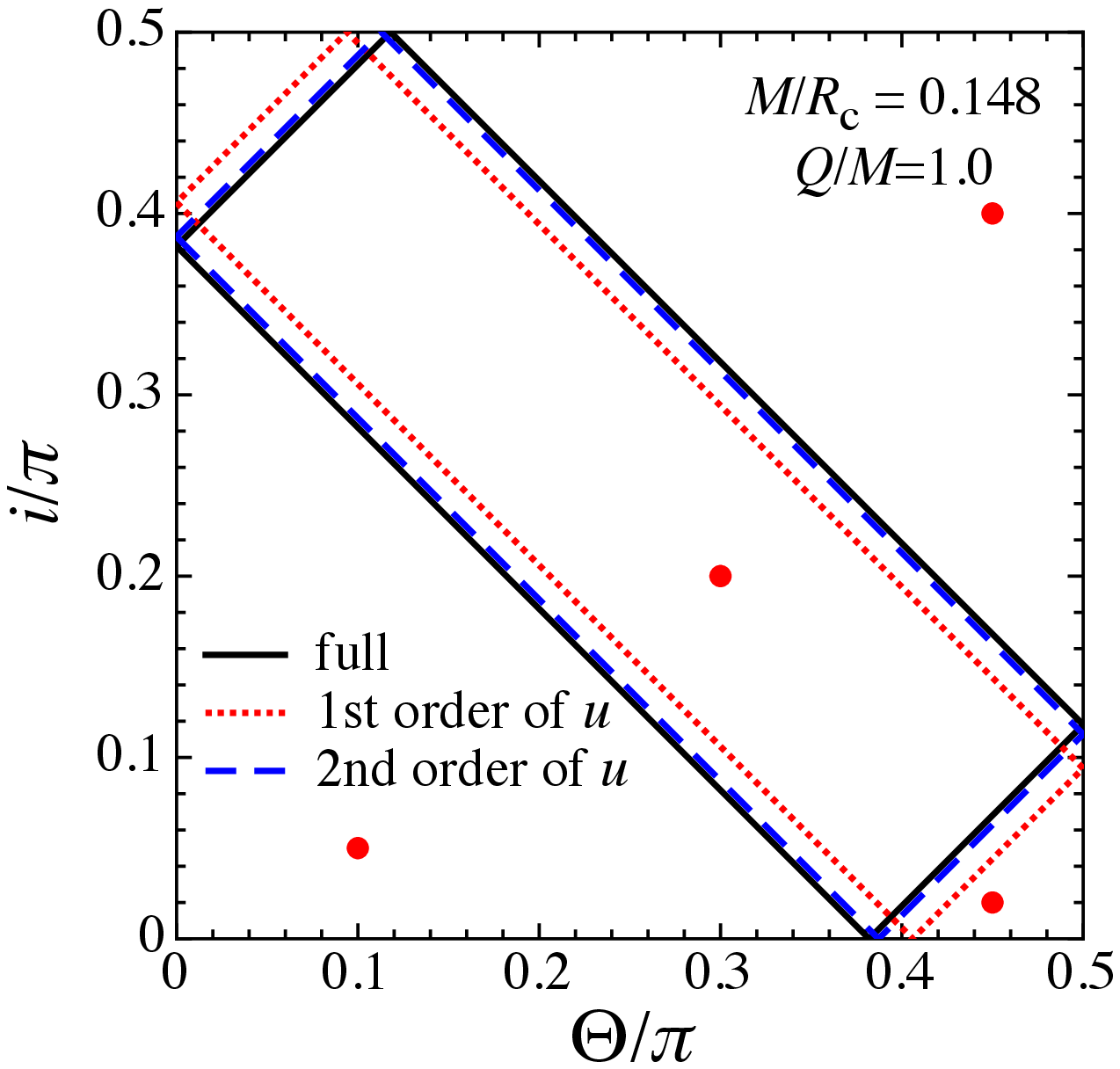} &
\includegraphics[scale=0.4]{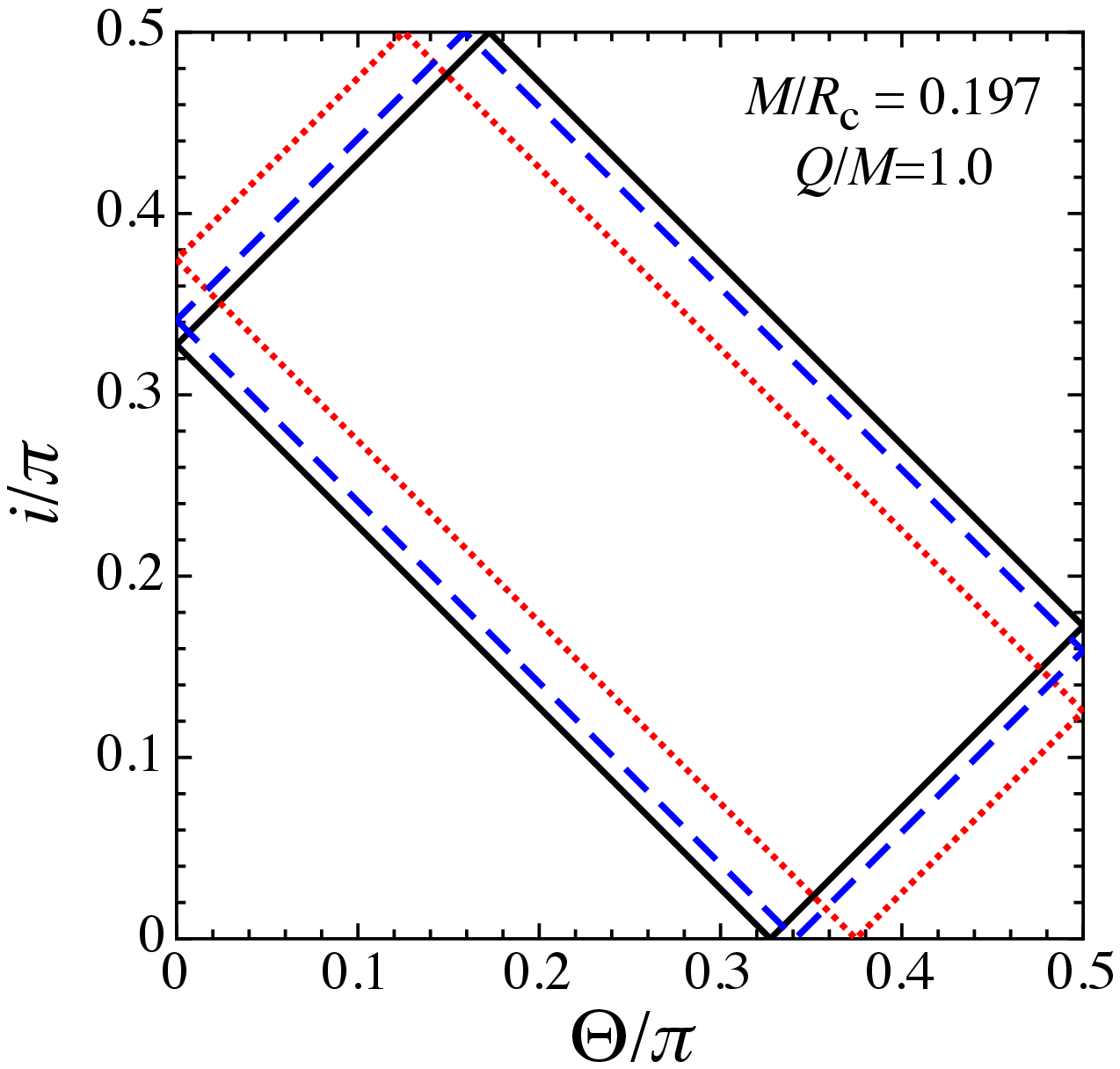} &
\includegraphics[scale=0.4]{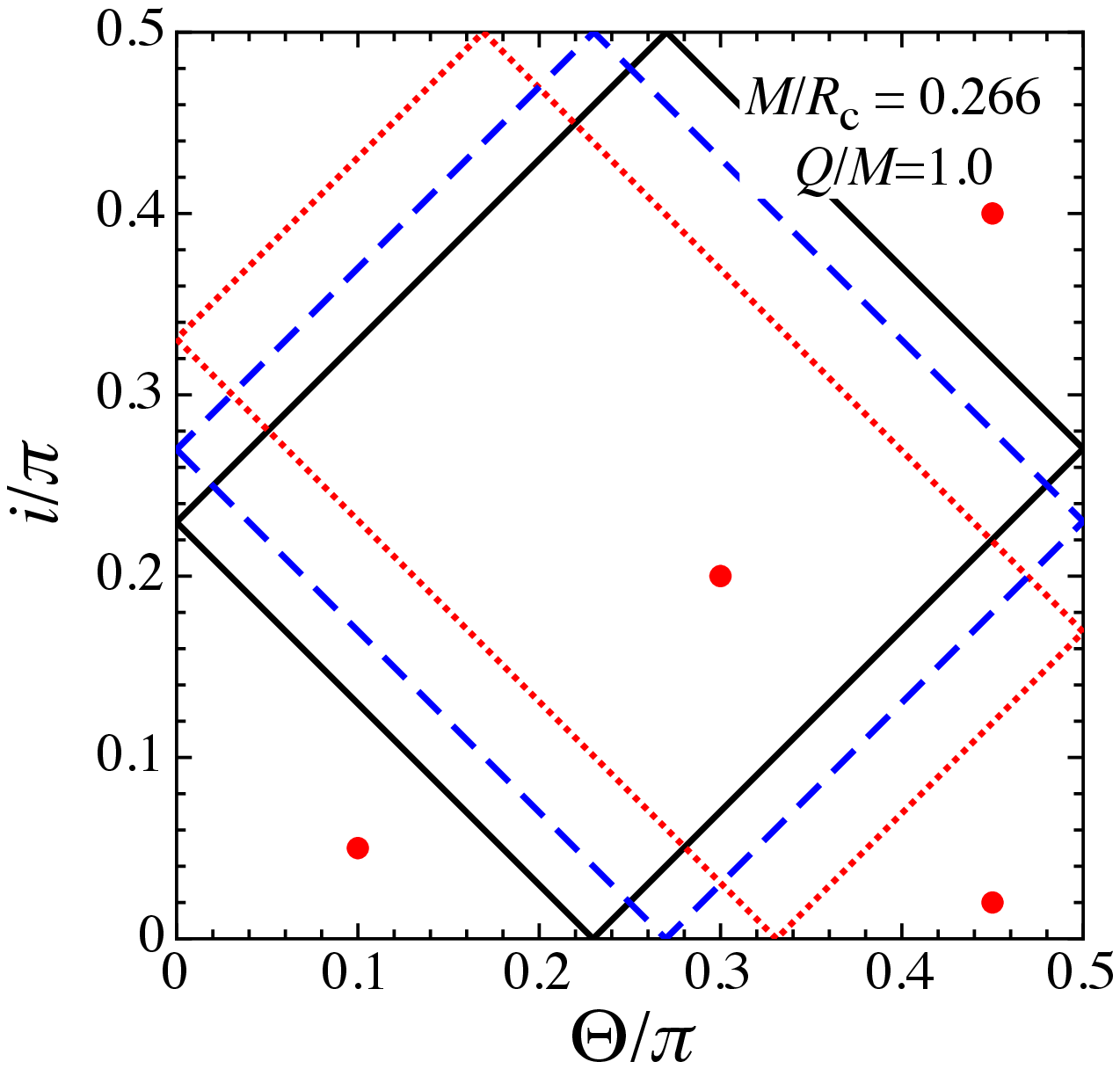}
\end{tabular}
\end{center}
\caption{%%
Same as Fig.~\ref{fig:ithetaS}, but for the Reissner-Nordstr\"{o}m spacetime. The upper and lower panels correspond to the cases for $Q/M=0.5$ and 1.0, respectively.
}%%
\label{fig:ithetaRN}
\end{figure*}

With the value of $\psi_{\rm cri}$ obtained for each stellar model, one can draw a similar figure to Fig.~\ref{fig:itheta0}. As an example, in Fig.~\ref{fig:ithetaRN} we show the classification whether the two hot spots are observed or not depending on the set of angles $(\Theta/\pi,i/\pi)$ for the stellar models with $M/R_c=0.148$, $0.197$, and $0.266$, where the upper and lower panels correspond to the cases with $Q/M=0.5$ and $1.0$. As shown in Fig.~\ref{fig:critical_RN},  since the deviation in $\psi_{\rm cri}$ between the results with the full order numerical integration and with the approximate relations becomes small as $Q/M$ increases for a given stellar model, the deviation in the classification whether the two hot spots are observed or not in $(\Theta/\pi,i/\pi)$-planes also becomes small as $Q/M$ increases, although that is still significant especially for the stellar model with higher compactness.

%%%%%%%%%%%%%%%%%%%%%%%%%%%%%%%%%%%
% Figure 17
%%%%%%%%%%%%%%%%%%%%%%%%%%%%%%%%%%%
%%%%%%%%%%%%%%%%%%%%%%%%%%%%%%%%%%%

\begin{figure*}
\begin{center}
\begin{tabular}{c}
\includegraphics[scale=0.4]{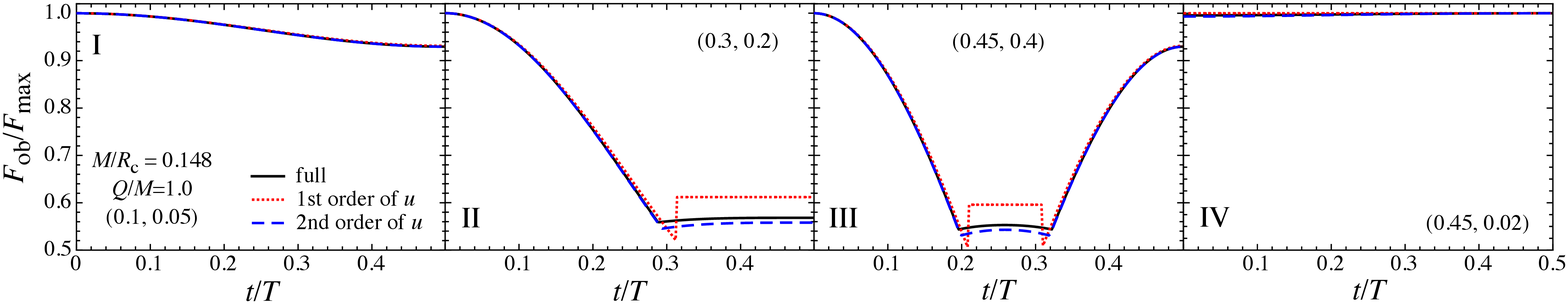} \\ 
\includegraphics[scale=0.4]{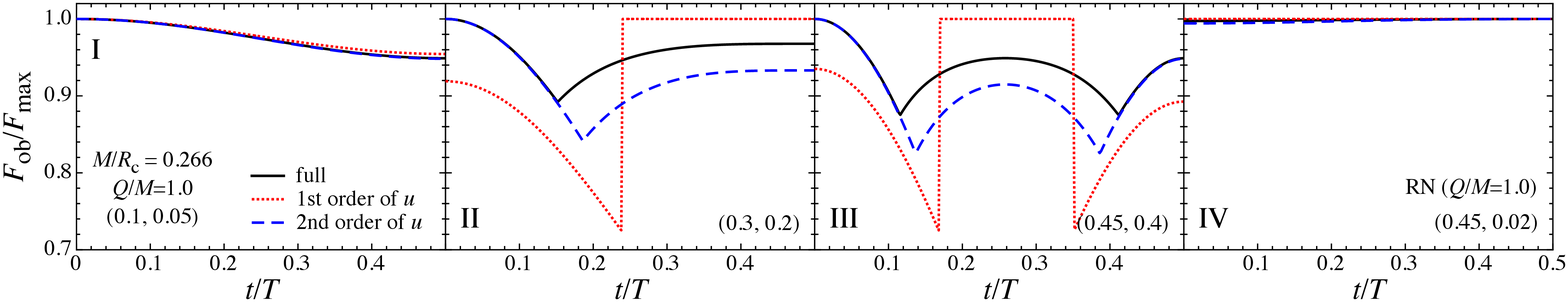} 
\end{tabular}
\end{center}
\caption{%%
Same as Fig.~\ref{fig:pulse-S}, for the Reissner-Nordstr\"{o}m spacetime with $Q/M=1.0$. 
}%%
\label{fig:pulse-RN10}
\end{figure*}

With the respect to the stellar models denoted in Fig.~\ref{fig:ithetaRN} with the dots, the observed flux is shown in Fig.~\ref{fig:pulse-RN10}, where the upper and lower panels correspond to the results for the stellar models with $M/R_c=0.148$ and $0.266$, while the panels from left to right for each stellar model correspond to the profiles with $(\Theta/\pi,i/\pi)=(0.1, 0.05)$, $(0.3, 0.2)$, $(0.45, 0.4)$, and $(0.45, 0.02)$. We remark that the case with $Q/M=1.0$ is considered here, because this is the extreme case. In the same way as in the Schwarzschild spacetime, one can observe the jump in the results obtained with the 1st order approximation in the panels for $(\Theta/\pi,i/\pi)=(0.3, 0.2)$ and $(0.45, 0.4)$, which corresponds to the moment when the antipodal spot comes into the visible zone on the stellar surface. Additionally, we find that even for the stellar model with $M/R_c=0.266$, the pulse profile with the 2nd order approximation is qualitatively similar to that with the numerical integration.

%%%%%%%%%%%%%%%%%%%%%%%%%%%%%%%%%%%
% Figure 18
%%%%%%%%%%%%%%%%%%%%%%%%%%%%%%%%%%%
%%%%%%%%%%%%%%%%%%%%%%%%%%%%%%%%%%%

%%%%%%%%%%%%%%%%%%%%%%%%%%%%%%%%%%%%%%%%%%%%%%%%
\subsection{Garfinkle-Horowitz-Strominger spacetime}
\label{sec:V-c}
%%%%%%%%%%%%%%%%%%%%%%%%%%%%%%%%%%%%%%%%%%%%%%%%

\begin{figure*}
\begin{center}
\begin{tabular}{cc}
\includegraphics[scale=0.5]{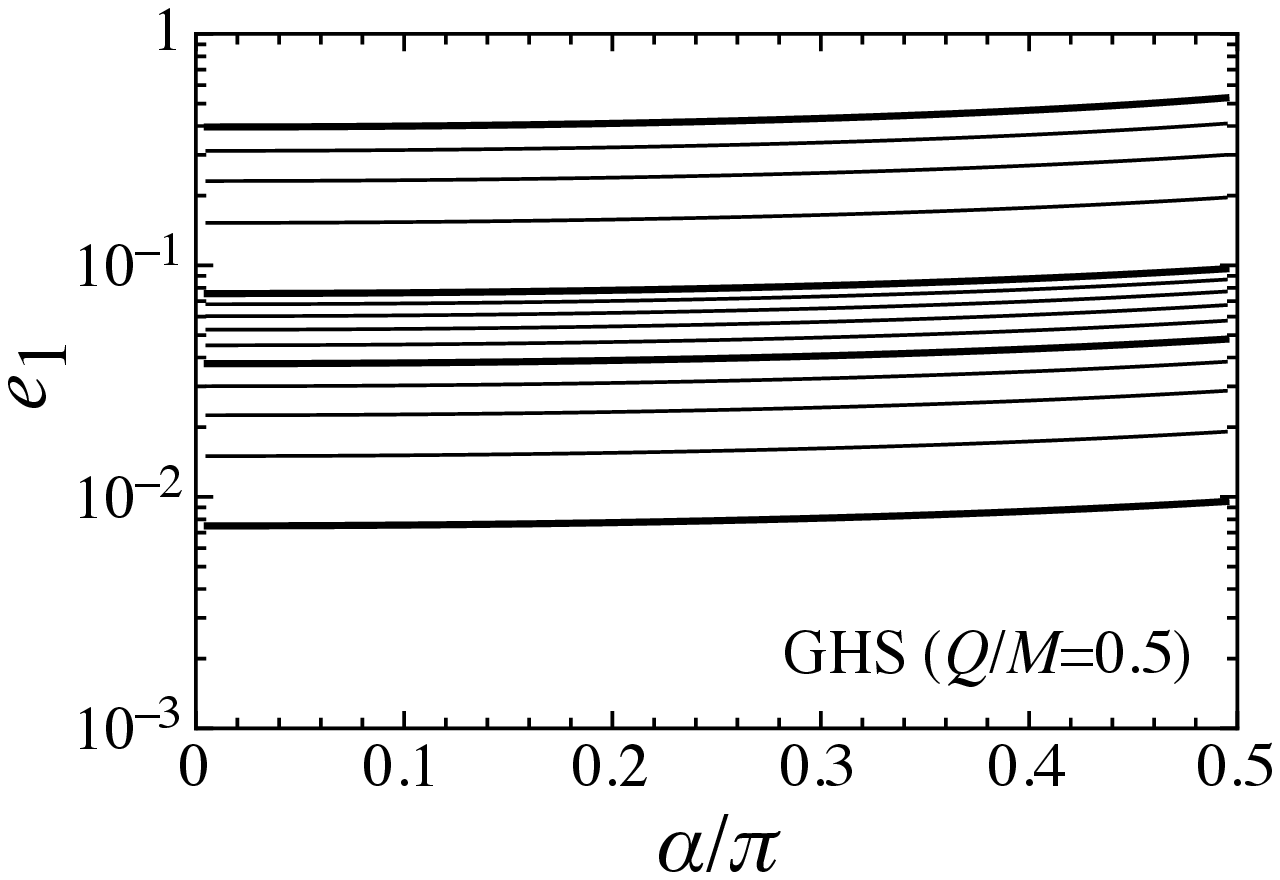} &
\includegraphics[scale=0.5]{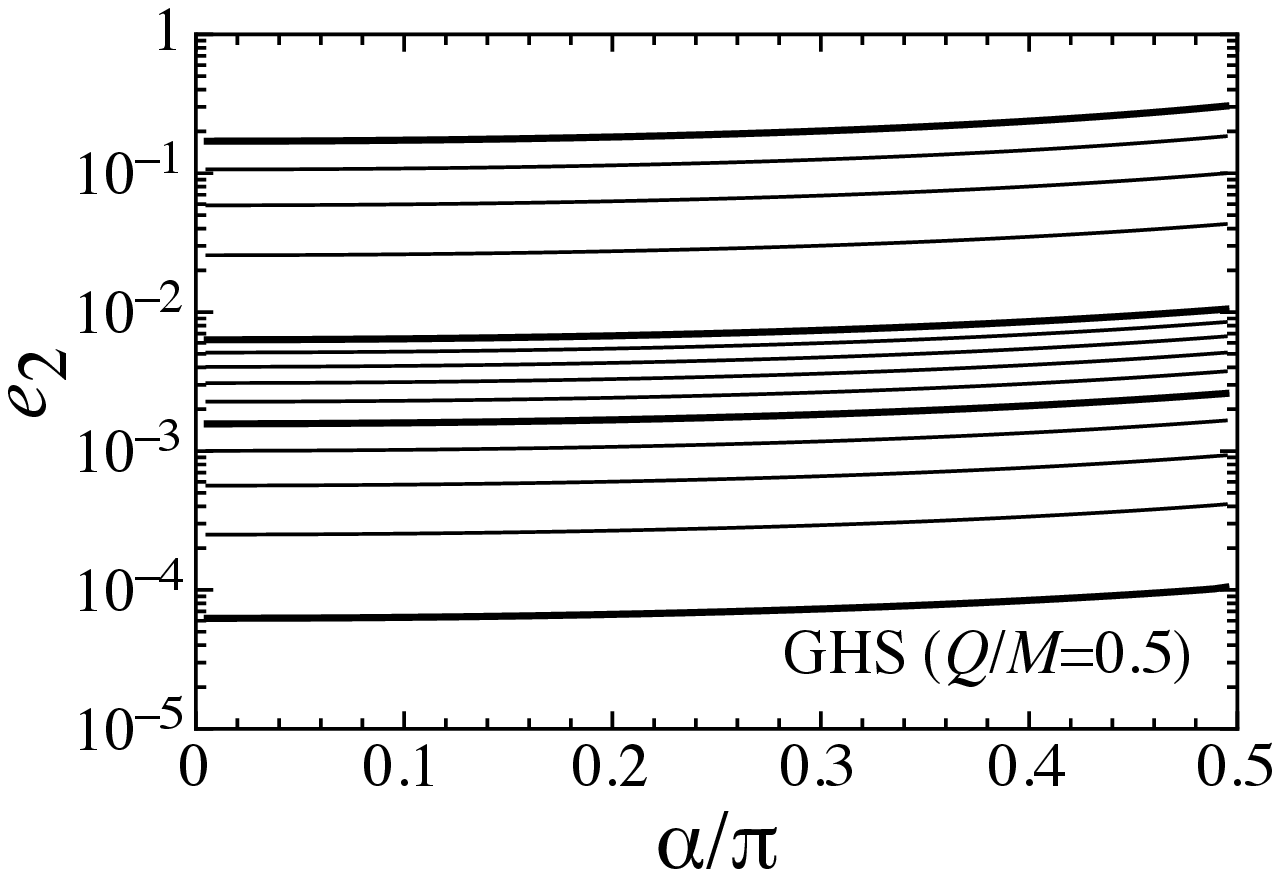} \\
\includegraphics[scale=0.5]{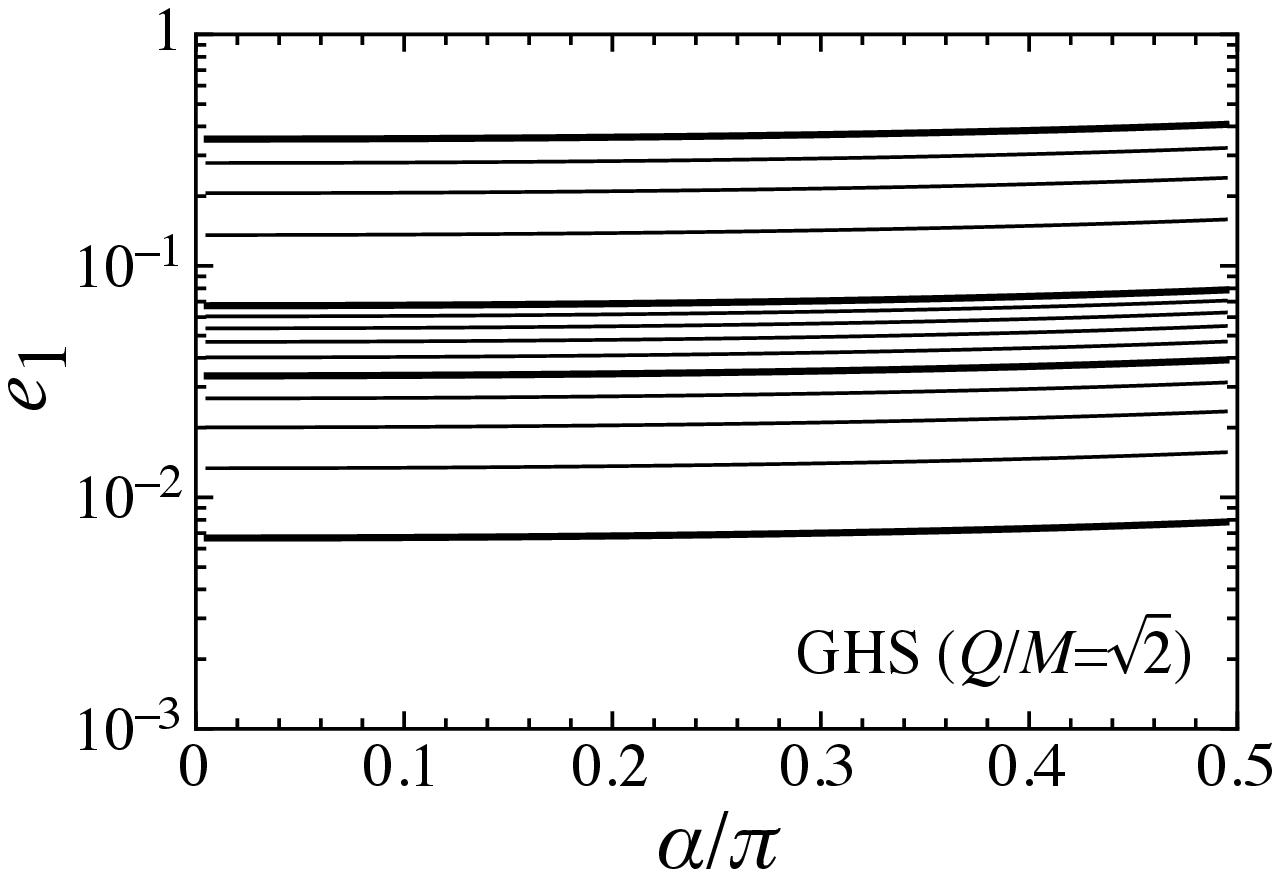} &
\includegraphics[scale=0.5]{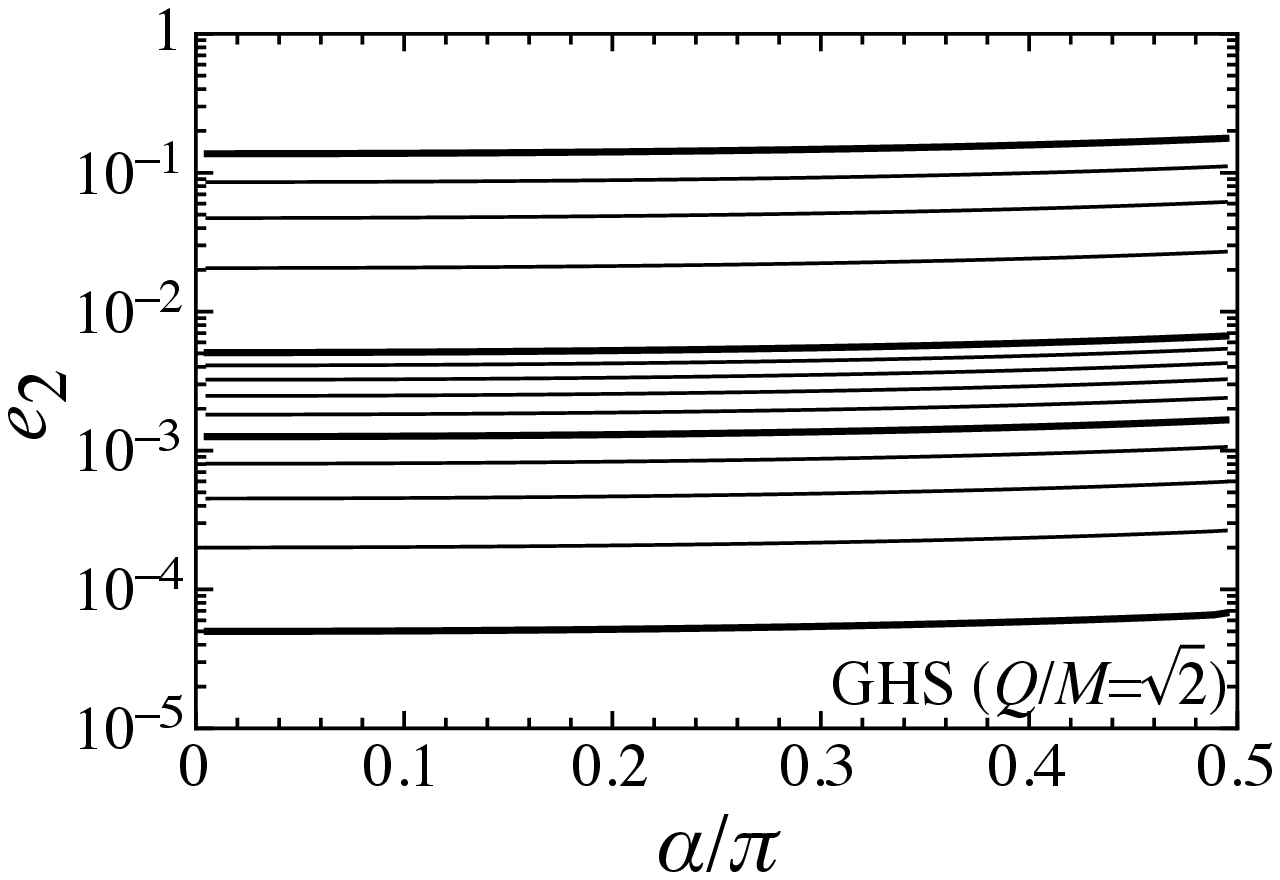} \\
\end{tabular}
\end{center}
\caption{%%
Same as Fig.~\ref{fig:aeS}, but for the Garfinkle-Horowitz-Strominger spacetime. The upper and lower panels correspond to the cases for $Q/M=0.5$ and $\sqrt{2}$, respectively.
}%%
\label{fig:aeGHS}
\end{figure*}

From the metric form, the gravitational radius is given by $r_g=2M$. As in the Schwarzschild and the Reissner-Nordstr\"{o}m spacetimes, we show the relative error in the bending angle $\beta$ for the Garfinkle-Horowitz-Strominger spacetime as a function of $\alpha$ in Fig.~\ref{fig:aeGHS}, where the upper and lower panels correspond to the results with $Q/M=0.5$ and $\sqrt{2}$. $e_1$ and $e_2$ correspond to the relative error of the results with the 1st order approximation and the 2nd order approximation from that with the full order numerical integration. Since the relative error in the bending angle for a typical neutron star with $M=1.4M_\odot$ and $R_c=12$ km, which corresponds to $u=0.337$ for $Q/M=0.5$ and $u=0.290$ for $Q/M=\sqrt{2}$, $e_1\sim 30\%$ and $e_2\sim 10\%$ for $Q/M=0.5$, while $e_1\sim 22\%$ and $e_2\sim 5\%$ for $Q/M=\sqrt{2}$, the relative errors decrease with $Q/M$.

%%%%%%%%%%%%%%%%%%%%%%%%%%%%%%%%%%%
% Figure 19
%%%%%%%%%%%%%%%%%%%%%%%%%%%%%%%%%%%
%%%%%%%%%%%%%%%%%%%%%%%%%%%%%%%%%%%

\begin{figure*}
\begin{center}
\begin{tabular}{cc}
\includegraphics[scale=0.5]{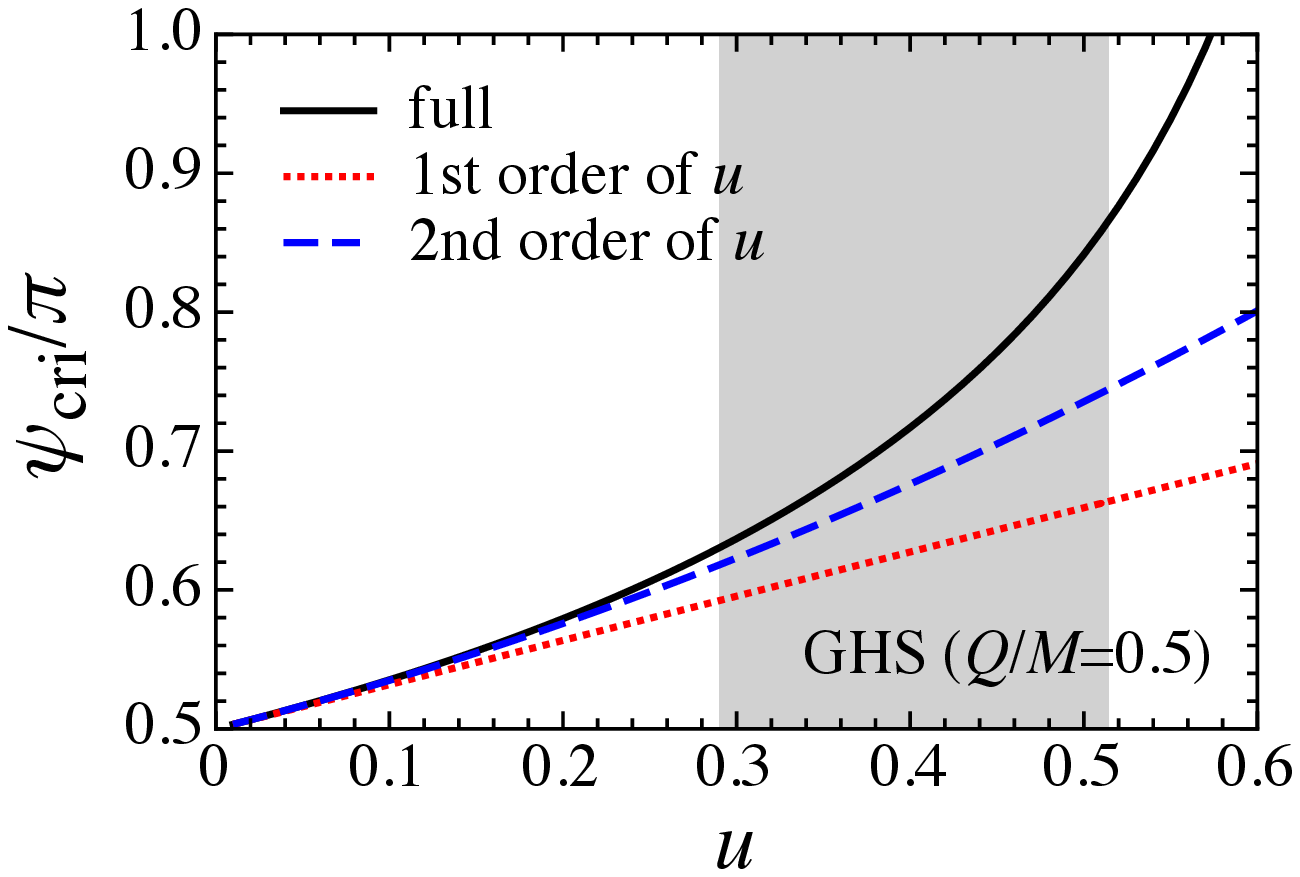} &
\includegraphics[scale=0.5]{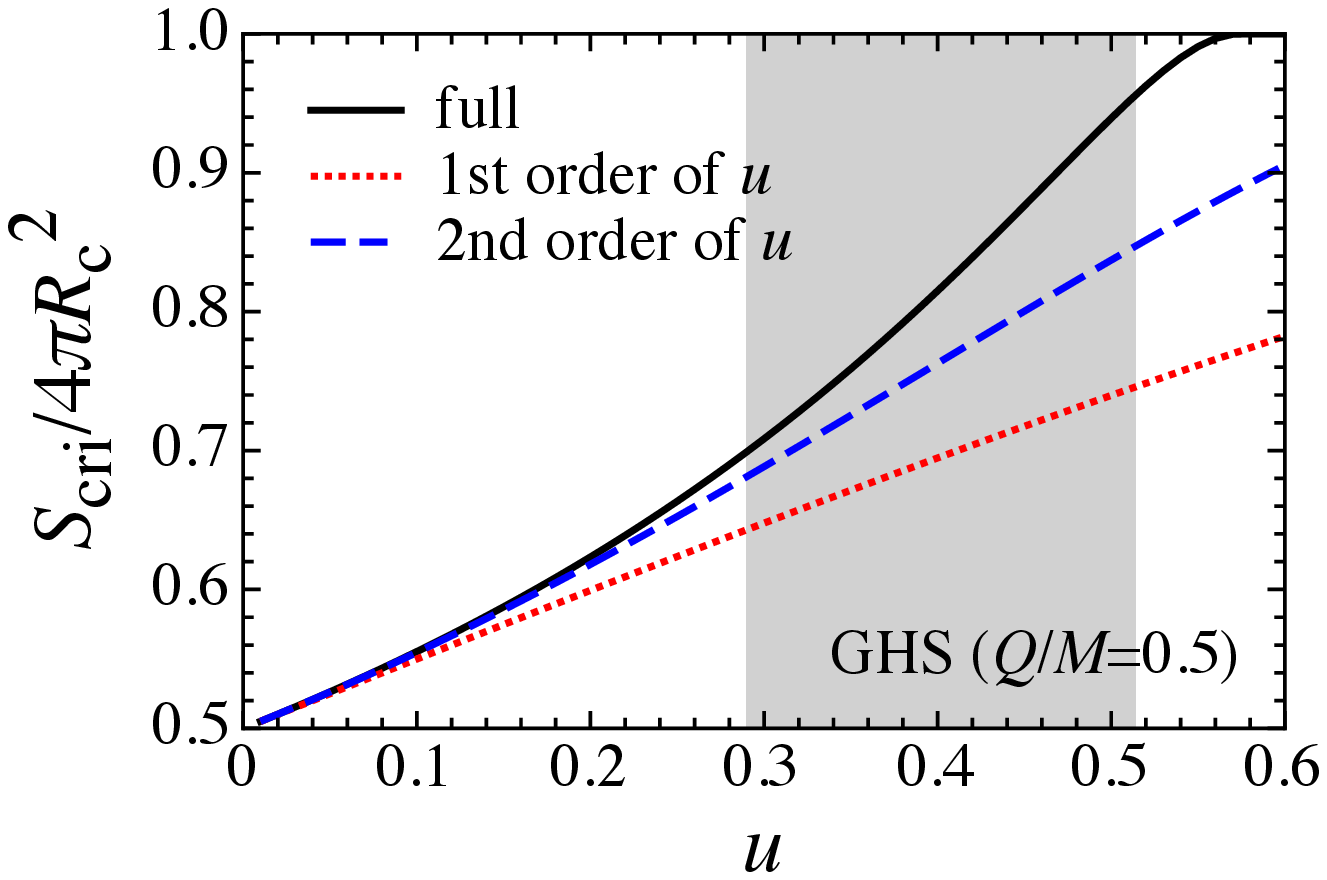} \\ 
\includegraphics[scale=0.5]{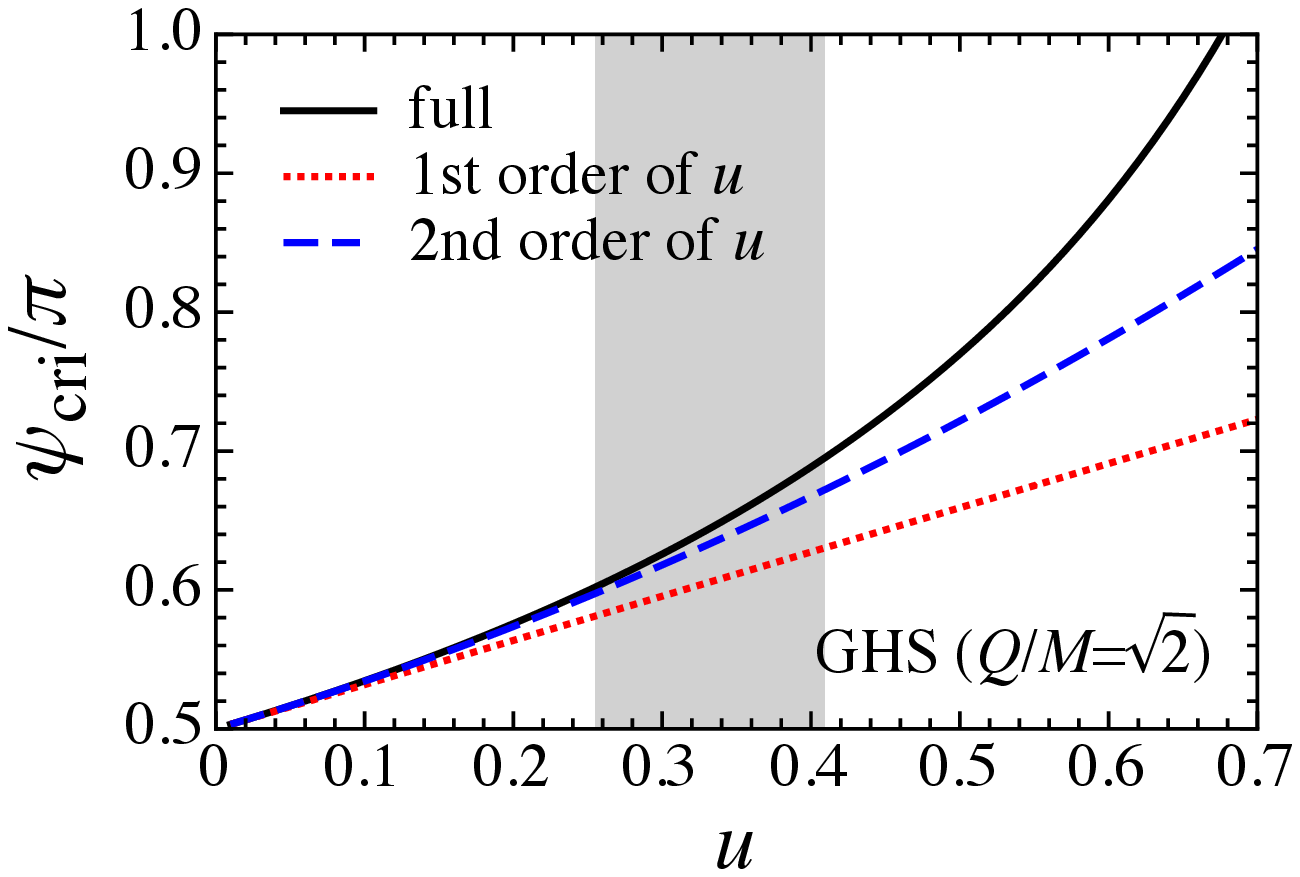} &
\includegraphics[scale=0.5]{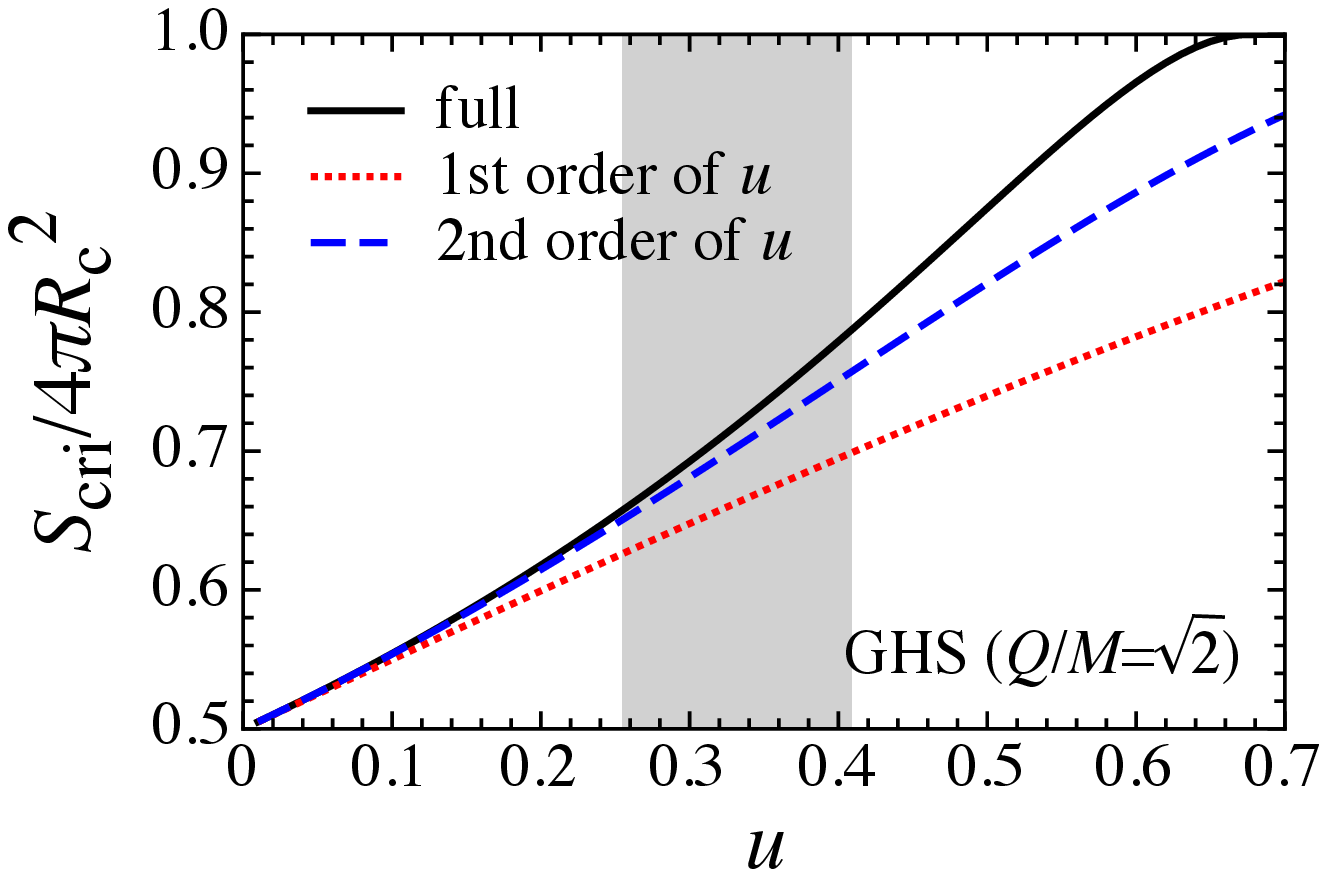} 
\end{tabular}
\end{center}
\caption{%%
Same as Fig.~\ref{fig:critical_S}, but for the Garfinkle-Horowitz-Strominger spacetime. The upper and lower panels correspond to the cases for $Q/M=0.5$ and $\sqrt{2}$, respectively.
}%%
\label{fig:critical_GHS}
\end{figure*}

In Fig.~\ref{fig:critical_GHS}, the critical value of $\psi$ and the visible fraction of stellar surface are shown as a function of $u$, where the upper and lower panels correspond to the cases for $Q/M=0.5$ and $\sqrt{2}$. For reference, the shaded region denotes the value of $u$ for the neutron star models with $R_c=10-14$ km and $M=1.4-1.8M_\odot$, i.e., $u$ becomes in the range of $0.290\le u\le0.514$ for $Q/M=0.5$ and $0.255\le u\le0.409$ for $Q/M=\sqrt{2}$. From this figure, one can observe that the critical value of $u$ where $\psi_{\rm cri}$ is $\pi$ is 0.573 and 0.676 for $Q/M=0.5$ and $\sqrt{2}$, respectively. That is, this critical value $u$ increases as $Q/M$ increases, while the value of $u$ for a specific stellar model decreases as $Q/M$ increases, which leads to the results that $\psi_{\rm cri}$ decreases as $Q/M$ increases.

%%%%%%%%%%%%%%%%%%%%%%%%%%%%%%%%%%%
% Figure 20
%%%%%%%%%%%%%%%%%%%%%%%%%%%%%%%%%%%
%%%%%%%%%%%%%%%%%%%%%%%%%%%%%%%%%%%

\begin{figure*}
\begin{center}
\begin{tabular}{ccc}
\includegraphics[scale=0.4]{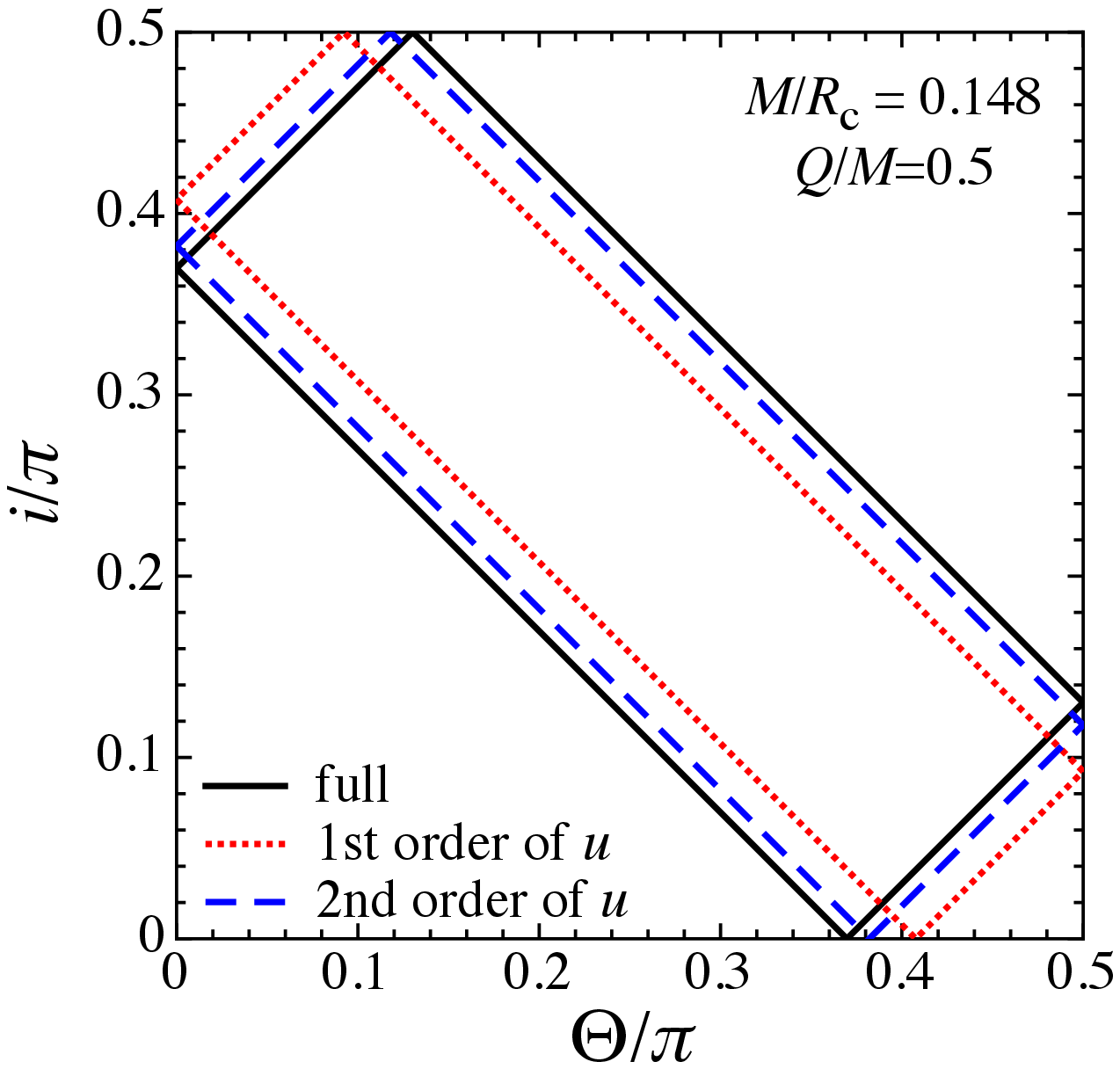} &
\includegraphics[scale=0.4]{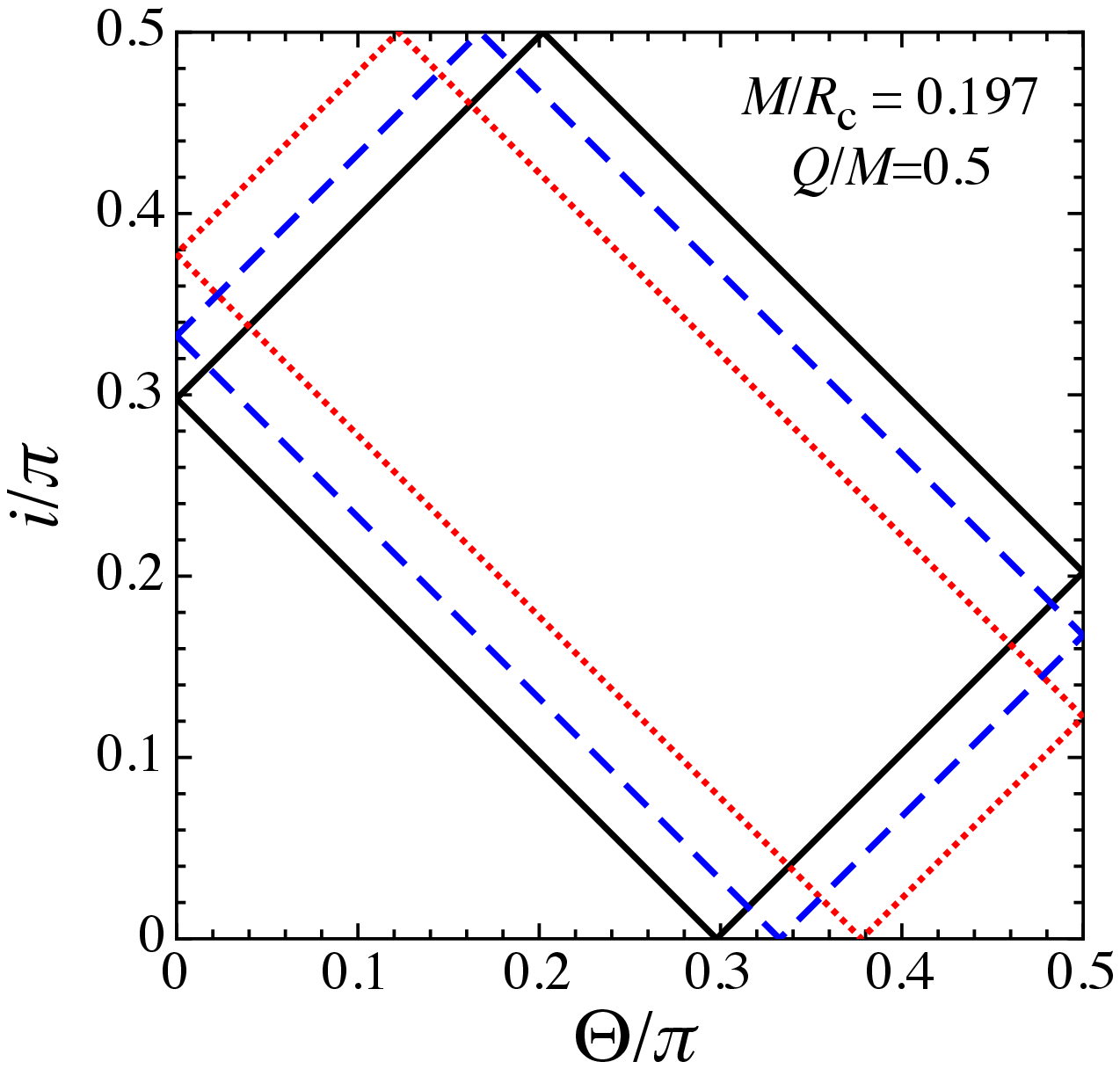} &
\includegraphics[scale=0.4]{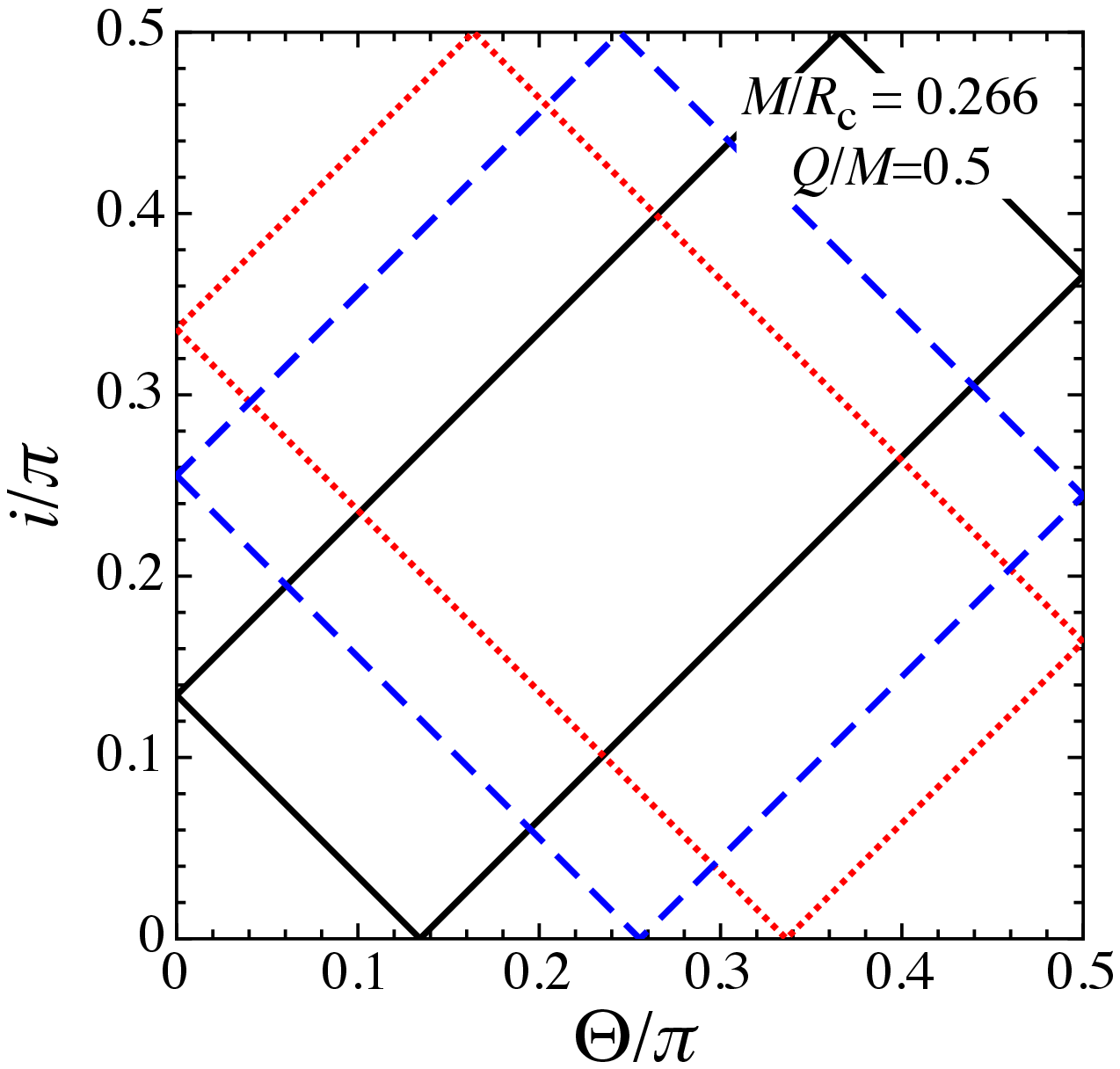} \\
\includegraphics[scale=0.4]{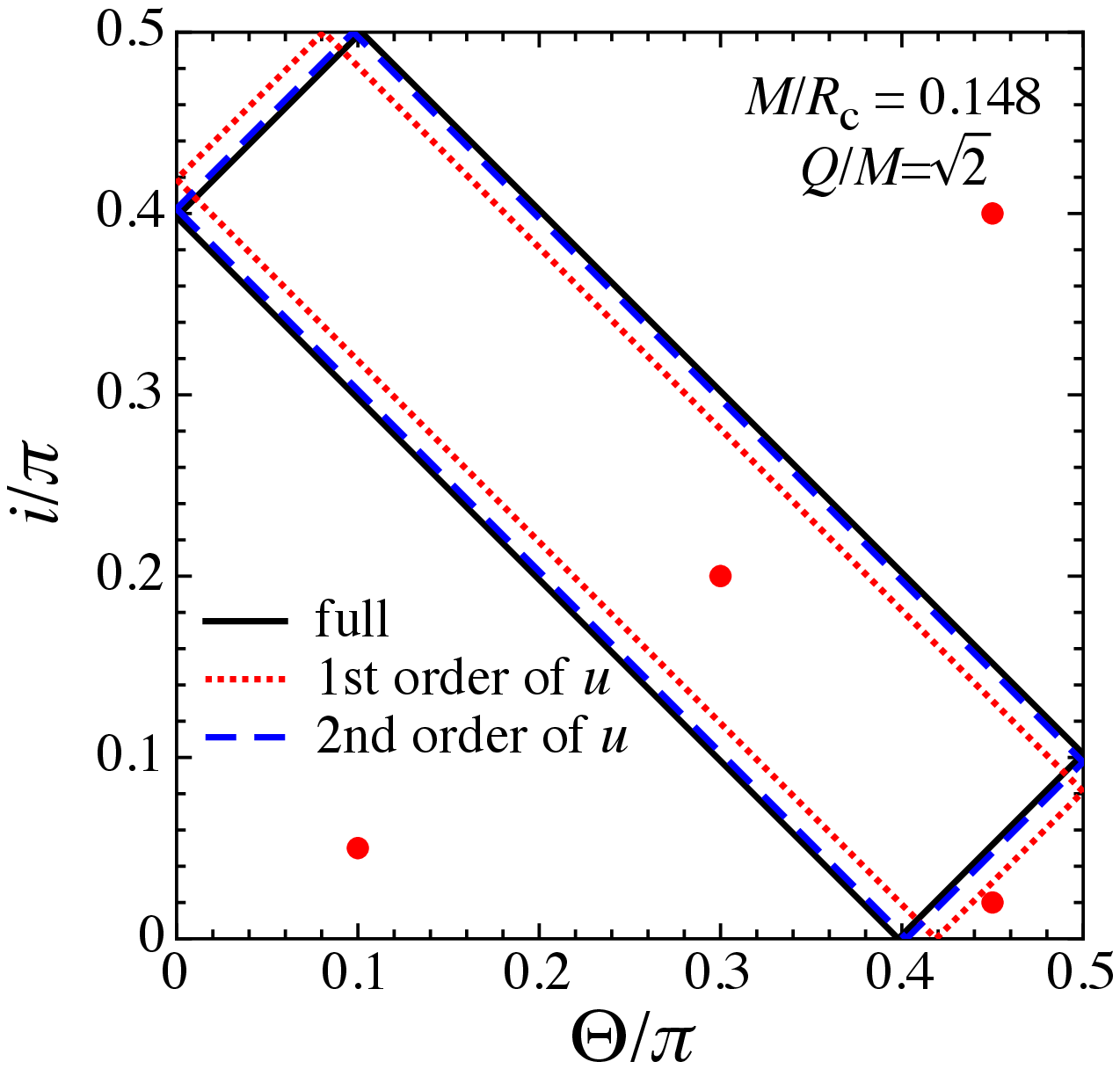} &
\includegraphics[scale=0.4]{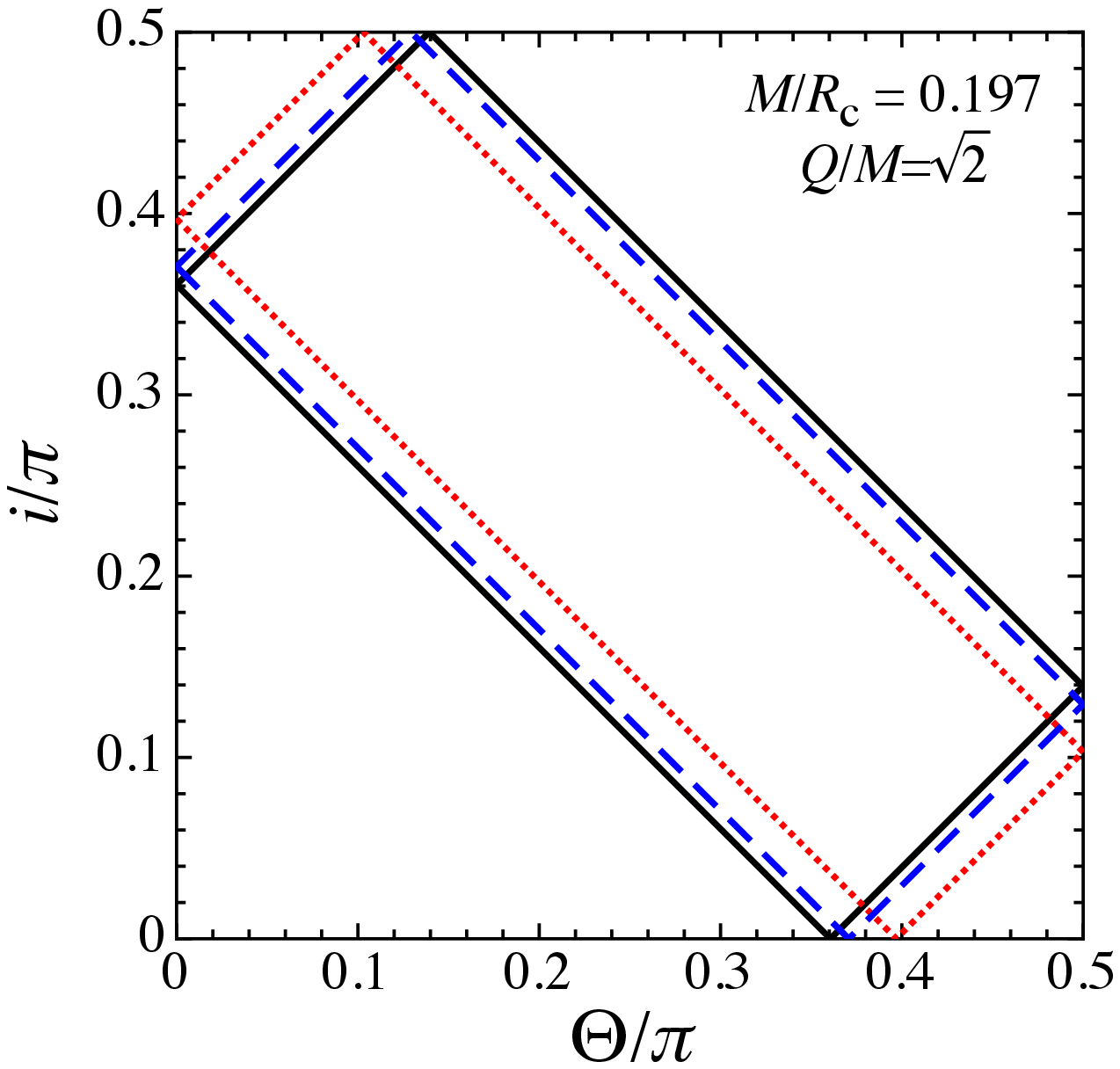} &
\includegraphics[scale=0.4]{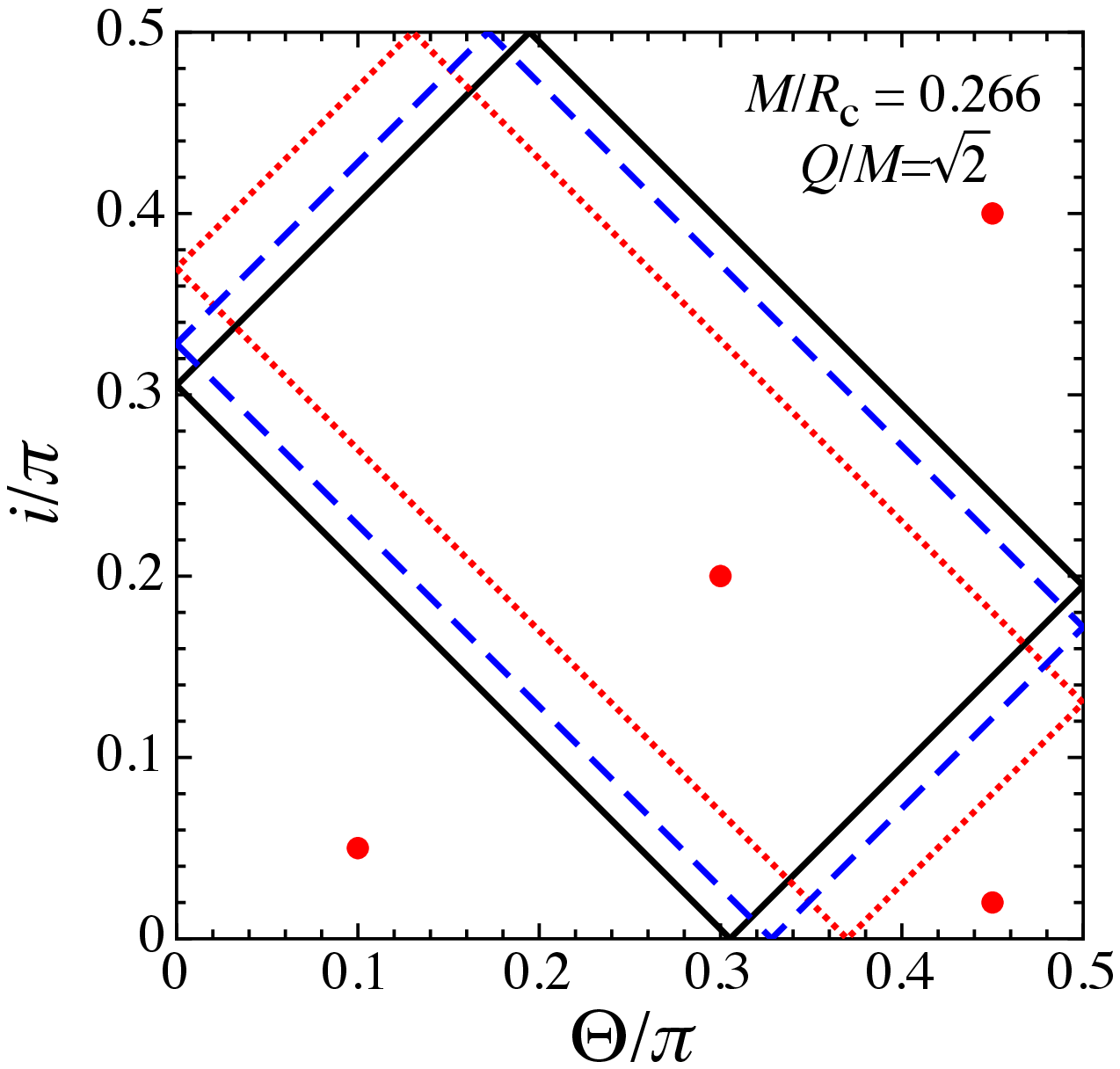}
\end{tabular}
\end{center}
\caption{%%
Same as Fig.~\ref{fig:ithetaS}, but for the Garfinkle-Horowitz-Strominger spacetime. The upper and lower panels correspond to the cases for $Q/M=0.5$ and $\sqrt{2}$, respectively.
}%%
\label{fig:ithetaGHS}
\end{figure*}

With the obtained value of $\psi_{\rm cri}$, in Fig.\ref{fig:ithetaGHS} we show the classification whether the two hot spots are observed or not, depending on the angles of $\Theta$ and $i$. The upper and lower panels correspond to the results with $Q/M=0.5$ and $\sqrt{2}$, while the panels for each value of $Q/M$ from left to right correspond to the results for the stellar models with $M/R_c=0.148$, $0.197$, and $0.266$. As for Reissner-Nordstr\"{o}m spacetime, the deviation between the results with the approximations and with the full order numerical integration decreases as $Q/M$ increases.

%%%%%%%%%%%%%%%%%%%%%%%%%%%%%%%%%%%
% Figure 21
%%%%%%%%%%%%%%%%%%%%%%%%%%%%%%%%%%%
%%%%%%%%%%%%%%%%%%%%%%%%%%%%%%%%%%%

\begin{figure*}
\begin{center}
\begin{tabular}{c}
\includegraphics[scale=0.4]{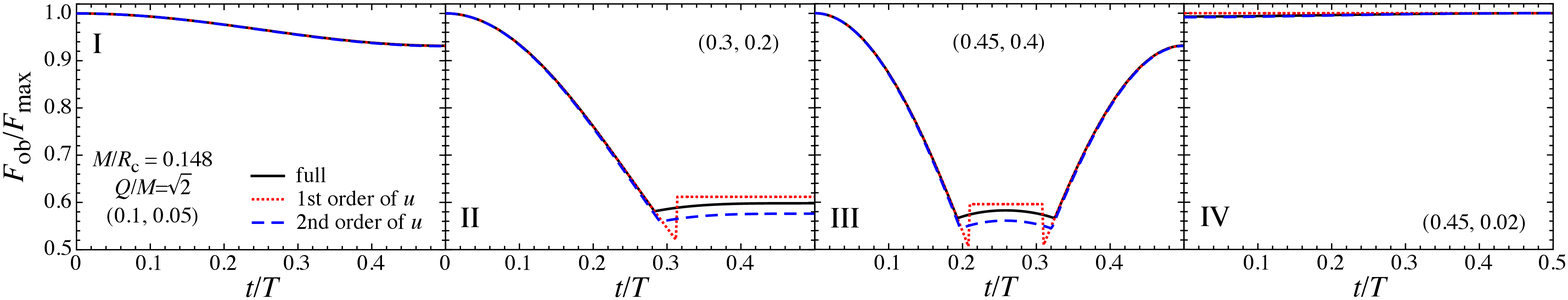} \\
\includegraphics[scale=0.4]{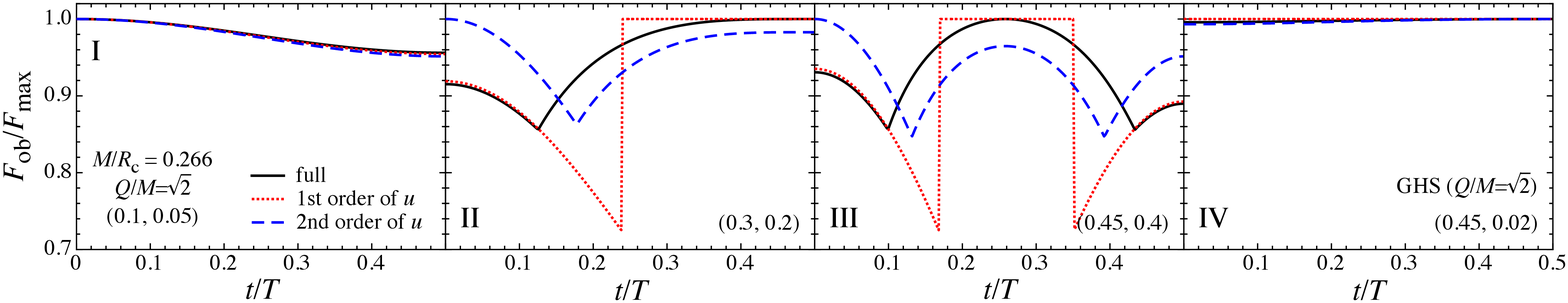}
\end{tabular}
\end{center}
\caption{%%
Same as Fig.~\ref{fig:pulse-S}, for the Garfinkle-Horowitz-Strominger spacetime with $Q/M=\sqrt{2}$. 
}%%
\label{fig:pulse-GHS14}
\end{figure*}

The pulse profiles from the stellar models denoted in Fig.~\ref{fig:ithetaGHS} with the dots, are shown in Fig.~\ref{fig:pulse-GHS14}. Again, one can observe the jump in the profile with the 1st order approximation at the moment when the antipodal spot comes into the visible zone. For the compact stellar model with $M/R_c=0.266$, since the amplitude when the both spots are visible is not at all constant with the full order numerical integration, the deviation of the result with the 1st order approximation from that with the full order integration becomes large. On the other hand, the 2nd order approximation more or less expresses well the pulse profiles.

%%%%%%%%%%%%%%%%%%%%%%%%%%%%%%%%%%%
% Figure 22
%%%%%%%%%%%%%%%%%%%%%%%%%%%%%%%%%%%
%%%%%%%%%%%%%%%%%%%%%%%%%%%%%%%%%%%

As a short summary via the examinations with a few different spacetimes, we find that the relative error in the bending angle $\beta$ becomes $e_1\sim 20-30\%$ and $e_2\sim 5-10\%$ for the typical neutron star model with $M=1.4M_\odot$ and $R_c=12$ km. The deviation in the critical value of $\psi$ for $\alpha=\pi$ obtained by the approximate relations and by the full order integration, increases with $u$, i.e., the stellar compactness. As a result, one can see the significant deviation in the classification whether the two hot spots are observed ot not for the stellar model with higher compactness. In any way, one can observe the jump in the pulse profile obtained with the 1st order approximation at the moment when the antipodal spot comes into the visible zone.
This is because the observed flux with the 1st order approximation becomes constant independently of angle $\psi$ as Eq.~(\ref{eq:Fob_1}), when the both hot spots are visible. That is, at the moment when the antipodal hot spot comes in the visible zone, the flux from the antipodal spot is not zero but has a finite value already.
On the other hand, the shape of pulse profile with the 2nd order approximation seems to be qualitatively similar to that with the full order integration, when the angles of $\Theta$ and $i$ are selected in such a way that the classification with the full order integration is the same as that with the 2nd order approximation.

%%%%%%%%%%%%%%%%%%%%%%%%%%%%%%%%%%%%%%%%%%%%%%%%

\end{document}